%% file: main.tex
\documentclass{article}

\PassOptionsToPackage{dvipsnames,table}{xcolor}
\PassOptionsToPackage{section,below}{placeins}
\PassOptionsToPackage{inline}{enumitem}
\PassOptionsToPackage{most}{tcolorbox}

\usepackage[arxiv1col,nonatbib,notheorems]{styles/paperkit}

\input{math_commands.tex}

\usepackage{url}
\usepackage{common} %

\newcommand{\Lara}{\textsc{Lara}\xspace}

\newcommand{\casestudycodesize}{\scriptsize}

\newcommand{\stJustified}{\textsf{justified}\xspace}
\newcommand{\stGap}{\textsf{gap}\xspace}
\newcommand{\stDefeated}{\textsf{defeated}\xspace}
\newcommand{\stContested}{\textsf{contested}\xspace}
\newcommand{\stEvidenceBlocked}{\textsf{evidence-blocked}\xspace}

\newcommand{\upAddLeaf}{\textsf{add-leaf}\xspace}
\newcommand{\upTighten}{\textsf{tighten}\xspace}
\newcommand{\upAddAttack}{\textsf{add-attack}\xspace}
\newcommand{\upAddInstance}{\textsf{add-instance}\xspace}

\newcommand{\atRebut}{\textsf{rebut}\xspace}
\newcommand{\atUndercut}{\textsf{undercut}\xspace}
\newcommand{\atUndermine}{\textsf{undermine}\xspace}

\newcommand{\lkObserved}{\textsf{observed}\xspace}
\newcommand{\lkAttested}{\textsf{attested}\xspace}
\newcommand{\lkAssumed}{\textsf{assumed}\xspace}
\newcommand{\lkCertified}{\textsf{certified}\xspace}

\newcommand{\supports}{\ensuremath{\mathsf{supports}}}
\newcommand{\concl}{\operatorname{concl}}
\newcommand{\leaves}{\operatorname{leaves}}
\newcommand{\certDeps}{\operatorname{certDeps}}
\newcommand{\nf}{\operatorname{nf}}
\newcommand{\Sig}{\ensuremath{\Sigma}}
\newcommand{\Pol}{\ensuremath{\Pi}}
\newcommand{\Ctx}{\ensuremath{\Gamma}}
\newcommand{\Reg}{\ensuremath{R}} %
\newcommand{\enc}[1]{\ensuremath{\mathsf{encode}_{#1}}}
\newcommand{\chk}[1]{\ensuremath{\mathsf{check}_{#1}}}
\newcommand{\backend}{\ensuremath{\beta}}

\newcommand{\rulename}[1]{\textsc{#1}}
\newcommand{\kw}[1]{\ensuremath{\mathsf{#1}}}
\newcommand{\judgbox}[1]{\framebox{\ensuremath{#1}}}
\newcommand{\gor}{\ \mid\ }

\newcommand{\ndjudg}{\ensuremath{\Delta \vdash e : \phi}}
\newcommand{\obl}{\mathrel{\vartriangleright}}
\newcommand{\supjudg}{\ensuremath{\Sig;\Pol;\Ctx;\Reg \vdash w : \supports(p) \obl O}}
\newcommand{\bmodels}{\ensuremath{\models_{\backend}}}
\newcommand{\attacks}{\ensuremath{\mathsf{attacks}}}

\makeatletter
\providecommand{\@proofnamefont}{\scshape}
\providecommand{\@proofindent}{\indent}
\newenvironment{proofof}[1]{\par
  \pushQED{\qed}%
  \normalfont \topsep6\p@\@plus6\p@\relax
  \trivlist
  \item[\@proofindent\hskip\labelsep]\ignorespaces
  {\@proofnamefont Proof of #1\@addpunct{.}}\hskip\labelsep\ignorespaces
}{%
  \popQED\endtrivlist\@endpefalse
}
\makeatother

\lstdefinelanguage{Lara}{%
  alsoletter={-},
  morekeywords={artifact, policy, use, backends, claim, leaf, arg, undercut, undermine, rebut, status, comparison, comparison-scheme, measurand, let},
  morekeywords=[2]{at, by, supports, challenges, discharge, with, open, as, on, where, cell, from, cert},
  morekeywords=[3]{nl, formal, binding, kind, provenance, refs, relation, recheck, bridge, result, baseline, claims, assurance},
  morekeywords=[4]{observed, attested, assumed, certified, ai-executed, reviewed, user, higher-is-better, lower-is-better, strictly-better, at-least-as-good},
  keywordstyle=\color{NavyBlue}\bfseries,
  keywordstyle=[2]\color{RoyalPurple},
  keywordstyle=[3]\color{OliveGreen},
  keywordstyle=[4]\color{Mahogany},
  string=[b]",
  showstringspaces=false,
  sensitive=true,
}

\usepackage[
    backend=biber,
    style=authoryear,
    natbib=true,
    maxcitenames=2,
    maxbibnames=5,
    sortcites=true,
    uniquename=false,
    uniquelist=false
]{biblatex}
\AtEveryBibitem{%
  \iffieldundef{doi}{%
    \iffieldundef{eprint}{}{\clearfield{url}}%
  }{\clearfield{url}}%
}

\graphicspath{{figures/}}

\papertitle[Beyond Natural Language: An Agent-Native Language for Autonomous Science]{%
  Beyond Natural Language:\\[2pt]
  An Agent-Native Language for Autonomous Science%
}
\paperrunningtitle{Beyond Natural Language: An Agent-Native Language for Autonomous Science}
\paperauthors{Yifeng He and Jiachen Liu}
\paperaffiliations{ARA Lab}
\paperabstract{\input{src/abstract}}
\papercorrespondence{%
  Yifeng He (\href{mailto:yfhe.cs@gmail.com}{yfhe.cs@gmail.com}),
  Jiachen Liu (\href{mailto:amber@ara-commons.com}{amber@ara-commons.com})%
}
\paperlogo{}{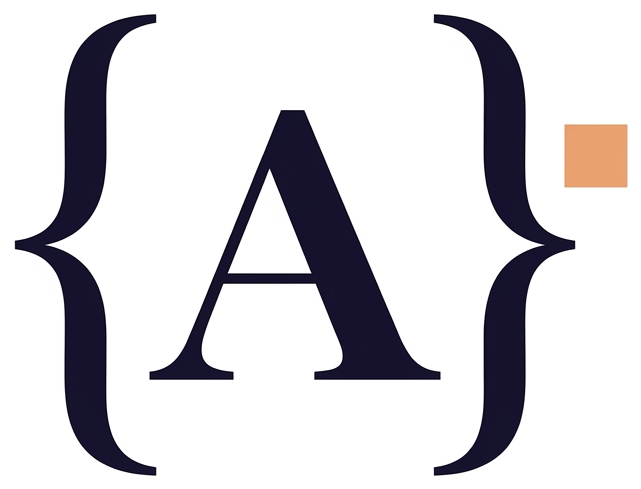}
\paperlogoheight{30pt}
\paperbrand[11pt]{figures/ara-logo.png}{ARA Lab}

\papercode{\href{https://github.com/ARA-Labs/Lara}{github.com/ARA-Labs/Lara}}

\setlist[enumerate]{leftmargin=*}
\setlist[itemize]{leftmargin=*}

\begin{document}

\makepaperheader

\input{src/intro}
\input{src/overview}
\input{src/language}
\input{src/semantics}
\input{src/worlds}
\input{src/metatheory}
\input{src/evaluation}
\input{src/related}
\input{src/conclusion}

\printbibliography

\clearpage
\appendix
\appendixcontents
\input{src/mechanization}

\input{src/supplement}
\clearpage
\input{src/corpus}
\input{src/casestudy}
\input{src/linking}
\input{src/realizability}
\input{src/surface}
\input{src/observations}
\input{src/proofs}
\input{src/appendix}

\end{document}

%% file: math_commands.tex
\usepackage{amsmath,amsfonts,bm}

\def\eqref#1{equation~\ref{#1}}

\def\1{\bm{1}}

\DeclareMathAlphabet{\mathsfit}{\encodingdefault}{\sfdefault}{m}{sl}
\SetMathAlphabet{\mathsfit}{bold}{\encodingdefault}{\sfdefault}{bx}{n}

%% file: src/abstract.tex
As autonomous AI agents take on every stage of scientific inquiry,
research output is expanding far beyond human review capacity.
Yet scientific communication still relies on natural-language prose:
an informal medium prone to ambiguity, hidden assumptions, and untracked
limitations that machines cannot reliably audit.
We introduce \Lara, a machine-checkable language and protocol for checking
and revising support for research claims.
By turning research arguments into executable artifacts, \Lara provides an
epistemic kernel for autonomous science: it enables automated validation
pipelines for research agents, lets declared bridges connect
arguments across papers into an auditable network, and allows both humans and
machines to recheck the standing of an encoded claim in milliseconds.
In a \Lara program, authors explicitly declare their claims,
supporting evidence and assumptions, and known objections or limitations.
A lightweight, deterministic checker adjudicates these interactions,
assigning each claim a reproducible status: \stJustified, \stDefeated,
\stContested, or \stGap, which marks a claim whose support is incomplete and
locates the unanswered question.
Case studies cover empirical review, a philosophical debate without
measurements, and the loss of support when an assumed axiom is withdrawn.
We establish the metatheory of claim checking and cross-context argument
transport, and mechanize the semantic guarantees in Lean~4
(roughly \num{117000} lines), leaving three arguments on paper.
The audited public metatheory is \textsf{sorry}-free and uses only
Lean's three standard axioms; some executable examples additionally
trust native evaluation.

%% file: src/intro.tex
\section{Introduction}
\label{sec:intro}

Scientific progress depends on accumulating verifiable knowledge through
rigorous peer critique. For more than three centuries, the research paper has
served as the universal medium for this critique. The genre assumes a human on
each side: an author compresses an investigation into a linear narrative that
states claims and cites selected evidence, while a reader reconstructs the
inferential links that the narrative leaves implicit.
\citet{Medawar_1996} famously argued that this genre misrepresents inquiry
itself: the published narrative preserves the successful path and discards the
dead ends and iterative revisions that produced it
\citep{Howitt_2014, Tregoning_2026}. The genre nonetheless
endured because human judgment bridged the gap: reviewers read the prose,
weighed the tables, and decided holistically whether the argument held together.
Such a verdict is un-auditable: it cannot locate a missing premise, and it
cannot distinguish a claim that was contested from a claim that was never
supported.

The rise of autonomous AI agents shatters this human-centric equilibrium.
Language-model agents now carry projects from hypothesis generation to
experiment execution and manuscript drafting
\citep{lu2024aiscientistfullyautomated, Ayg_n_2026, tang2026farsfullyautomatedresearch}. Inside OpenAI's research
organization, agents already perform 3.1 workdays of research for every
workday of human labor \citep{openai2026acceleration}. Beyond the lab, annual
submissions to flagship AI venues now exceed \num{20000} papers each, forcing
peer review to automate in response: venues now deploy machine reviews across
major tracks \citep{biswas2026aiassistedpeerreviewscale} or mandate end-to-end AI authorship and
review \citep{Bianchi_2025}. In this regime, natural language
fails as a communication protocol: it is an informal medium
prone to ambiguity, unstated assumptions, and untracked failures that machines
cannot reliably audit. When machines write claims and machines judge them
through prose, no step in the loop can be checked.

The same shift supplies the remedy. Software
engineering absorbed the same surge in productivity when code generation
became cheap: verification migrated from manual code inspection into automated
type checkers, test suites, and continuous integration pipelines. Autonomous
scientific agents now capture the complete computational substrate of
discovery: structured datasets, execution logs, code repositories, and failed
attempts \citep{liu2026humanwrittenpaperagentnativeresearch}. The full record of an investigation is now
digital and machine-readable. We no longer need to
force machines to compress structured discoveries into informal prose only for
other machines to guess their meaning; instead, we can formalize research
validation itself as an executable computing discipline.

Realizing executable verification for science requires resolving one
obstacle. In formal mathematics, proof certificates are factive and
monotonic \citep{Necula_1997, miller2015foundational}: once a proof checker
accepts a theorem, the conclusion holds under its axioms, and no
subsequent discovery can retract it while its assumptions remain fixed.
Research arguments can also be defeasible. A later
experiment can invalidate a scientific
conclusion even when its original premises and inference steps remain
well-formed. For instance, an inequality between two measured benchmark scores
($0.74 > 0.71$) remains an arithmetic fact; yet if a subsequent
audit exposes unaddressed confounders or a shifted test distribution
\citep{pmlr-v97-recht19a}, that arithmetic ceases to justify the claim that one
method outperforms another. A formal system for research claims must therefore
obey a governing design principle: it must separate \emph{checking the
validity of a support term} from \emph{accepting the claim it
supports}.

We introduce \Lara, a machine-checkable language and protocol for checking
and revising support for research claims that operationalizes this principle.
Its premises may be measurements, attributed statements, or explicit
assumptions. A philosophical objection can defeat an inference, and a
deductive argument can lose its reusable support when a target context
withholds an assumption. The same distinction between checked support and
accepted claims applies in each case. By transforming research
arguments into executable, checkable programs, \Lara enables four capabilities:
\begin{enumerate}
  \item \textbf{Continuous integration for autonomous science}: AI research
    agents can run automated regression tests on their own hypotheses, based
    on their declared claims and support, obtaining deterministic verdicts from
    a checker rather than relying on noisy prompt-based evaluations. \Lara thus
    serves as an epistemic kernel for research agents, ensuring they build
    subsequent research on verified findings.
  \item \textbf{Auditable links across the literature}: Instead of
    treating papers as isolated documents, declared bridges across contexts
    connect findings into an auditable network; when a later world declares a
    distribution shift or counterexample as a typed attack on a transported
    argument, a modal query from the earlier world reports that the claim is
    no longer justified in every accessible context. A literature-scale
    knowledge graph over checked claims is a possible downstream application.
  \item \textbf{Objections and limitations as first-class assets}: Authors
    and agents declare challenges to premises, conclusions, or inference
    steps as typed attacks. A failed experiment or a philosophical objection
    can therefore defeat the support it targets, while a claim with no
    complete support stays distinct from one whose support is defeated.
  \item \textbf{Separating checked deductions from revisable claims}:
    Strict backends certify steps relative to declared premises and theories.
    The surrounding argument records how those steps support a claim and
    whether that support survives challenges or a change of assumptions.
    A certified numerical comparison, for example, can remain valid after
    an objection defeats the claimed empirical improvement.
\end{enumerate}

At a high level, a \Lara program acts as a research claim certificate. Authors
declare their claims (pairing natural-language prose with formal
targets), evidence leaves tagged with provenance, inference steps, and typed
attacks. Inferences split into two tiers: strict deductive steps delegate proof
obligations to registered backends (such as rational arithmetic or natural
deduction), while defeasible steps instantiate policy-governed schemes
that must explicitly answer mandatory \emph{critical questions} (such as
randomization, sample power, or the scope of an assumed premise) or face
immediate checker rejection. Attacks
target specific structural positions: rebutting conclusions, undermining
evidence leaves, or undercutting inference rules when exception conditions
trigger. A support term that checks does not yet settle its claim: another
argument can defeat it, and a third can reinstate it by defeating the
attacker. A lightweight, deterministic checker therefore compiles accepted
programs into a
Dung argumentation framework \citep{Dung_1995}, whose grounded
evaluation assigns each claim a status: \stJustified when complete
support survives all challenges, \stDefeated when all support is overturned,
\stContested when arguments reach an undecided deadlock, or \stGap when no
complete support argument exists. Across distinct papers or environments, a
conservative possible-world semantics models each accepted program as a world,
employing declared structural bridges to transport support terms while relying on
an executable attack-bisimulation checker to decide when claim statuses survive
the translation.

A status is relative to the signature, policy, backend registry, and
admitted evidence. The checker decides whether the declared support stands;
it does not establish that a formal atom faithfully represents a prose claim
or that an admitted measurement is accurate. Transport is guarded in the same
way: structural bridges preserve checked support, but a target world may
contain additional attackers, and a mechanized example transports support
while the claim changes from \stJustified to \stDefeated.

The case studies make this scope concrete (\appautoref{app:casestudy}).
Alongside empirical review and cross-paper comparisons, a philosophy-of-mathematics
debate uses only assumed and attested premises to exhibit unresolved
disagreement and reinstatement. An axiom-withdrawal example combines
certified hypothesis reuse with a Lean witness showing why support cannot
cross into a context that withholds its premise. These demonstrations
instantiate the existing calculus without adding domain-specific kernel rules.

We make the following contributions:
\begin{itemize}
  \item \textbf{A typed claim-support calculus}: We formulate the core
    language and protocol with explicit critical-question obligations, typed
    positional attacks, and isolated strict backends (\autoref{sec:language}).
  \item \textbf{Metatheory and status preservation}: We prove status
    preservation between direct source semantics and compiled argumentation
    frameworks, establish dependency accountability and backend
    replacement, and develop an update calculus governing status transitions
    (\autoref{sec:metatheory}).
  \item \textbf{Cross-context possible-world semantics}: We formalize modal
    observations over checked contexts, establish exact support transport, and
    provide a sound and complete executable decider for status-preserving attack
    bisimulations (\autoref{sec:worlds} and \autoref{sec:cross-context-guarantees}).
  \item \textbf{Mechanization and empirical evaluation}: We mechanize the
    semantic results in Lean~4 across 144 files and roughly
    \num{117000} lines. The audited public metatheory uses only Lean's
    standard three axioms; some executable examples additionally trust
    native evaluation, and three arguments remain on paper
    (\autoref{sec:metatheory}). We validate a Haskell checker through
    differential testing on 673 verdict-bearing and 66 malformed inputs,
    including systematic mutation suites (\autoref{sec:evaluation}).
\end{itemize}

%% file: src/overview.tex
\section{Background and overview}
\label{sec:overview}

A paper reports that a new method works better than an existing one.
Before building on that result, another scientist needs to understand
how the comparison was made, whether any objections remain unanswered,
and whether the result applies in the setting they care about. \Lara
makes these relationships explicit so that a checker can follow them,
drawing on three ideas: research records that agents read and write,
formal argumentation over reasons and objections, and comparison
across possible worlds. A running example then shows how a \Lara
program expresses them, and a trust architecture sets out what the
checker guarantees.

\subsection{Background}
\label{sec:background}

\subsubsection{Agent-native research: records that agents read and write}
A scientist continuing someone else's investigation needs access to
the work behind its conclusions: the data, the analysis, the experimental
conditions, and the reasons for abandoning earlier approaches. An AI
research agent, a system that uses tools to carry out research tasks,
needs this information too. We call a record \emph{agent-native} when
it organizes the information for an agent to find and use directly.
A claim has a name, its evidence points to specific results, and the
record links those results to the procedures that produced them.
The record serves the role of a laboratory notebook whose entries
connect directly to the relevant data and code.

Agents now write these records as well as read them.
\citet{lu2024aiscientistfullyautomated} report a system that generates a research idea,
writes the code, runs the experiment, and drafts the manuscript, then
repeats the cycle on its own output. Earlier tools helped a human with a
single step; this system runs the whole loop. Each pass builds on what an
earlier pass recorded, which places the record between two machines.

\citet{liu2026humanwrittenpaperagentnativeresearch} propose the Agent-Native Research Artifact (ARA)
format for such records. An ARA brings together claims and methods,
code and configurations, evidence, and a history of the investigation,
including failed experiments and abandoned approaches. For example,
it can preserve both a reported improvement and a later experiment
where the improvement disappeared. A scientist or an agent building on
that result can then inspect the conditions of both experiments instead
of seeing only the successful one.

The record stores both experiments but leaves their relationship
unstated. The later experiment might expose a flaw in the original
comparison, or it might concern a different population. Whoever
continues the work has no declared answer to check.
\Lara lets an author state the proposed relationship and makes its
consequences checkable. ARA supplies one source of these records;
authors and agents can also write \Lara programs directly. The checker
reads a program the same way whichever of them wrote it.

\subsubsection{Formal argumentation: reasons and objections}
A measured improvement, a claim about that improvement, and a reason
to believe the claim are different things. The measurement is
\emph{evidence}. The statement that the new method works better is a
\emph{claim}. An \emph{argument} connects the evidence to the claim,
for example by explaining why the two methods were compared under
conditions that make the result meaningful.
Scientific review tests these connections. A reviewer might object
that the two methods were tested under different conditions, so their
reported scores do not establish the claimed improvement. The author
might respond with records showing that the reviewer consulted an
outdated protocol. If that response defeats the objection, the original
argument can stand again. This is \emph{defeasible} reasoning: a
reason for accepting a claim can lose its force when a challenge
appears, and regain it when that challenge is answered.

Formal argumentation represents this exchange as a diagram of arguments
and the objections between them~\citep{Dung_1995}. Each
node represents an argument; an arrow points from an objection to the
argument it challenges. Such a challenge is called an \emph{attack}.
An \emph{abstract argumentation framework} records this pattern while
setting aside the arguments' internal details. Its \emph{semantics}
specifies the rules for deciding which arguments to accept. \Lara uses
\emph{grounded semantics}: begin with arguments that face no declared
attack, mark the arguments they attack as defeated, and accept further
arguments once all their attackers have been defeated. Continue until
no decision changes. Two arguments that only attack each other remain
undecided; the rules do not choose a winner without further grounds.

\Lara adds a check before this evaluation: each argument must contain
the support its stated inference rule requires, and each attack must
be allowed by the declared rules. The checker then reports one of
four statuses. A claim is \stJustified when at least one complete
support argument survives, \stDefeated when complete support exists
but every such argument is defeated, and \stContested when no support
argument is accepted and some remain undecided. A claim with no
complete support argument at all is \stGap.
The premises need not be measurements: an argument may instead use an
explicit assumption or an attributed position. The policy specifies which
inferences and objections these premises license; the philosophy example in
\appautoref{app:cs-philmath} exercises this case.
These statuses describe the declared case for a claim. They do not
establish that its measurements are accurate or that the claim is true.

\subsubsection{Possible worlds: comparing research settings}
A claim may have convincing support in one research record and face
an unanswered objection in another. A scientist comparing them needs
to keep track of which evidence and assumptions each conclusion rests
on. \emph{Possible-world semantics} provides a way to do this: evaluate
a statement separately in each specified situation, called a
\emph{world}, and then ask how its standing differs across
worlds~\citep{kripke1963semantical}. Logicians study the languages
that describe such situations as \emph{modal
logic}~\citep{Blackburn_2001}.

In \Lara, a world is a program that passes the checker. It may express a
research record or an argument under a chosen set of assumptions.
It contains evidence, arguments, and challenges,
together with the rules used to assess them. A world can combine
material from several papers; it need not correspond to one paper.
To compare worlds, a researcher declares a \emph{bridge} that specifies
which comparison to make and how a claim in one record corresponds to
a claim in the other. These connections form the model's
\emph{accessibility relation}: the set of comparisons its queries use.
Choosing an appropriate comparison remains the
researcher's responsibility.

A researcher can now put two questions to the checker. Is the claim justified in
\emph{some} compared record? Is it justified in \emph{every} compared
record? Modal logic calls these the \emph{diamond} and \emph{box} queries.
If one compared record justifies the claim and another defeats it,
the answer to the first question is yes and to the second is no.
The comparison leaves the original record's own verdict intact.
The second question tests robustness, and a claim cannot pass it
without at least one declared comparison.
We depict both operations in \autoref{fig:background}: following reasons
and objections within a record, and comparing claim statuses across
records. The formal definitions follow in \autoref{sec:semantics} and
\autoref{sec:worlds}.

\input{figures/background}
\FloatBarrier

\subsection{A running example}

We now encode the distinction between missing and defeated support.
An experiment report asserts that method $M$ improves accuracy on
distribution $D$, on the strength of one measured effect over a
baseline $M_0$; the same
artifact's exploration trace records a dead end in which the effect
vanished under a distribution shift. We give the artifact's support for
that claim as a \Lara program in \autoref{fig:example}, in two runs:
run 1 leaves the mandatory external-validity question undischarged,
and run 2, shown in full, answers it. This policy requires
off-distribution evidence even for the stated claim on $D$; that
requirement is a policy choice. Both programs and their expected
reports ship in the artifact.

\input{figures/example}

The program declares exactly what the checker judges. Lines 1--3 fix
the replay identity: the artifact digest, the versioned claim-support
policy, and the backend registry. Each claim then pairs a
natural-language sentence with a formal target atom, and the checker does not establish that the signed
binding between them is faithful
(\autoref{sec:leaves}). The leaves admit evidence with its kind,
provenance tag, and source reference, the recorded dead end \kw{e4}
among them. Admitting \kw{e4} does not make it an attack; it counts
only once a typed attack cites it (\autoref{sec:attacks}). Of the two
arguments, \kw{s1} re-checks the ordered comparison of the score cells
under a certificate the checker replays, while \kw{a1} instantiates
the policy's controlled-experiment scheme (\autoref{fig:scheme}). A
support argument must conclude the claim's formal target
syntactically, after normalization, and must account for every
critical question: \kw{e2} and \kw{e3} discharge randomization and
power, and the off-distribution result \kw{e6}, absent from run 1,
discharges external validity. The last declarations turn the dead end
into the typed attack \kw{d1}, which undercuts \kw{a1} at that rule
occurrence because it concludes the scheme's declared exception.

\input{figures/report}

Both programs are accepted (\autoref{fig:report}). Run 1 omits
\kw{a1}: declaring it with an unanswered mandatory question would
cause rejection. Thus $\mathrm{supp}(\kw{c1})$ is empty and \kw{c1}
is \stGap. Run 2 declares the additional evidence and complete support
argument, together with the undercut that makes \kw{c1} \stDefeated.
The undercut challenges the inference's applicability; it does not
prove the empirical claim false. Adding \kw{e6} alone would not cause
this transition: run 2 also declares \kw{a1}, \kw{d1}, and the attack.

The attack on \kw{a1} leaves \kw{s1} and \kw{c2} justified in both
runs. Strict rule occurrences cannot be rebutted or undercut, although
an attack on a premise can still defeat a containing argument. The two claims
stand apart here because the program declares no inference carrying
the arithmetic into the empirical claim. We give a program whose
certified arithmetic does feed a defeasible claim, and outlives its
defeat, in \appautoref{lst:ord-under-attack}.

\paragraph{Comparing nearby settings}
Reusing support from another paper requires identifying the relevant
differences (\autoref{tab:settings}). New evidence may change a world's
status; a changed premise may instead invalidate the transport contract.
The possible-world semantics makes these comparisons explicit through
declared bridges (\autoref{sec:worlds}).

\input{tables/settings}

\subsection{Trust architecture}

Following proof-carrying code~\citep{Necula_1997}, an untrusted
producer supplies the program and its backend certificates. The trusted
pipeline (\autoref{fig:architecture}) checks them against the signature,
versioned policy, and backend registry. Each verdict records these
versions, the core version, and the artifact digest for replay, together
with its evidence and backend dependencies
(\autoref{thm:accountability}). Choosing the policy and assessing the
evidence remain separate judgments.

\input{figures/architecture}

\paragraph{Scope}
Acceptance establishes that the declared support and attacks pass the
checker; a claim's status then describes its support in the compiled
framework. Both are relative to the admitted evidence, policy, and
backend registry. A modal answer ranges over the declared comparisons
alone, and says nothing about a setting that no bridge reaches. The
checker does not establish the truth of the encoded claims,
the accuracy of admitted evidence, the suitability of the policy or
backend theory, or the faithfulness of the prose-to-formal binding
(\autoref{sec:leaves}, \autoref{thm:isolation}).

%% file: figures/background.tex
\begin{figure}[t]
\centering
\begin{minipage}[t]{0.46\textwidth}
\centering
\textbf{Reasons within one record}\par\medskip
\begin{tikzpicture}[
  font=\small,
  arg/.style={draw, rounded corners=2pt, align=center,
              text width=4.0cm, inner sep=5pt},
  attack/.style={-{Stealth[length=2mm]}, semithick}
]
\node[arg, fill=blue!6] (c) {Response to the objection\\\emph{accepted}};
\node[arg, fill=orange!12, below=8mm of c] (b)
  {Objection to the comparison\\\emph{defeated}};
\node[arg, fill=blue!6, below=8mm of b] (a)
  {Argument for the improvement\\\emph{accepted}};
\draw[attack] (c) -- node[right, font=\scriptsize] {challenges} (b);
\draw[attack] (b) -- node[right, font=\scriptsize] {challenges} (a);
\end{tikzpicture}
\end{minipage}\hfill
\begin{minipage}[t]{0.51\textwidth}
\centering
\textbf{Comparisons between records}\par\medskip
\begin{tikzpicture}[
  font=\small,
  world/.style={draw, rounded corners=2pt, align=center,
                text width=2.45cm, inner sep=5pt},
  bridge/.style={-{Stealth[length=2mm]}, semithick,
                 dashed, blue!65!black}
]
\node[world] (w0) at (0,0) {Source\\research record};
\node[world, fill=orange!12] (w1) at (-1.7,-2.0)
  {Record 1\\claim: \stDefeated};
\node[world, fill=blue!6] (w2) at (1.7,-2.0)
  {Record 2\\claim: \stJustified};
\draw[bridge] (w0.south west) -- (w1.north);
\draw[bridge] (w0.south east) -- (w2.north);
\node[font=\scriptsize, align=center] at (0,-1.0)
  {declared\\comparisons};
\node[align=center, font=\small] at (0,-3.2)
  {justified in \emph{some} compared record: yes\\
   justified in \emph{every} compared record: no};
\end{tikzpicture}
\end{minipage}
\caption{Following a debate and comparing research records are different
operations. Left: each box is an argument. With no further challenges,
the response defeats the objection, so the original argument is accepted
again. Right: each box is a whole research record. The same claim is
justified in one compared record and defeated in the other, so it is
justified somewhere but not everywhere among these comparisons. Comparing
records does not rewrite the original record's verdict.}
\label{fig:background}
\end{figure}
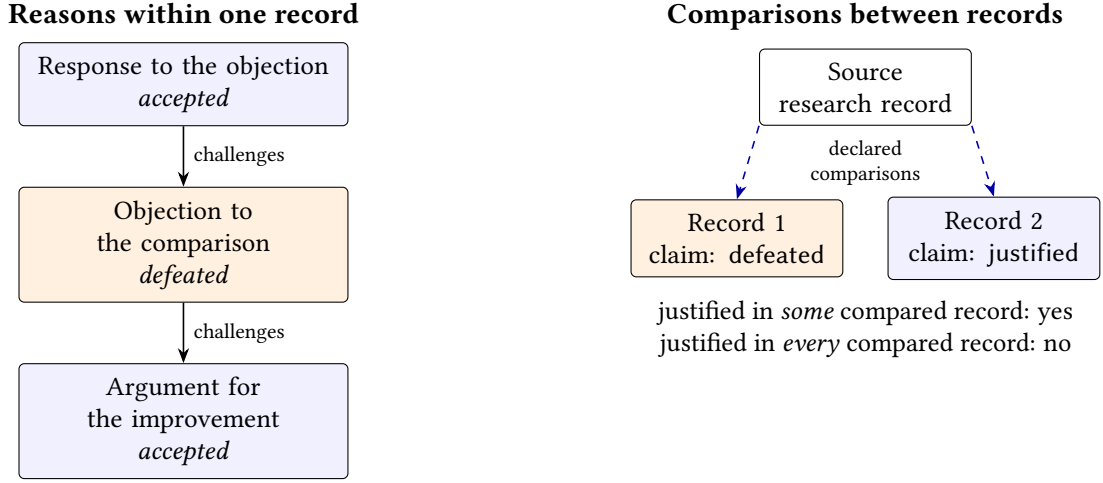

%% file: figures/example.tex
\begin{figure}
\begin{lstlisting}[language=Lara,basicstyle=\ttfamily\scriptsize,numbers=left]
artifact paper_17 at sha256:aaaa...
policy empirical-v3
use backends [nd@1, ord@1]
let base_score = 0.71
let ours_score = 0.74
claim c1
  nl      = "Method M improves accuracy on distribution D"
  formal  = improves(M, accuracy, D)
  binding = { author = alice, audit-status = reviewed }
claim c2
  nl      = "The baseline accuracy {base_score} is strictly below {ours_score}"
  formal  = num_lt(base_score, ours_score)
  binding = { author = alice, audit-status = reviewed }
leaf e1 : reports(exp_3, effect(M, accuracy, D, positive))
  kind       = observed
  provenance = ai-executed
  refs       = [evidence/table_2.csv#row=mean]
leaf base : reports(exp_3, score_cell(M0, accuracy, D, base_score))
leaf ours : reports(exp_3, score_cell(M, accuracy, D, ours_score))
leaf e4 : distribution_shift(M, accuracy, D)
arg s1 : supports(c2) by lt_recheck from [base, ours]
  assurance = cert(ord@1, sha256:empv3-t0, (ordcmp (prem base) (prem ours)))
leaf e6 : generalizes(M, accuracy, D)
arg a1 : supports(c1) by controlled_experiment from [e1]
  discharge randomization     with e2
  discharge adequate_power    with e3
  discharge external_validity with e6
arg d1 : challenges(external_validity(a1)) by leaf(e4)
undercut d1 a1.rule
status c1
status c2
\end{lstlisting}
\caption{Run 2 of the running example, from the artifact. Run 1 omits
lines 23--29. We elide \kw{e2} and \kw{e3}, and \kw{base}, \kw{ours},
\kw{e4}, and \kw{e6} keep only their declaration line; their metadata
has the form shown for \kw{e1}. Elaboration infers rule substitutions
and resolves named certificate premises to numeric slots.}
\label{fig:example}
\end{figure}

%% file: figures/report.tex
\begin{figure}
\begin{lstlisting}[basicstyle=\ttfamily\scriptsize,numbers=none]
$ lara check run1/example.lara   # without lines 23-29
(verdict (replay-id ...) accept (labels (0 in)) (edges)
  (statuses (status (atom improves (con M) (con accuracy) (con D)) gap)
            (status (atom num_lt (num 0.71) (num 0.74)) justified)))

$ lara check run2/example.lara   # the program as shown
(verdict (replay-id ...) accept (labels (0 in) (1 out) (2 in)) (edges (2 1))
  (statuses (status (atom improves (con M) (con accuracy) (con D)) defeated)
            (status (atom num_lt (num 0.71) (num 0.74)) justified)))
\end{lstlisting}
\caption{Checker reports from the artifact's pinned running example.
Replay identities are elided, paths abbreviated, and verdicts wrapped
for the page. Acceptance exits 0; rejection returns a nonzero exit
code and a located diagnostic.}
\label{fig:report}
\end{figure}

%% file: tables/settings.tex
\begin{table}[t]
\caption{Motivating comparisons between nearby research settings.
These are illustrative scenarios, not evaluated case studies.
A small difference does not by itself establish comparability.}
\label{tab:settings}
\small
\centering
\begin{tabular}{@{}p{0.25\linewidth}p{0.275\linewidth}p{0.415\linewidth}@{}}
\toprule
Shared setting & Difference & What must be checked \\
\midrule
Task and classifier & New test population & Whether the evidence still supports the claim \\
\addlinespace
Inference-scheme family & New premise or obligation & Whether the support-transport contract holds \\
\addlinespace
Claim and existing support & Additional attacker & Whether justification survives the attacker \\
\addlinespace
Argument structure & Renamed vocabulary & Whether support and attacks are preserved \\
\bottomrule
\end{tabular}
\end{table}

%% file: figures/architecture.tex
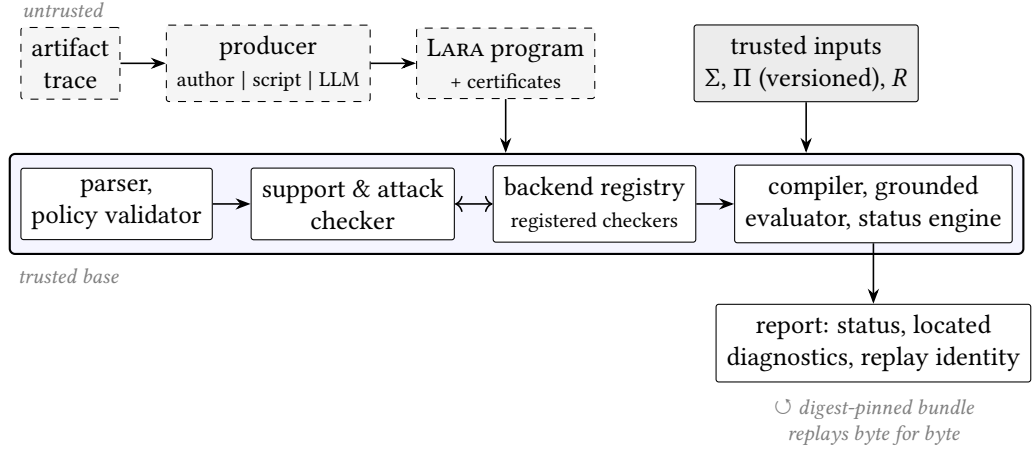
\begin{figure}
\centering
\small
\begin{tikzpicture}[
  font=\small,
  box/.style={draw, rounded corners=1pt, align=center, inner sep=4pt,
              minimum height=1.9em},
  untrusted/.style={box, fill=gray!8, dashed},
  trusted/.style={box, fill=white},
  input/.style={box, fill=gray!15},
  flow/.style={-{Stealth[length=2mm]}, semithick},
  lbl/.style={font=\scriptsize\itshape, gray}
]
\node[untrusted] (trace)    {artifact\\trace};
\node[untrusted, right=6mm of trace] (producer)
  {producer\\{\scriptsize author $\mid$ script $\mid$ LLM}};
\node[untrusted, right=6mm of producer] (program)
  {\Lara program\\{\scriptsize + certificates}};
\node[input, right=13mm of program] (policy)
  {trusted inputs\\$\Sig$, $\Pol$ (versioned), $\Reg$};
\node[trusted, below=9mm of trace.south west, anchor=north west]
  (parser) {parser,\\policy validator};
\node[trusted, right=5mm of parser] (checker) {support \& attack\\checker};
\node[trusted, right=5mm of checker] (backends)
  {backend registry\\{\scriptsize registered checkers}};
\node[trusted, right=5mm of backends] (engine)
  {compiler, grounded\\evaluator, status engine};
\node[box, below=8mm of engine] (report)
  {report: status, located\\diagnostics, replay identity};
\begin{scope}[on background layer]
\node[draw, thick, rounded corners=2pt, fill=blue!4,
      fit=(parser)(checker)(backends)(engine), inner sep=4pt] (tcb) {};
\end{scope}
\node[lbl, anchor=north west, yshift=-0.5mm] at (tcb.south west) {trusted base};
\node[lbl, anchor=south west, xshift=-1mm] at (trace.north west) {untrusted};
\draw[flow] (trace) -- (producer);
\draw[flow] (producer) -- (program);
\draw[flow] (program.south) -- (program.south |- tcb.north);
\draw[flow] (parser) -- (checker);
\draw[flow, <->] (checker) -- (backends);
\draw[flow] (backends) -- (engine);
\draw[flow] (policy.south) -- (policy.south |- tcb.north);
\draw[flow] (engine) -- (report);
\node[lbl, below=1mm of report, align=center]
  {$\circlearrowleft$ digest-pinned bundle\\replays byte for byte};
\end{tikzpicture}
\caption{The trust architecture. Only the producer's output is checked:
the signature \Sig, versioned policy \Pol, and registry \Reg\ reach the
trusted base directly (shaded node), never through the producer. The
emitted status is one of four core values or \stEvidenceBlocked.}
\label{fig:architecture}
\end{figure}

%% file: src/language.tex
\section{The claim-support language}
\label{sec:language}

A \Lara program is a claim-support certificate: it declares evidence
leaves, policy-rule instances, the obligations those instances leave
open, and typed attacks. The checker interprets it against four fixed
inputs: a logical signature \Sig, a versioned claim-support policy
\Pol, a strict-backend registry \Reg, and an artifact snapshot at a
recorded digest, whose versions and digest form the replay identity.
We give the abstract syntax in \autoref{fig:syntax}. The JSON front end
decodes directly to it; the presentation surface reaches it through a
validated elaborator that expands \texttt{comparison} blocks, labelled
rule premises, and \texttt{\{cell \dots\}} interpolation before the
core-calculus boundary.

\input{figures/syntax}

\subsection{Propositions, leaves, and claims}
\label{sec:leaves}

Support-level propositions are ground first-order atoms over a fixed
vocabulary in \Sig. Programs may instantiate this vocabulary but may
not extend it. The policy declares \Sig{} as a many-sorted signature
over two built-in base sorts, and the checker sorts the policy's
patterns and each rule instance's substitution range; a substitution
lemma carries well-sortedness to every atom an accepted program
reaches. Conflict at the support level comes from a declared
\emph{contrary} relation rather than from negation to falsum, and an
implication is reified as a named rule rather than asserted as a
proposition. An evidence leaf declares a proposition together with its
kind (\lkObserved, \lkAttested, \lkAssumed, or
\lkCertified), provenance, and source references.

A claim combines a natural-language string, a formal target atom, and
an untrusted, human-signed binding between them. A support term
\emph{supports} the claim exactly when its conclusion equals the
formal target under the normalized-identity relation $\equiv$.
Support is positional identity rather than entailment, so no solver
enters the trusted base. An auditor must assess whether the formal atom expresses the prose;
normalized identity checks only the formal targets, not their meaning. The presentation surface narrows one source of
error: \texttt{\{cell e\}} inserts the named leaf's value into the
prose, so the quoted number equals the value stored in the leaf.

\subsection{Proposition normalization}
\label{sec:normalization}

The identity relation $\equiv$ is the trusted-base equality on
propositions, so its definition stays minimal. It operates on the core abstract
syntax that both front ends reach (the JSON wire directly, the
presentation surface after elaborator expansion), so whitespace,
field order, encoding differences, and surface sugar are gone before
$\equiv$ applies. A normal form $\nf(\cdot)$ canonicalizes literals and
recurses structurally, but it reorders no arguments: the core calculus
treats no symbol as associative or commutative. Then $p \equiv q$ iff
$\nf(p) = \nf(q)$, which is decidable, total, an equivalence relation,
and linear in term size.

\subsection{Claim-support policies and obligations}
\label{sec:policy}

A claim-support policy defines named inference schemes. A strict scheme
is deductive; a defeasible scheme provides presumptive support and lists
critical questions, in the argument-scheme
tradition~\citep{Walton_2008}, that an instance must discharge or
leave explicitly open. The policy also declares the \emph{contrary} relation (the only
source of propositional conflict) and \emph{exception} conditions that
license undercuts. Policies are versioned trusted inputs: an artifact
program instantiates schemes but never defines or modifies them at
runtime. Whether a rule is strict or defeasible is a property of its
policy entry, not separate program syntax.

Each rule instance carries an explicit ground substitution. The checker
therefore performs no unification; it applies the substitution and
checks syntactic identity. An instance must account for every declared
critical question, either with a discharging subterm or with an
explicit hole. Only optional questions may remain open in an accepted
program. If an instance leaves a mandatory question open, the checker
rejects the program and names both the argument and the question. The
ablation in \autoref{sec:ablation-summary} removes this gate. An instance with an unanswered mandatory question cannot support a
claim in an accepted program. The claim is \stGap only if it has no
other complete support argument.

Leaf admission precedes evaluation and returns \emph{admit},
\emph{quarantine}, or \emph{reject} from a policy table over
\emph{(leaf kind, provenance)} keys (\autoref{sec:semantics} fixes the
judgment). A \emph{quarantine} decision retains the declaration for
audit but removes arguments that use it. Quarantine creates no attack,
but it can remove one; the conservative public reporting of
\autoref{sec:semantics} prevents any resulting promotion.

The support judgment $\supjudg$ records unmet mandatory obligations
$O$ for a term $w$ supporting proposition $p$
(\autoref{fig:supportrules}). An argument is complete exactly when
$O = \varnothing$; optional questions can remain open without
contributing to $O$. The defeasible scheme in \autoref{fig:scheme}
supplies the shape the running example uses. For \kw{a1}, the
\rulename{Rule} form in \autoref{fig:supportrules} derives
$\Sig;\Pol;\Ctx;\Reg \vdash \kw{a1} :
\supports(\mathit{improves}(M, \mathit{accuracy}, D)) \obl
\varnothing$. The premise leaf \kw{e1} has no obligations, and the
three discharged questions contribute empty sets. If
external validity were instead an explicit hole, the rule would derive
$\{\mathit{external\_validity}\}$. Such a term fails the program-level
gate.

\input{figures/supportrules}
\input{figures/scheme}

\subsection{Strict-certificate backends}
\label{sec:strict}

A strict rule instance has no critical questions or holes. The policy
either marks it explicitly trusted or requires an opaque certificate
for a fixed, versioned backend in \Reg. Backend formulas, proof terms,
axioms, and model theory remain outside the source calculus. A registry
entry supplies only an encoding $\enc\backend$ from source
propositions to backend formulas and a checker $\chk\backend$.

\input{figures/ndrules}

For a strict scheme $P_1,\dots,P_n \Rightarrow C$ at substitution
$\theta$, the instance checks only when $\chk\backend$ accepts
$\enc\backend(C\theta)$ against the encoded premise conclusions
$\enc\backend(P_i\theta)$ and the declared, digest-addressed theory.
The checker handles the premise terms normally and propagates their
obligations; only their normalized conclusions cross the interface.
That interface runs one way, so no backend formula or proof term can be
eliminated into a source truth judgment and the support judgment stays
non-factive even when a backend is factive. Nothing else about a backend
reaches the calculus, so replacing a backend preserves statuses when the replacement accepts
the same strict instances (\autoref{thm:replacement}). A registry
entry owes the calculus four laws, which the metatheory cites by name.

Different strict instances may use different registered backends.
Each certificate establishes a consequence in its own backend's logic
(\autoref{thm:hetero}).

\begin{definition}[Registry entry]
\label{def:registry-entry}
A registry entry for a backend \backend{} pairs an encoding
$\enc\backend$ from source propositions to backend formulas with a
certificate checker $\chk\backend$ satisfying four laws:
\begin{enumerate}[label=\textbf{B\arabic*}, ref=B\arabic*, leftmargin=3.2em]
\item\label{law:replay} \emph{Deterministic replay.} $\chk\backend$ is
  a total function of the theory, premises, goal, and certificate:
  replay is deterministic and terminating.
\item\label{law:reflect} \emph{Normalization reflection.}
  $p \equiv q$ iff $\enc\backend(p) = \enc\backend(q)$.
\item\label{law:sound} \emph{Certificate soundness.} If
  $\chk\backend(T, \Delta, \phi, \kappa) = \mathit{accept}$, then
  $T ; \Delta \bmodels \phi$ for the backend consequence relation
  $\bmodels$.
\item\label{law:deps} \emph{Dependency accounting.} Every reported
  slot of an accepted certificate names an in-range premise or theory
  entry. Replay gives the same verdict on contexts that agree at all
  reported slots, and the conclusion is a $\bmodels$-consequence of
  the reported entries alone. These laws permit over-reporting;
  exact consultation requires a backend-specific result.
\end{enumerate}
\end{definition}

An unwitnessed strict instance
is accepted only when its policy entry says so explicitly; reports
distinguish such a trusted-policy step from a certified one.
The reference backend is intuitionistic natural deduction for
implication and falsum (\autoref{fig:ndrules}). Its certificates use de
Bruijn indices, so the checker follows structural recursion without
proof search, and it encodes each proposition as an atom over an
injective serialization of $\nf(p)$, which makes backend formula
equality coincide with source $\equiv$. The Logic of Proofs is an
optional adapter behind the same interface, its $t : F$ judgments and
realization internal to it. The corpus study records which strict steps
carry checked certificates (\appautoref{app:corpus}).

\input{figures/attackrules}

\subsection{Typed positional attacks}
\label{sec:attacks}

An attack names a position in a support term, and the three kinds match
the three kinds of position. A \atRebut targets the conclusion of a
defeasible instance and requires a declared contrary pair. An
\atUndercut targets a defeasible rule occurrence and requires the
attacker to conclude the rule's declared exception. An \atUndermine
targets a leaf and requires the attacker to conclude a contrary of the
leaf's proposition. Strict rules expose neither exceptions nor
rebuttable conclusions, so they can be neither undercut nor rebutted.
We state the attack judgment $\Sig;\Pol;\Ctx;\Reg \vdash k : \attacks(w,
u@\pi)$ in \autoref{fig:attackrules}; the position $\pi$ doubles as the
source location a diagnostic reports. A recorded dead end therefore
defeats nothing until one of the three forms cites it: a dead-end tag,
confidence score, or provenance downgrade cannot attack on its own.
Whole-program checking ends with a completeness scan: each ordered
contrary pair of complete arguments with a leaf or defeasible root as
target must be covered by a declared attack, including closure onto
containing arguments (\autoref{fig:acceptance}). The consistency half of \autoref{thm:rationality}
rests on this attack-completeness postcondition, and an ablation in
\autoref{sec:ablation-summary} removes the scan.

%% file: figures/syntax.tex
\begin{figure}
\centering
\small
\[
\begin{array}{@{}r@{\;}c@{\;}l@{\qquad}l@{}}
\multicolumn{4}{@{}l}{\textit{Propositions and leaves}}\\[2pt]
g      & \Coloneqq & k \gor k(g_1,\dots,g_n)            & \text{ground term over \Sig} \\
p      & \Coloneqq & \mathit{pred}(g_1,\dots,g_n)       & \text{atomic proposition} \\
\ell   & \Coloneqq & \kw{leaf}\ l : p                   & \text{evidence leaf} \\
\kappa & \Coloneqq & \lkObserved \gor \lkAttested \gor \lkAssumed \gor \lkCertified & \text{leaf kind} \\[6pt]
\multicolumn{4}{@{}l}{\textit{Support terms}\;\; (\text{mode of } r \text{ is fixed by } \Pol)}\\[2pt]
w      & \Coloneqq & \kw{leaf}\ l                        & \text{evidence leaf} \\
       & \mid      & r\langle \theta;\, w_1,\dots,w_n;\, \{q \mapsto w_q\};\, \{o\};\, \alpha\rangle & \text{rule instance} \\
\theta & \Coloneqq & \{X_1 \mapsto g_1,\dots,X_m \mapsto g_m\} & \text{ground substitution} \\
\alpha & \Coloneqq & \kw{none} \gor \kw{trusted} \gor \kw{cert}(\backend, h, \kappa_\backend) & \text{assurance} \\[6pt]
\multicolumn{4}{@{}l}{\textit{Strict-certificate backend}\;\;(\text{registered in } \Reg,\ \text{opaque to the source})}\\[2pt]
\enc\backend  & : & \mathit{prop} \to \mathit{Form}_\backend & \text{proposition encoding} \\
\chk\backend  & : & \mathit{Theory}_\backend \times [\mathit{Form}_\backend] \times \mathit{Form}_\backend \times \mathit{Cert}_\backend \to \{\mathit{accept}, \mathit{reject}\} & \text{certificate checker} \\[6pt]
\multicolumn{4}{@{}l}{\textit{Claims and typed attacks}}\\[2pt]
c      & \Coloneqq & \kw{claim}\ (\mathit{nl},\, p,\, \mathit{binding}) & \text{claim} \\
\pi    & \Coloneqq & \varepsilon \gor \pi.i \gor \pi.q                 & \text{position in a term} \\
k      & \Coloneqq & \atRebut\ w\ u \gor \atUndercut\ w\ u@\pi \gor \atUndermine\ w\ u@\pi & \text{attack} \\
\end{array}
\]
\caption{Abstract syntax of \Lara. A single category of support term $w$
carries the assurance annotation $\alpha$: defeasible instances use
\kw{none}, strict ones are \kw{trusted} by policy or discharged by an opaque
certificate for a backend registered in \Reg\ at theory digest $h$. Backend
formulas, proof terms, and theories ($\enc\backend$, $\chk\backend$, \dots)
are not source constructs and never enter a support term, which keeps a
backend's factivity off the support level. Metavariables: $l$ leaves, $r$
rule identifiers, $X$ rule parameters, $q$ critical questions, $o$ declared
holes.}
\label{fig:syntax}
\end{figure}

%% file: figures/supportrules.tex
\begin{figure}
\centering
\judgbox{\supjudg}\quad\text{\small $w$ supports $p$ with open obligations $O$}
\[
\begin{array}{c}
\inferrule*[left=\rulename{Leaf}]
  {\Ctx(l) = p}
  {\Sig;\Pol;\Ctx;\Reg \vdash \kw{leaf}\ l : \supports(p) \obl \varnothing}
\\[20pt]
\inferrule*[left=\rulename{Rule}]
  {r\langle X_1,\dots,X_m\rangle \in \Pol \quad
   (P_1,\dots,P_n \Rightarrow C) = r \\
   \mathrm{dom}(\theta) = \{X_1,\dots,X_m\} \qquad |\vec w| = n \\
   \Sig;\Pol;\Ctx;\Reg \vdash w_i : \supports(p_i) \obl O_i \quad p_i \equiv P_i\theta \quad (1 \le i \le n) \\
   \{q_j : A_j\} = \mathrm{cq}(r) \qquad
   \{q_j\} = D \uplus H \\
   \Sig;\Pol;\Ctx;\Reg \vdash w_q : \supports(p_q) \obl O_q \quad p_q \equiv A_q\theta \quad (q \in D) \\
   \text{assurance ok (\autoref{sec:strict})} \\
   O = \textstyle\bigcup_i O_i \cup \bigcup_q O_q \cup \big(H \cap \mathrm{mand}(r)\big)}
  {\Sig;\Pol;\Ctx;\Reg \vdash r\langle\theta; \vec{w}; \{q \mapsto w_q\}; H\rangle
     : \supports(C\theta) \obl O}
\\[24pt]
\inferrule*[left=\rulename{Supports}]
  {\Sig;\Pol;\Ctx;\Reg \vdash w : \supports(\concl(w)) \obl O \\
   \concl(w) \equiv c.\mathit{formal}}
  {w \mathrel{\supports} c}
\end{array}
\]
\caption{The support judgment. Substitutions and question maps have unique
keys, $H$ is a set, and every displayed instantiation must be defined. A
strict-mode instance discharges its assurance side condition (trusted by
policy or accepted by a registered backend certificate,
\autoref{sec:strict}) and carries an empty question map and hole set; a
defeasible instance carries no strict assurance. Only mandatory holes enter
$O$, so an optional hole adds no obligation.}
\label{fig:supportrules}
\end{figure}

%% file: figures/scheme.tex
\begin{figure}
\centering
\vspace{-5pt}
\small
\begin{tabular}{@{}ll@{}}
\multicolumn{2}{@{}l}{\textbf{Scheme} $\mathsf{controlled\_experiment}(M, Y, D)$ \quad(mode: \kw{defeasible})}\\
\midrule
Premise      & $\kw{leaf}\ e : \mathit{reports}(\mathit{exp}, \mathit{effect}(M, Y, D, \delta))$ \\
Conclusion   & $\mathit{improves}(M, Y, D)$ \\
\midrule
$q_1$ & \emph{randomization}: were units assigned at random? \hfill (mandatory) \\
$q_2$ & \emph{adequate\_power}: was the sample adequately powered? \hfill (mandatory) \\
$q_3$ & \emph{external\_validity}: does the effect hold off distribution $D$? \hfill (mandatory) \\
\midrule
\multicolumn{2}{@{}l}{$\kw{exception}\ \mathsf{controlled\_experiment} : \mathit{distribution\_shift}(M,Y,D)$}\\
\end{tabular}
\caption{A defeasible inference scheme with its critical questions. An
instance must discharge every mandatory question with a subterm; one that
leaves such a question open as a hole is rejected as incomplete. The
\kw{exception} clause turns the ``shown otherwise'' failure mode of $q_3$
into an \atUndercut target, so a defeater concluding the exception atom
attacks instances of this scheme at rule position
(\autoref{fig:attackrules}).}
\label{fig:scheme}
\end{figure}

%% file: figures/ndrules.tex
\begin{figure}
\centering
\judgbox{\ndjudg}\quad\text{\small under assumptions $\Delta$, term $e$ checks at backend formula $\phi$}
\[
\begin{array}{c}
\phi \Coloneqq a \gor \bot \gor \phi \to \phi
\qquad
e \Coloneqq \kw{hyp}\ i \gor \kw{lam}\ \phi\ e \gor \kw{app}\ e\ e \gor \kw{abort}\ \phi\ e
\\[16pt]
\inferrule*[left=\rulename{Hyp}]
  {\Delta(i) = \phi}
  {\Delta \vdash \kw{hyp}\ i : \phi}
\qquad
\inferrule*[left=\rulename{Abs}]
  {\Delta, \phi_1 \vdash e : \phi_2}
  {\Delta \vdash \kw{lam}\ \phi_1\ e : \phi_1 \to \phi_2}
\\[18pt]
\inferrule*[left=\rulename{App}]
  {\Delta \vdash e_1 : \phi_1 \to \phi_2 \\
   \Delta \vdash e_2 : \phi_1}
  {\Delta \vdash \kw{app}\ e_1\ e_2 : \phi_2}
\qquad
\inferrule*[left=\rulename{Abort}]
  {\Delta \vdash e : \bot}
  {\Delta \vdash \kw{abort}\ \phi\ e : \phi}
\end{array}
\]
\caption{The reference strict backend: intuitionistic natural deduction for
implication and falsum. The assumption context $\Delta$ holds the encoded
premise conclusions, which a certificate indexes by de Bruijn
position. Induction on the
derivation gives soundness and exact dependency accounting. Because source
propositions are atomic, the backend certifies only the propositional
consequences its encoding and declared theory make visible: no domain law is
silently relabelled a tautology.}
\label{fig:ndrules}
\end{figure}

%% file: figures/attackrules.tex
\begin{figure}[!tb]
\centering
\judgbox{\Sig;\Pol;\Ctx;\Reg \vdash k : \attacks(w, u@\pi)}
\small
\[
\begin{array}{c}
\inferrule*[left=\rulename{Rebut}]
  {\Sig;\Pol;\Ctx;\Reg \vdash w : \supports(p_w) \obl O_w \\
   \mathrm{mode}(\mathrm{root}(u)) = \kw{defeasible} \quad
   \mathrm{Contrary}_{\Pol}(p_w,\concl(u))}
  {\Sig;\Pol;\Ctx;\Reg \vdash \atRebut\ w\ u : \attacks(w, u@\varepsilon)}
\\[12pt]
\inferrule*[left=\rulename{Undercut}]
  {\Sig;\Pol;\Ctx;\Reg \vdash w : \supports(p_w) \obl O_w \\
   u@\pi = r\langle\theta;\dots\rangle \quad \mathrm{mode}(r) = \kw{defeasible} \\
   \kw{exception}\ r : E \in \Pol \quad p_w \equiv E\theta}
  {\Sig;\Pol;\Ctx;\Reg \vdash \atUndercut\ w\ u@\pi : \attacks(w, u@\pi)}
\\[12pt]
\inferrule*[left=\rulename{Undermine}]
  {\Sig;\Pol;\Ctx;\Reg \vdash w : \supports(p_w) \obl O_w \\
   u@\pi = \kw{leaf}\ l \quad \Ctx(l) = p \quad
   \mathrm{Contrary}_{\Pol}(p_w,p)}
  {\Sig;\Pol;\Ctx;\Reg \vdash \atUndermine\ w\ u@\pi : \attacks(w, u@\pi)}
\end{array}
\]
\caption{Typed positional attacks.
$\mathrm{Contrary}_{\Pol}(p,q)$ holds when some declared
$\kw{contrary}\ A\ B$ and ground substitution $\rho$ satisfy
$p\equiv A\rho$ and $q\equiv B\rho$. Rule lookup, occurrence lookup,
and instantiation must succeed. The local judgment permits open
$O_w$; whole-program acceptance requires both endpoints to be declared
complete arguments (\autoref{fig:acceptance}).}
\label{fig:attackrules}
\end{figure}

%% file: src/semantics.tex
\section{Compilation and claim-support semantics}
\label{sec:semantics}

A well-formed \Lara program compiles to a finite Dung argumentation
framework, and the framework's grounded labelling, aggregated per claim,
is the four-state core status.

\paragraph{Source admission} The trusted source boundary derives the
checking context from the policy's admission table $\Pi_A$, which maps
\emph{(leaf kind, provenance)} keys to decisions and defaults to
\kw{admit} (\autoref{fig:admissionrules}). Duplicate policy keys,
duplicate leaf identifiers, or a metadata view that departs from the
declared leaf table make the source invalid before any leaf decision. A
\kw{reject} decision stops the source at the first rejected leaf in
declaration order (R8) and produces no checked unit, while a
\kw{quarantine} decision removes the leaf from the checking context,
as do inconsistent duplicate-report groups. Admission runs after
structural elaboration has reconstructed every support tree and before
the checker, so one combined pruning pass sees the full support graph:
it removes every argument that uses a removed leaf and every attack
whose raw endpoint no longer resolves, and emits the canonical audit.
Identity follows declared argument identifiers rather than support-term
equality, so equal terms can still name distinct arguments, and group
consistency always reads the original leaf table, so quarantining one
member cannot hide a conflict.

\input{figures/admissionrules}

\paragraph{Whole-program acceptance}
Admission supplies the context for the local support and attack judgments.
The checker then accepts exactly the units satisfying
\autoref{fig:acceptance}: every argument is complete, every attack is
licensed, and the required conflicts are covered. An accepted program
can still contain defeated arguments; the judgment permits status
computation and does not predetermine its outcome.

\input{figures/acceptance}

\paragraph{Compilation} The arguments of the framework are the complete
checked support terms; the attacks are exactly the compiled typed
attacks, closed under subarguments in the \aspic manner. An attack on an
occurrence compiles to an edge onto every argument that contains that
occurrence, so defeating a subterm defeats every complete term built on
it. Certificate payloads stay attached to nodes for replay but are not
term positions or attack targets. Relabelling them injectively preserves
argument names, rule instances, conclusions, obligations, and
positions; collapsing distinct payloads to one \lkCertified marker can
identify subterms and add closure edges (\autoref{thm:replacement}). The grounded labelling (\autoref{def:grounded}) then maps
each argument to \emph{in}, \emph{out}, or \emph{undec}; the definition
states the grounded semantics of \citet{Dung_1995} in the
form the mechanization uses.

\begin{definition}[Compiled framework]
\label{def:compile}
For a checked program $P$, let $A(P)$ be its set of complete checked
terms: the declared terms $w$ with
$\Sig;\Pol;\Ctx;\Reg \vdash w : \supports(\concl(w)) \obl \varnothing$.
The compiled framework is $\mathit{AF}(P) = (A(P), \rightsquigarrow)$,
where $v \rightsquigarrow w$ iff $P$ declares a checked attack
$\Sig;\Pol;\Ctx;\Reg \vdash k : \attacks(v, u@\pi)$ and the occurrence
$u@\pi$ is contained in $w$.
\end{definition}

\begin{definition}[Grounded labelling]
\label{def:grounded}
An argumentation framework $F = (\mathcal{A}, \rightsquigarrow)$ is a
finite set of arguments with an attack relation. A set
$S \subseteq \mathcal{A}$ \emph{defends} $a$ when every attacker of
$a$ is attacked by some member of $S$. The \emph{characteristic
function} maps $S$ to the set of arguments $S$ defends:
$\mathcal{F}_F(S) = \{\, a \in \mathcal{A} \mid S \text{ defends } a
\,\}$. This function is monotone on the subset lattice. Its least
fixpoint $\mathsf{G}(F)$, reached from $\varnothing$ in at most
$\lvert\mathcal{A}\rvert$ iterations, is the \emph{grounded extension}.
The \emph{grounded labelling} $L$ maps an argument $a$ to
\emph{in} when $a \in \mathsf{G}(F)$, to \emph{out} when some member
of $\mathsf{G}(F)$ attacks $a$, and to \emph{undec} otherwise.
\end{definition}

\paragraph{Two monotonicities} For a fixed framework, the
characteristic function in \autoref{def:grounded} is monotone, so
evaluation is deterministic and terminating and cycles produce
\emph{undec} rather than nontermination. \Lara is non-monotonic one
level up, where the input framework can grow: a new checked argument
may attack an argument labelled \emph{in} and retract a claim's
\stJustified status, even though the original support term remains well
typed. Deletion is not a weakening operation either, since removing an
attacker may move its target from \stContested or \stDefeated to
\stJustified. The internal transfer operator is thus monotone on a
finite-height lattice, while the external map from frameworks to
accepted claims is not, as defeasible reasoning requires.

\paragraph{Arguments as terms, not proofs} That split locates \Lara
relative to the Curry--Howard correspondence. The judgment
$\Sig;\Pol;\Ctx;\Reg \vdash w : \supports(p) \obl O$ is a typing
judgment: the checker decides whether a supplied term inhabits its declared
support type. It does not search for an inhabitant. \Lara realizes the correspondence only at the
strict seam, where a certificate is a proof term, its backend checks
the proof, and acceptance is a monotone property of that instance
(\autoref{thm:strict-soundness}). Under \autoref{def:status}, by
contrast, an inhabitant of $\supports(p)$ settles nothing: it enters a
framework where other terms may attack it, and a later argument can
move the claim to \stDefeated while $w$ stays well typed. Typing a
support term is arguing, not proving; otherwise the dead end of
\autoref{sec:overview} could never overturn its claim.

\paragraph{Why grounded semantics} We use grounded semantics for the default report. Its extension is unique (the least complete
one), so evaluation is deterministic; it is polynomial to compute
(\autoref{prop:complexity}); and it is skeptical, labelling an
argument \emph{in} only when the argument is defended against every
attacker. Preferred and stable semantics can admit multiple extensions, and
stable extensions may fail to exist. Deterministic reporting is still
possible by quantifying over all extensions, as we do in
\autoref{sec:observation}, but the computational cost changes
(\autoref{sec:cost}). Grounded semantics leaves an unsupported cycle of
mutual attacks \emph{undec}, so claims supported only by those arguments
remain \stContested.

\paragraph{Four-state aggregation} A claim's support comprises the
complete checked arguments whose conclusions equal its target, and the
aggregation follows a fixed priority: \stGap when the claim has no
support, \stJustified when some support argument is labelled \emph{in},
\stContested when none is \emph{in} but some are \emph{undec}, and
\stDefeated when support exists and every support argument is
\emph{out}. Three of these lift the labels of
\citet{Caminada_2006} from arguments to claims. The fourth,
\stGap, has no labelling analogue because it records absent support
before labelling begins, the difference between absence and refutation
that a three-state or scored account collapses.

\begin{definition}[Claim status]
\label{def:status}
Let $L$ be the grounded labelling of $\mathit{AF}(P)$ and, for a claim
$c$, let $\mathrm{supp}(c) = \{\, w \in A(P) \mid \concl(w) \equiv
c.\mathit{formal} \,\}$. Then
\[
\mathrm{status}(c) =
\begin{cases}
\stGap & \text{if } \mathrm{supp}(c) = \varnothing\\
\stJustified & \text{if } L(w) = \mathit{in} \text{ for some } w \in \mathrm{supp}(c)\\
\stContested & \text{if none is \emph{in} and } L(w) = \mathit{undec} \text{ for some } w \in \mathrm{supp}(c)\\
\stDefeated & \text{otherwise.}
\end{cases}
\]
An incomplete term never enters $A(P)$: the checker rejects a program
that declares one (\autoref{sec:policy}).
\end{definition}

\paragraph{Conservative public reporting}
Quarantine forms $F$ from the declared framework $G$ by removing affected
arguments and attacks. Let $B$ be the forward closure in $G$ of removed
arguments and retained arguments that lost an incoming edge. A claim with
declared complete support in $B$ publishes \stEvidenceBlocked; its status in
$F$ is only a \textsf{conditional} diagnostic. Other claims publish their
core status. The forward closure suffices because the grounded labelling
reads only transitive attackers; with nothing blocked, verdict bytes are
unchanged. The public report therefore flags claims whose status may depend on
removed evidence rather than silently publishing a new justification.

\paragraph{The running example, compiled}
Compiling both runs of the program from \autoref{sec:overview} shows
what the aggregation reads (\autoref{fig:af}). The undischarged
question keeps \kw{a1} out of $A(P)$ altogether in run 1, so the
dead-end leaf is admitted but has no argument to target; discharging it
in run 2 admits \kw{a1} and turns the declared undercut into the edge
that defeats it. The lower panels show why reusing a checked term in
another world need not preserve its status: the target can contain an
additional attacker (\autoref{sec:worlds}).

\input{figures/af}

\paragraph{Consistency} With an arbitrary contrary relation and no
transposition of strict rules, grounded semantics guarantees only
closure under declared subarguments. Direct and indirect consistency can fail: two
contrary claims may both become \stJustified when their conflict
appears only behind unattackable strict inferences. The calculus
therefore imposes a compile-time restriction on
policies (\autoref{def:strict-wf}): under it every conflict is rebuttable at a
defeasible step, so strict closure introduces no new conflict and the
two consistency notions coincide by construction. The rationality
theorem assumes this restriction (\autoref{thm:rationality}). The cost
is an expressiveness limit: strict chains may target only uncontested
claims. Lifting this restriction would require a different consistency
argument.

\begin{definition}[Strict-chain well-formedness]
\label{def:strict-wf}
A proposition pattern is \emph{strict-reachable} in \Pol when it is
the conclusion pattern of a declared strict rule; this finite set is
closed under strict premises-to-conclusion steps, since every such step
ends in a strict conclusion. \Pol is \emph{strict-chain well-formed}
when no strict-reachable pattern may overlap either side of a declared
\kw{contrary} pair, where two patterns may overlap when they could
share a canonically equivalent ground instance (a conservative,
decidable check).
\end{definition}

%% file: figures/admissionrules.tex
\begin{figure}
\centering
\judgbox{\Pi_A \vdash \langle M, U \rangle \leadsto \mathit{outcome}}\quad\text{\small source admission over ordered leaf metadata $M$ and the declared unit $U$}
\small
\[
\begin{array}{c}
\inferrule*[left=\rulename{Decide}]
  {(k, p) \in \Pi_A}
  {\mathrm{decide}_{\Pi_A}(k, p) = \Pi_A(k, p)}
\qquad
\inferrule*[left=\rulename{Default}]
  {(k, p) \notin \Pi_A}
  {\mathrm{decide}_{\Pi_A}(k, p) = \kw{admit}}
\end{array}
\]
\[
\begin{array}{r@{\;=\;}l}
\Gamma_{\mathrm{policy}} & \{\, l \text{ declared} \mid \mathrm{decide}_{\Pi_A}(\mathrm{kind}(l), \mathrm{prov}(l)) = \kw{admit} \,\}\\
Q_{\mathrm{pol}} & \{\, l \text{ declared} \mid \mathrm{decide}_{\Pi_A}(\mathrm{kind}(l), \mathrm{prov}(l)) = \kw{quarantine} \,\}\\
Q_{\mathrm{grp}} & \{\, l \mid l \text{ is a member of an inconsistent duplicate-report group} \,\}\\
Q & Q_{\mathrm{pol}} \cup Q_{\mathrm{grp}}\\
\Gamma_{\mathrm{checked}} & \{\, l \text{ declared} \mid l \notin Q \,\}
\end{array}
\]
\smallskip

\parbox{0.92\linewidth}{\small The prune removes the leaves in $Q$, every
argument whose support tree uses a removed leaf, and every attack with a
removed raw endpoint. $\mathit{audit}$ is its canonical projection: removed
leaf rows with their causes (policy first, then one cause per inconsistent
containing group, in group declaration order), removed argument ids, and
removed raw attacks, all in declaration order.}
\[
\begin{array}{c}
\inferrule*[left=\rulename{Reject}]
  {\mathrm{valid}(\Pi_A,M,U) \\
   l \text{ is the first declared leaf with } \mathrm{decide}_{\Pi_A}(\mathrm{kind}(l), \mathrm{prov}(l)) = \kw{reject}}
  {\Pi_A \vdash \langle M, U \rangle \leadsto \mathrm{R8}(l)}
\\[20pt]
\inferrule*[left=\rulename{Admit}]
  {\mathrm{valid}(\Pi_A,M,U) \\
   \mathrm{decide}_{\Pi_A}(\mathrm{kind}(l), \mathrm{prov}(l)) \neq \kw{reject} \text{ for every declared leaf } l}
  {\Pi_A \vdash \langle M, U \rangle \leadsto \langle \Gamma_{\mathrm{policy}}, \Gamma_{\mathrm{checked}}, Q, \mathit{audit} \rangle}
\end{array}
\]
\caption{Policy admission at the source boundary, over the admission table
$\Pi_A$. Because \rulename{Admit} requires that no declared leaf rejects,
$\Gamma_{\mathrm{checked}}$ equals
$\Gamma_{\mathrm{policy}} \setminus Q_{\mathrm{grp}}$: the policy-admitted
leaves minus the members of inconsistent duplicate-report groups.}
\label{fig:admissionrules}
\end{figure}

%% file: figures/acceptance.tex
\begin{figure}
\centering
\small
\[
\inferrule*[left=\rulename{Accept}]
 {\mathrm{Static}(\Sig,\Pol,P) \quad
  \mathrm{NoDup}(\mathrm{ruleIds}(\Pol)) \quad \mathrm{NoDup}(P.\mathit{args}) \\
  \forall w\in P.\mathit{args},\ \exists p.\ \Sig;\Pol;\Ctx;\Reg\vdash w:\supports(p)\obl\varnothing \\
  \forall k\in P.\mathit{atts},\ \exists w,u,\pi.\ \Sig;\Pol;\Ctx;\Reg\vdash k:\attacks(w,u@\pi)
  \ \land\ w,u\in P.\mathit{args} \\
  \mathrm{AttackComplete}(P.\mathit{args},P.\mathit{atts})}
 {\Sig;\Pol;\Ctx;\Reg\vdash P\ \mathsf{ok}}
\]
\caption{Whole-program acceptance after decoding and admission.
$\mathrm{Static}$ requires a well-formed signature, well-sorted policy,
ground inputs and arguments, in-scope policy variables, and strict-chain
well-formedness (\autoref{def:strict-wf}). In the attack premise, $w$
and $u@\pi$ are the endpoints and occurrence named by $k$.
$\mathrm{AttackComplete}$ requires a declared attack from each source to
an occurrence contained in each contrary target whose root is a leaf or
a defeasible instance. Pairs are ordered and include self-pairs.
Decoding requires unique argument identifiers and declared endpoints;
$\mathrm{NoDup}(P.\mathit{args})$ also excludes equal support terms.}
\label{fig:acceptance}
\end{figure}

%% file: figures/af.tex
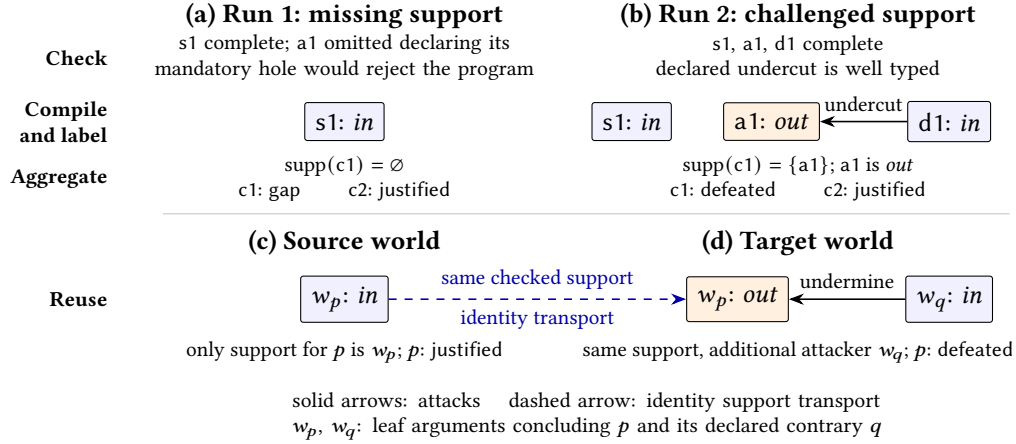
\begin{figure}
\centering
\begin{tikzpicture}[
  font=\small,
  arg/.style={draw, rounded corners=1pt, align=center, inner sep=4pt},
  accepted/.style={arg, fill=blue!6},
  defeated/.style={arg, fill=orange!12},
  attack/.style={-{Stealth[length=2mm]}, semithick},
  note/.style={font=\scriptsize, align=center},
  stage/.style={font=\scriptsize\bfseries, anchor=east, align=right}
]
\node[font=\small\bfseries] at (3,0) {(a) Run 1: missing support};
\node[font=\small\bfseries] at (9,0) {(b) Run 2: challenged support};
\node[stage] at (0,-0.55) {Check};
\node[note, text width=5.6cm] at (3,-0.55)
  {\kw{s1} complete; \kw{a1} omitted declaring its\\mandatory hole would reject the program};
\node[note, text width=5.6cm] at (9,-0.55)
  {\kw{s1}, \kw{a1}, \kw{d1} complete\\declared undercut is well typed};
\node[stage] at (0,-1.4) {Compile\\and label};
\node[accepted] (s1a) at (3,-1.4) {\kw{s1}: \emph{in}};
\node[accepted] (s1b) at (6.8,-1.4) {\kw{s1}: \emph{in}};
\node[defeated] (a1b) at (8.65,-1.4) {\kw{a1}: \emph{out}};
\node[accepted] (d1b) at (11,-1.4) {\kw{d1}: \emph{in}};
\draw[attack] (d1b) -- node[note, above] {undercut} (a1b);
\node[stage] at (0,-2.15) {Aggregate};
\node[note] at (3,-2.15)
  {$\mathrm{supp}(\kw{c1})=\varnothing$\\\kw{c1}: \stGap\qquad \kw{c2}: \stJustified};
\node[note] at (9,-2.15)
  {$\mathrm{supp}(\kw{c1})=\{\kw{a1}\}$; \kw{a1} is \emph{out}\\\kw{c1}: \stDefeated\qquad \kw{c2}: \stJustified};
\draw[gray!45] (0.6,-2.62) -- (11.8,-2.62);
\node[font=\small\bfseries] at (3,-3.0) {(c) Source world};
\node[font=\small\bfseries] at (9,-3.0) {(d) Target world};
\node[stage] at (0,-3.75) {Reuse};
\node[accepted] (src) at (3,-3.75) {$w_p$: \emph{in}};
\node[defeated] (tgt) at (8.2,-3.75) {$w_p$: \emph{out}};
\node[accepted] (extra) at (11,-3.75) {$w_q$: \emph{in}};
\draw[attack, dashed, blue!65!black] (src) --
  node[note, above] {same checked support}
  node[note, below] {identity transport} (tgt);
\draw[attack] (extra) -- node[note, above] {undermine} (tgt);
\node[note] at (3,-4.45) {only support for $p$ is $w_p$; $p$: \stJustified};
\node[note] at (9,-4.45) {same support, additional attacker $w_q$; $p$: \stDefeated};
\node[note, text width=11cm, anchor=north] at (6.2,-4.85)
  {solid arrows: attacks \quad dashed arrow: identity support transport \quad
   $w_p$, $w_q$: leaf arguments concluding $p$ and its declared contrary $q$};
\end{tikzpicture}
\caption{From checked support to claim status. Both running-example programs
(a--b) are accepted; the empirical claim changes from missing support to
defeated support, while the separate arithmetic claim stays justified.
In the mechanized transport witness (c--d), the source declares only $w_p$,
the target also $w_q$ and its undermine. Transport preserves the complete
support term and its empty obligation set, yet the additional attacker
changes $p$'s status.}
\label{fig:af}
\end{figure}

%% file: src/worlds.tex
\section{Cross-context comparison semantics}
\label{sec:worlds}

Research papers are often comparable without using identical checking
contexts. A vocabulary may be renamed, a policy extended, or additional
evidence admitted. Comparing their claims requires stating what carries
across and what must be checked again. We now formalize the
possible-world account introduced in \autoref{sec:background}: each
accepted program supplies a world with local claim statuses, and
declared, applicable bridges determine which worlds a modal query
compares. This layer preserves the local calculus.

\subsection{Worlds and observations}

\paragraph{A review scenario}
Suppose an earlier paper's encoded claim $c$ is \stJustified. A reviewer
transports its supporting argument into a world that also records
evidence from a new paper. The argument may still check, but an additional
attacker can make the translated claim \stDefeated. This is the shape
of the mechanized transport counterexample in
\autoref{sec:cross-context-guarantees}. If instead a proposed
translation fails to preserve an admitted premise or policy rule,
the structural-bridge contract fails before status preservation can be
claimed. A difference in settings is therefore neither automatically
irrelevant nor automatically a refutation.

In the checked-program instance, a world is an accepted program, not a
paper identifier. Several papers
may contribute to one world; distinct worlds may share a context.
Changes in evidence or attacks can distinguish worlds without changing
their signature, policy, or registry. Context changes additionally
require relating those checking parameters. In either case a bridge
specifies the comparison.

\begin{definition}[Outer model]
\label{def:outer-model}
An \emph{outer frame} carries a set of contexts, each with a type of
worlds and queries. A \emph{bridge} between two contexts carries a
candidate relation on worlds, an applicability judgment, and a partial
query translation.
An edge is \emph{accepted} when the candidate relation and the
applicability judgment both hold, independently of the formula being
evaluated. The modal language has status atoms over a world's
queries, negation, conjunction, and one box and diamond per bridge;
satisfaction is Kripke-style over accepted edges. The valuation is a
parameter assigning propositions to status atoms at worlds.

In the \emph{checked-program instance}, a context fixes a signature,
policy, registry, and checker parameters, and its worlds are accepted
checked units. We read this instance under either the direct source
observation or compiled evaluation. Both valuations assign exactly one
status to each query; the generic frame model imposes no such constraint.
\end{definition}

For a bridge $b$ with claim translation $\tau_b$, the formula
$\langle b\rangle\mathsf{Justified}(\tau_b c)$ asks whether some
accepted successor justifies the translated claim; the corresponding
box asks whether every accepted successor does. A robust-justification
query also requires a successor to exist, so an empty comparison set
cannot establish robustness vacuously. Accessibility ranges over declared,
applicable comparisons, not every imaginable setting.
The one-world, no-bridge instance recovers local semantics by
definitional equality.

\subsection{Structural bridges}

\begin{definition}[Structural bridge]
\label{def:structural-bridge}
A \emph{structural bridge} between two contexts sharing a
canonicalizer comprises a partial symbol translation on the
predicate and constructor namespaces, a total renaming of evidence
leaves, and three clauses: an admitted leaf is admitted at the
translated proposition; the target policy carries, at the same rule
identifier, exactly the translated rule, with mode, parameters,
premise order, question keys, and certifier allowlist preserved
verbatim; and certificate acceptance survives translation of the
encoded step. Everything else (variables, literals, rule
identifiers, question keys, hole sets, assurances) is carried across
verbatim, and a lift is undefined exactly at an out-of-vocabulary
symbol.
\end{definition}

A structural bridge specifies how support terms translate. Preserving a
claim's status additionally requires every source argument to have a
target transport, every target argument to arise from a source argument,
and compiled attacks to correspond in both directions. The executable
status checker tests source coverage, target coverage, forward attacks,
and backward attacks. The support and status guarantees are stated in
\autoref{sec:cross-context-guarantees}.

\paragraph{Scope of transport}
The structural contract supports exact translations: it preserves rule
identifiers, question keys, and obligations. A near match that changes
a rule's assumptions does not satisfy that contract merely because its
vocabulary translates. Such a comparison requires additional evidence
or a different bridge argument. General bridge applicability is a
parameter; the executable status checker decides the stated structural
correspondence conditions, not relevance to the domain.

Undefined claim translations produce an incomparability reason; a
defined query with no support produces \stGap. The layer supplies no
approximation theory for limits or numerical error and no presentation
syntax for the outer language. Its modal formulas describe support
statuses in accepted worlds, preserving the non-factivity of the local
calculus.

%% file: src/metatheory.tex
\section{Metatheory}
\label{sec:metatheory}

A research artifact should let a reader inspect why a claim is
supported, compute its standing among declared challenges, and determine
what can be reused when evidence or context changes. The metatheory
establishes guarantees for these operations. Checking makes support
decidable and its dependencies auditable. The central local result,
status preservation, connects the artifact's declared arguments to the
evaluator's reports. Certificate soundness identifies what strict
inference establishes, while update and transport results characterize
how support and claim status behave under change.

These guarantees make the distinction between checked support and claim
justification precise. A well-formed argument can be defeated, and a
transported argument can encounter new challenges. All results are
relative to the declared checking inputs; they do not establish the
accuracy of admitted measurements or the adequacy of the chosen policy.
We retain the specification's labels G1--G6 for decidability,
accountability, compilation soundness with status preservation,
strict-certificate soundness, backend replacement, and consistency. Cross-context results follow the
local results in \autoref{sec:cross-context-guarantees}.

The semantic results are mechanized in Lean~4. Three arguments remain
on paper: the pipeline cost bound (\autoref{prop:complexity}), the
machine-model bookkeeping for NP-completeness behind
\autoref{thm:np-hardness}, and the closure and indirect-consistency
clauses of \autoref{thm:rationality}. The Haskell implementation is validated by differential
tests, which do not prove equivalence on arbitrary inputs
(\autoref{sec:evaluation}). \appautoref{app:mechanization} records the
mechanized scope; \appautoref{app:proofs} supplies the proofs. Details
of realizability, extension-semantics distinctions, linking, and surface
correctness appear in the appendix.

\subsection{Checking}

An artifact producer supplies an explicit support term so that a reader
can check the proposed inference independently. This requires a
terminating acceptance procedure and an unambiguous account of what the
term supports. G1 provides the procedure; conclusion uniqueness supplies
the latter property and makes checking attacks well-defined.

\begin{theorem}[Decidability (G1)]
\label{thm:decidability}
Program and attack checking terminate. For any \Sig, \Pol, \Ctx, and
candidate term or attack, the checker decides acceptance.
\end{theorem}

\begin{proposition}[Conclusion uniqueness and support adequacy]
\label{prop:support}
A support term determines its conclusion and its obligation set: two
derivations for the same $w$ agree on both. Deciding whether $w$
supports a claim therefore reduces to one run of the support checker
followed by a normalized identity test on that unique conclusion, and
the relation respects $\equiv$.
\end{proposition}

Uniqueness is what lets the attack rules name ``the'' conclusion of an
attacker, so the proof of G1 consumes this proposition rather than
merely restating it.

Auditing also requires knowing what the accepted support depends on.
For example, a reader investigating a disputed measurement must be able
to find it among the leaves used by an argument; a strict certificate
must account for the premises and theory entries it uses. G2 gives exact
leaf accounting and coverage of backend dependencies.

\begin{theorem}[Dependency accountability (G2)]
\label{thm:accountability}
The reported leaf-dependency set of a checked term $w$ equals
$\leaves(w)$, every member is declared in \Ctx, and the reported
backend dependencies $\certDeps(w)$ contain every free premise and
theory entry each accepted certificate uses.
\end{theorem}

\subsection{Compilation and claim statuses}

The artifact's claim report must reflect the support and challenges
expressed in its source. A compilation error could instead introduce a
spurious attacker or direct an attack at an unrelated argument.
Compilation soundness rules out malformed compiled structure and
misdirected closure edges. Determinism then ensures that fixed checking
inputs yield a reproducible report. Status preservation establishes the
central semantic guarantee: that report agrees with a direct reading of
the source.

\begin{theorem}[Compilation soundness (G3)]
\label{thm:compilation}
No untyped node or attack enters the compiled framework; subargument
closure adds edges only to arguments containing the attacked
occurrence.
\end{theorem}

\begin{theorem}[Determinism and termination of evaluation]
\label{thm:determinism}
Grounded evaluation and claim aggregation are deterministic and
terminating on the finite compiled framework.
\end{theorem}

Termination and determinism alone do not connect the report to the
source's meaning. To state that connection, we define claim statuses
directly from checked source attacks, without building the framework.
The mutually inductive predicates below provide this independent side
of the status-preservation theorem.

\begin{definition}[Direct claim status]
\label{def:direct-status}
For a checked program $P$, let \emph{directly in} and \emph{directly
out} be the least predicates on $A(P)$ satisfying two conditions. A
term $w$ is directly in when every term $v$ that attacks an occurrence
in $w$ through a checked attack declared in $P$ is directly out. A term
$v$ is directly out when some directly-in term attacks an occurrence in $v$.

The \emph{direct status} of a claim $c$ is \stGap when
$\mathrm{supp}(c) = \varnothing$, and \stJustified when some member of
$\mathrm{supp}(c)$ is directly in. It is \stContested when no member is
directly in and some member is not directly out. It is \stDefeated when
no member is directly in and every member is directly out.
\end{definition}

\begin{theorem}[Status preservation (G3)]
\label{thm:preservation}
The direct source semantics and the compiled-framework semantics
agree: for every claim, the direct status of
\autoref{def:direct-status} is exactly the status of
\autoref{def:status}, in both directions.
\end{theorem}

Status preservation lets a reader interpret the computed report in
terms of the source arguments, without inspecting the compiled graph.
It does not by itself ensure that the source semantics treats strict
inference and conflict coherently. G6 establishes that additional
property. Its strict-chain hypothesis matters because otherwise a
conflict can occur behind strict inferences that cannot be rebutted
(\autoref{def:strict-wf}).

\begin{theorem}[Rationality postulates (G6)]
\label{thm:rationality}
Closure under declared subarguments holds without the strict-chain
hypothesis, and so does closure under
declared strict steps: a declared strict instance whose immediate
premise arguments are declared and accepted is itself accepted. Under
strict-chain well-formedness (\autoref{def:strict-wf}), direct
consistency holds, so two contrary claims are never jointly
\stJustified, and closing the accepted conclusions under strict rules
introduces no new conflict, so indirect consistency coincides with
direct consistency.
\end{theorem}

Direct consistency is mechanized for the reference checker over the
complete claims of an accepted unit; the closure and
indirect-consistency clauses are short paper arguments over the
mechanized edge relation (\appautoref{app:proofs}).

Evidence admission creates a separate risk. Quarantining a questionable
measurement can remove an argument that attacks another claim; ordinary
grounded evaluation may then accept that claim. A public report should
not gain justification merely because relevant material was excluded
from checking. The conservative reporting rule prevents this promotion.

\begin{theorem}[Quarantine non-promotion]
\label{thm:nonpromotion}
If the production checker publishes \stJustified for a claim after
quarantine, then that claim is \stJustified in the declared framework
with the quarantined material reinstated.
\end{theorem}

The implication is one-way: status \emph{equality} would require equal
complete-support sets, which quarantine need not preserve
(\appautoref{app:proofs}).

\paragraph{Realizable frameworks}
A framework is realizable when some accepted program compiles to it,
up to an isomorphism preserving conclusions and attacks.
Every realizable framework satisfies range and conflict-completeness
invariants, but these conditions are insufficient for an arbitrary
fixed policy. \appautoref{sec:image} gives the definitions and
counterexample; the cost results below concern this realizable class.

\subsection{Semantics-parametric observation}
\label{sec:observation}

Grounded evaluation is one choice among Dung's semantics, so a fair
question is which status guarantees depend on that choice. This
subsection makes the semantics a parameter. Compilation never
inspects it: the compiled framework stays fixed, and only the status
observation varies.

\begin{definition}[Extension semantics]
\label{def:ext-semantics}
An \emph{extension semantics} pairs a specification, a relation
between a finite framework and a candidate extension, with an
enumeration, such that on any duplicate-free carrier the enumeration
lists exactly the carrier subsets satisfying the specification. We
instantiate five: \emph{grounded}, specified first-order as the least
complete extension, and \emph{complete}, \emph{preferred},
\emph{stable}, and \emph{semi-stable} in their standard
forms~\citep{Dung_1995}.
\end{definition}

\begin{definition}[Claim observation]
\label{def:observation}
Fix an extension semantics and a framework $F$. The
\emph{observation} of a claim $c$ is \stGap when
$\mathrm{supp}(c) = \varnothing$; the report \emph{no-extension} when
the enumeration of $F$ is empty; \stJustified when every extension
contains a member of $\mathrm{supp}(c)$; \stDefeated when every
extension attacks every member of $\mathrm{supp}(c)$; and
\stContested otherwise.
\end{definition}

The guards are evaluated in the stated order. Testing support first
makes \stGap independent of the semantics and prevents empty support from
satisfying the defeat condition vacuously. Testing extension existence
next prevents both universal conditions from succeeding when no
extension exists. For \stJustified, the supporting argument may differ
between extensions: every extension must contain some support, but no
single argument need belong to them all.

This separates two questions an artifact reader may ask: whether the
artifact supplies any complete support, and whether that support survives
a chosen treatment of conflict. The next theorem connects this general
interface to the grounded report used throughout the paper.

\begin{theorem}[The printed status is the grounded instance]
\label{thm:observe-grounded}
On a duplicate-free carrier, the grounded enumeration has one
element, and the observation under grounded semantics is the status
of \autoref{def:status}. Skeptical and credulous
argument acceptance collapse under grounded semantics.
\end{theorem}

The remaining results separate what depends on the extension semantics.
Only \stGap is semantics-independent. Stable semantics may have no
extension, which produces the explicit no-extension report; a source
observation transport theorem generalizes agreement under
attack-extensionality and bounded claim support. The statements,
separation examples, and necessary hypotheses appear in
\appautoref{app:observations}.

\subsection{Strict-certificate backends}

A research argument may combine a checked numerical comparison with a
logical derivation. The artifact must record what each proof tool
establishes without treating either as a proof of the source claim's truth.
Isolation enforces this source boundary; G4 supplies the positive
guarantee that an accepted certificate establishes the encoded
consequence in its backend's logic.

\begin{theorem}[Strict-certificate isolation and non-factivity]
\label{thm:isolation}
No support term, attack, or backend proof term crosses the
strict-certificate interface, and no backend formula or proof term is
eliminated into a source truth judgment. Programs cannot extend \Reg.
\end{theorem}

\begin{theorem}[Strict-certificate soundness (G4)]
\label{thm:strict-soundness}
For a certified strict instance of $P_1,\dots,P_n \Rightarrow C$ at
$\theta$, the encoded conclusion is a backend consequence of the encoded
premise conclusions and declared theory:
$T ; [\enc\backend(P_i\theta)] \bmodels \enc\backend(C\theta)$. The
result excludes \kw{trusted}-policy instances.
\end{theorem}

Each registered backend discharges the obligation in its own way:
\textsf{nd@1} by induction on the natural-deduction derivation,
\textsf{ra@1} by witness cancellation over exact integer
cross-multiplication, \textsf{ord@1} by the decided order on the
two canonical numerals, and \textsf{insp@1} by checking membership,
absence, or uniqueness in declared exhaustive inventories
(\appautoref{app:proofs}).

For the comparison in the introduction, G4 validates the inference
between encoded accuracy values under its declared premises. Whether
those values describe comparable experiments remains a source-level
support question. G5 addresses a different concern: changing the proof
tool should preserve claim reports when it accepts exactly the same
strict instances.

\begin{theorem}[Backend replacement (G5)]
\label{thm:replacement}
Source-identical programs related by an injective relabelling of
certificate payloads, whose backends accept the same strict instances,
compile to the same argumentation framework and assign every claim the
same status.
\end{theorem}

Replacement compares whole programs one backend at a time. Within one
artifact, different inference steps may need different proof tools.
Heterogeneous compositionality establishes that these steps can form one
checked support term while each retains its own consequence guarantee
and dependency account.
Call a node of a support term a \emph{certified occurrence} when it
instantiates a strict rule under a certificate assurance. The
occurrence records its rule, substitution, backend identity, theory
digest, and certificate reference; its premises and conclusion are
derived from the rule and substitution rather than stored, so an
occurrence cannot disagree with its own typing.

\begin{theorem}[Heterogeneous backend compositionality]
\label{thm:hetero}
Let $w$ be a checked support term. Every certified occurrence in $w$
names a strict policy rule that allows its backend identity and
theory digest, its instantiated premises and conclusion resolve, its
own registered checker accepts its certificate against them, and the
occurrence-local consequence holds: the encoded conclusion follows,
in that backend's own logic, from the selected encoded premises and
resolved theory. The backend, its formula type, its theory, and its
consequence relation are quantified independently at each
occurrence. Moreover, $\certDeps(w)$ is exactly the union of the
occurrence-local reports, each computed by its occurrence's own
backend.
\end{theorem}

A parent checks a child's conclusion and obligations without inspecting
its backend. Replacing a premise or discharge subterm by another with
the same conclusion and obligation set therefore preserves typing.
The theorem composes source support judgments; it requires no
translation or common model theory between backends.

\paragraph{Linking and certificate representation}
The replacement theorem also holds inside admissible linking contexts,
including contexts that fill critical-question holes. The extension
quantifies over all extension semantics and over partial-bijective
relations between certificates. We give the fragment calculus, the
precise admissibility conditions, and these results in
\appautoref{sec:context}.

\subsection{The source boundary}
Admission determines the checking context before core checking. Its
eight properties are stated in \appautoref{app:mechanization}.
Endpoint-safe pruning preserves declared argument identities and removes
attacks whose endpoints disappear (\autoref{pty:pruning}). Tightening
admission can only remove structure (\autoref{pty:monotonicity}), but
removing an attacker can change a retained argument's label.
Source non-promotion applies the conservative reporting rule at this
boundary: a published \stJustified claim was already justified with
the quarantined material reinstated (\autoref{pty:source-nonpromotion}).

\subsection{Source updates and status dynamics}
\label{sec:updates}

Research artifacts change as authors add evidence, supply an argument,
or record a challenge. These edits have different meanings: a new
measurement alone supplies no inference connecting it to a claim, while
a new attack can defeat existing support. We therefore ask which status
transitions each accepted source edit can cause. A closed, one-step
update calculus makes this question precise. Every reachable transition carries an accepted
witness program, every unreachable transition a theorem, and the
public reporting layer recovers non-promotion as a corollary of the
resulting matrix (\autoref{cor:matrix-nonpromotion}).

\begin{definition}[Source states and updates]
\label{def:source-update}
A \emph{source state} carries exactly the raw inputs from which
admission and whole-unit checking derive: signature, policy,
admission table, leaves and metadata, raw arguments, raw attacks,
duplicate groups, and ground input. It is \emph{accepted} when
admission accepts the raw carriers and the checker accepts the
pruned unit that admission produces. A \emph{source update} is one
of four constructors, each with a decidable side condition:
\upAddLeaf (a fresh leaf and metadata row), \upTighten (move an
admitted admission key to \kw{quarantine}), \upAddAttack (a raw
attack between declared arguments), and \upAddInstance (a fresh raw
argument). Applying an update checks the side condition, edits only
the corresponding raw carrier, and reruns admission and the
whole-unit checker; it returns the edited state only when both
accept, and otherwise one of a closed set of rejections.
\end{definition}

An edit must first leave an artifact that can still be checked. The
following sufficient conditions expose the obligations on an editing
tool, including freshness, endpoint retention, and attack completeness.
Preserving acceptance permits the claim statuses themselves to change.

\begin{theorem}[Acceptance preservation]
\label{thm:update-preservation}
Each constructor maps an accepted source to an accepted target under
explicit sufficient conditions: freshness and an admitted metadata
key for \upAddLeaf; a tightened table at least as restrictive,
valid, and rejection-free, with retained-leaf agreement and
surviving coverage witnesses, for \upTighten; resolution, typing,
and endpoint retention for \upAddAttack; and freshness,
well-sortedness, complete support, and extended attack completeness
for \upAddInstance. No constructor succeeds unconditionally.
\end{theorem}

\begin{theorem}[Grounded transition matrices]
\label{thm:update-matrices}
For grounded observations, of the 64 cells of update kind by source
status by target status, exactly 37 are reachable: 4 for \upAddLeaf,
13 for \upTighten, 10 for \upAddAttack, and 10 for \upAddInstance
(\appautoref{tab:updates}). Each reachable cell carries an accepted
source and target program with a successful update between them, and
a theorem excludes each unreachable cell under its constructor's
stated premises. In particular, \upAddLeaf fixes every core status,
and a fresh instance enters the old framework as an unattacked sink,
so \upAddInstance never maps \stJustified or \stContested to
\stDefeated.
\end{theorem}

These are individual edit operations. \upAddInstance adds no attack
declaration; adding an argument with attacks requires multiple
operations and falls outside its one-step matrix.

The matrices distinguish possible effects from guaranteed invariants;
they do not predict a particular edit's outcome without evaluating its
arguments and attacks. One distinction survives every choice of
extension semantics: supplying a leaf or a challenge cannot fill a
missing support argument. The next result makes this boundary explicit.

\begin{theorem}[The \stGap boundary, semantics-parametrically]
\label{thm:gap-boundary}
For every extension semantics: a successful \upAddAttack preserves a
source \stGap; no successful additive update (\upAddLeaf,
\upAddAttack, or \upAddInstance) moves a non-\stGap observation to
\stGap; and every non-instance update (\upAddLeaf, \upTighten, or
\upAddAttack) preserves a source \stGap. \upAddInstance is the one
constructor that can leave \stGap.
\end{theorem}

\paragraph{The public layer} Public reports take five values: the
four statuses plus \stEvidenceBlocked. Call an accepted run
\emph{clean} when its prune removes nothing. On a clean source the
public report equals the grounded core report, a successful additive
update preserves cleanliness, and additive public transitions are
therefore exactly the core transitions, never reaching
\stEvidenceBlocked. Cleanliness cannot be dropped: on a real
accepted, non-clean source, one added attack grows the blocked
closure and moves a public \stJustified to \stEvidenceBlocked while
the conditional core status reads \stDefeated. \upTighten may
prune, so its public matrix is separate: of its 20 cells, 13 are
reachable, three of them into \stEvidenceBlocked, and a public
\stJustified target is reachable only from a \stJustified source.

\begin{corollary}[Matrix non-promotion]
\label{cor:matrix-nonpromotion}
From a clean accepted source, if a successful \upTighten publishes
\stJustified for a claim, then that claim is \stJustified in the
source's declared framework, the framework with the quarantined
material reinstated. The proof reads the tighten column of the
public matrix and the clean-source identities rather than
\autoref{pty:source-nonpromotion}, so the reporting-layer guarantee
is recovered from the dynamics, not assumed.
\end{corollary}

\paragraph{Against belief revision} A belief-set projection exists,
the set of claims observed \stJustified, so AGM-style postulates are
statable. The two we probed both fail by mechanized counterexample:
success, because a fresh instance's sole support can be defeated on
arrival, and inclusion, because one added attack can remove a
previously justified claim. Update in \Lara is argumentation
dynamics, not belief revision, and we claim no AGM compliance.

\paragraph{Surface correctness}
An independent judgment for the parsed presentation syntax connects
authoring to the core: elaboration preserves checking and reflects it
on the supported fragment. Direct surface observations agree with
compiled observations for attack-extensional extension semantics.
\appautoref{sec:surface} states the results and their parser boundary.

\subsection{Cost}
\label{sec:cost}

A checkable artifact also needs a computational account of its reports.
The realizability definition in \appautoref{sec:image} fixes the class of
frameworks \Lara programs can denote; this subsection bounds status
queries on that class. Realizability matters here: a difficult arbitrary
framework would say little about artifact evaluation if no accepted
program could express it. The cost model counts attack-relation queries
in an instrumented mirror of the grounded evaluator whose value
provably equals the reference evaluator's, in the short-circuit
order the Boolean connectives actually evaluate. The bounds measure this evaluator, not a lower bound for every
algorithm implementing grounded semantics. The two bounds of
\autoref{thm:grounded-cost} apply to every input framework; the tightness witness and
the reduction of \autoref{thm:np-hardness} are statements about the
realizable class, and each carries its own realization in one fixed
context.

\begin{theorem}[Grounded query bounds]
\label{thm:grounded-cost}
On every finite framework with $n$ arguments, grounded evaluation
issues at most $n^3(1+n)$ and at least $n^2$ attack queries, and a
four-state status query issues at most $n^3(1+n) + 2n^2$. The upper
bounds are tight up to degree on the realizable class: a family of
accepted programs, realizable in one fixed signature, policy, and
registry, has size $3k$ and costs at least $k^4$ queries for every
$k \ge 2$.
\end{theorem}

The quantifiers differ by design: the quadratic lower bound is
universal, while the quartic cost is a worst case on a realizable
family; no instance-wise cubic floor holds, the two-node all-attacks
framework costing exactly four queries.

The preceding bound describes the reference grounded evaluator. Choosing
a richer acceptance query raises a separate question: do the language's
structural restrictions eliminate the difficulty of that query? The next
reduction shows that credulous complete acceptance remains hard even
when every constructed framework comes from an accepted artifact in one
fixed checking context.

\begin{theorem}[Hardness on the realizable class]
\label{thm:np-hardness}
Fix one explicit signature, policy, and registry. A computable
reduction maps each 3SAT formula $\varphi$ to a structured framework
that is realizable in that fixed context, with $2v + m + 1$ nodes
for $v$ variables and $m$ clauses and encoded size at most
$64\,(\lvert\varphi\rvert + 1)^2$, such that $\varphi$ is
satisfiable exactly when the framework's query claim is credulously
accepted under complete semantics. The statement is genuinely
restricted: the one-node self-attacking framework is not realizable
in the same context.
\end{theorem}

Credulous complete acceptance on \Lara-realizable frameworks is
therefore NP-complete: hardness by the reduction, membership by
guessing an extension and checking it in polynomial time. The
reduction, its realization, and its size bounds are mechanized; the
NP bookkeeping over encodings and the machine model stays on paper,
with the counting argument below.

\begin{proposition}[Polynomial core checking and evaluation]
\label{prop:complexity}
Let $N$ be the size of the explicit core checking input, including the
signature, policy, evidence tables, substitutions, and support trees.
Assume linear-time literal normalization and count backend encoding and
replay as opaque calls. Support and attack checking, whole-program
acceptance, compilation, and grounded claim-status computation perform
polynomially many syntactic operations and backend calls in $N$.
This bound excludes surface elaboration and the work inside backends;
it does not assert linear time for the list-based checker.
\end{proposition}

\subsection{Cross-context guarantees}
\label{sec:cross-context-guarantees}

The outer model and structural bridges of \autoref{sec:worlds} build on
the local calculus. Direct source observations and compiled evaluation
agree at every modal formula: \autoref{thm:preservation} gives agreement
at each world, and satisfaction is extensional in the valuation.
The practical question is what an artifact reader may carry from one
context into another. We separate an executable observation of declared
worlds, reuse of a checked support term, and preservation of the claim's
status. Each requires its own guarantee.

A bridge records an allowed comparison; its existence does not make the
comparison reversible or ensure that successive comparisons compose into
an available direct one. The modal laws respect this freedom. Comparison
adequacy additionally ensures that the executable report answers the
stated modal query and distinguishes failure to compare from a claim
status.

\begin{theorem}[Modal normality and comparison adequacy]
\label{thm:pw-normal}
The K axiom, necessitation, and box--diamond duality hold in every
generic outer model and hence in the checked-program instance.
Each of T, D, B, 4, and 5 has a countermodel with unrestricted valuations;
these establish neither checked-program counterexamples nor a complete
axiomatization. An
executable comparison decides the diamond: given a finite
presentation of a bridge's candidates and acceptance, its output
contains a status exactly when the model satisfies the
corresponding diamond formula, and an incomparable query is
reported by reason, an undefined translation, no candidate, or all
candidates rejected, never as a status.
\end{theorem}

Reusing an argument requires more than an accessible world. Its
translated steps must check, its obligations must survive, and its
certificates must replay in the target registry. Exact support
transport supplies this contract, including agreement between composite
and stepwise translation.

\begin{theorem}[Exact support transport]
\label{thm:support-transport}
Along a structural bridge, a checked support term whose translation
is defined checks in the target context with the translated
conclusion and the verbatim obligation set; completeness transports,
claim-level support survives normalized identity, and every
certified occurrence of the transported term replays against the
target registry, in the sense of \autoref{thm:hetero}. Transport
along a composite of bridges equals stepwise transport along the
path, and a chosen direct bridge agrees with a chosen path exactly
when their symbol and leaf maps are equal, a commuting condition
that is characterized, not assumed.
\end{theorem}

Transport is not status preservation. A mechanized witness
transports a source argument whose claim is \stJustified onto a
target world where the same translated claim is \stDefeated: the
target may hold attackers the source lacks. Preservation needs the
target's arguments and attacks matched, and that is the next
theorem.

\begin{theorem}[Conditional status preservation]
\label{thm:status-transport}
Let a structural bridge with an injective translation relate two
worlds such that (i) every source argument transports into the
target program; (ii) every target argument is a transport of some
source argument; and (iii) compiled attack edges correspond in both
directions along the transport. Then for every translatable query,
the target status of the translated claim equals the source status,
in all four states. The support correspondence follows from the bridge
contract; the argument and attack clauses supply a total attack
bisimulation. The executable checker decides four clauses soundly and
completely: source coverage, target coverage, forward attacks, and
backward attacks. These hypotheses compose along bridge paths.
With at least one accepted successor and every accepted successor so
related, each local status atom agrees with its boxed and diamonded
translation.
\end{theorem}

These conditions identify when a reader can reuse a claim's status as
well as its support. The checker makes the conditions testable, while
their composition permits reuse along a path of contexts. If they fail,
support transport may still apply, but the target status must be
established from the target's arguments and attacks.

%% file: src/evaluation.tex
\section{Mechanization and implementation}
\label{sec:evaluation}

We mechanize the semantic guarantees in Lean~4 and validate a Haskell
implementation against the executable Lean semantics. A worked example
shows how the implementation separates certified arithmetic from
defeasible empirical support.

\subsection{Mechanization}
At implementation revision \texttt{b62d33e0}, the Lean tree contains
144 files and roughly \num{117000} lines. The audited public
metatheory is \textsf{sorry}-free and uses only Lean's standard trio
(\textsf{propext}, \textsf{Classical.choice}, \textsf{Quot.sound}).
Some private executable examples additionally trust native evaluation
through \texttt{native\_decide}; \appautoref{app:mechanization} details
the audit boundary and workflow. We mechanize thirty-four of the
thirty-five specification results in full, including backend dependency
accounting for all four registered backends. Future backends must
discharge the same interface laws (\autoref{sec:metatheory}).

The cross-context proofs and examples establish the behavior of declared
bridges (\autoref{sec:cross-context-guarantees}); they do not evaluate
bridge discovery or automated literature-based review.

\subsection{Implementation conformance}
The Haskell implementation covers the checking pipeline and four strict
backends. A differential harness uses the compiled Lean semantics as
its reference interpreter and requires byte-identical
reports and exit codes from both drivers. At \texttt{b62d33e0}, the core
differential agrees on 673/673 verdict-bearing inputs, including
accepted programs and semantic rejections, and 66/66 malformed inputs.
These finite suites
check implementation conformance; they do not prove equivalence of the
two implementations. The surface elaboration theorems start at a parsed
AST (\appautoref{sec:surface}); tests validate the Haskell parser and
elaborator.

The trusted base, further conformance checks, and timings
appear in \appautoref{app:measurements}. An exploratory study of
machine-annotated research artifacts appears in \appautoref{app:corpus}.
Its results describe the supplied encodings under a chosen policy;
the semantic guarantees hold regardless.

\subsection{Ablation study: effects of removing checks}
\label{sec:ablation-summary}
At revision \texttt{7d0f7384}, we reran 655 inputs with selected acceptance
checks disabled (\autoref{fig:acceptance}, \autoref{tab:ablation}).
Removing obligation checking admits 18 incomplete-support programs;
removing coverage admits five missing-attack programs; disabling attack
typing as well admits 32 further invalid-target or unlicensed-attack
cases. These seeded defects demonstrate detection of the intended
failures, not their prevalence in research artifacts. The ablation,
conformance, and timing revisions are recorded separately in
\appautoref{app:measurements}.

\input{tables/ablation}

\subsection{Arithmetic and empirical support}
The worked-example suite covers all four statuses and three attack
kinds, with expected reports replayed by the differential harness.

One example draws the boundary between arithmetic and empirical
meaning. \textsf{ord@1} certifies only a ground inequality between
canonical decimal numerals, while a defeasible, attackable bridge rule
states that those numerals measure the systems the claim names. One
\texttt{comparison} block specifies the systems, measurand, dataset,
relation, and evidence leaves, and a validated elaborator emits all
three pieces: the strict arithmetic check, the defeasible bridge, and
the arithmetic subclaim. Sibling programs show why the split matters.
In a tie, arithmetic blocks the stronger claim without reviewer
judgment: equal cells admit only the non-strict comparison certificate,
replay rejects the strict version, and the bridge can conclude
\emph{at least as good} but never \emph{beats}. Under attack, a
settings audit undermines the bridge's binding premise, the attestation
that the two cells are comparable, so the bridge and the headline claim
become \stDefeated while the certified inequality remains \stJustified
(\appautoref{lst:ord-under-attack}). The certificate established the
arithmetic and nothing more, which makes the non-factivity of
\autoref{thm:isolation} visible in the verdict.

\subsection{Demonstrations beyond empirical claims}
Two later demonstrations exercise the same calculus with assumed and
attested premises (\appautoref{app:cs-philmath} and
\appautoref{app:cs-axiom-withdrawal}). A philosophical debate checks
contestation and reinstatement without measurements or a strict backend.
An axiom-withdrawal example checks source admission and certified
hypothesis reuse, with a separate executable Lean witness for the
structural transport boundary. We cite their sources at revision
\texttt{0f2a6dce}; these examples supplement the earlier conformance counts
reported above and introduce no new kernel rule.

\subsection{Limitations}
The prose-to-formal binding is untrusted by design, so a valid
certificate can sit on an unfaithful formalization
(\autoref{sec:leaves}). Strict-chain well-formedness excludes conflicts
involving strict-rule conclusions (\autoref{sec:semantics}). The only
producer exercised end to end so far is the deterministic lowering:
how well other untrusted producers fare against the checker is untested,
though no soundness result depends on the answer.

Lowering itself also has a limit. ARA is an artifact format rather
than a programming language, so an arbitrary artifact has no
canonical compile step: the ARA Compiler already uses
a language model to reconstruct an artifact from a finished
paper~\citep{liu2026humanwrittenpaperagentnativeresearch}, and lowering such an artifact to \Lara
inherits the same untrusted, model-mediated step. The architecture
absorbs this by design, since the checker alone decides acceptance,
but a rejected program then locates a fault in the proposed lowering,
not necessarily in the research. The clean path runs through
auto-research itself: an agent that already records claims, evidence,
and attacks can emit \Lara programs alongside them as one more
intermediate artifact, so support becomes checkable at authoring time
instead of reconstructed after the fact.

We design the surface syntax only far enough to exercise the calculus end
to end and left their design as an authoring language unexamined,
because the contribution is the core calculus and semantics behind the elaborator
boundary, which any surface can reach. Future work treats that surface
as a target for language models: which syntax an agent emits most
reliably and efficiently, and how emitting \Lara programs fits into a model-driven
research workflow.

%% file: tables/ablation.tex
\begin{table}
\caption{Removing checks admits malformed programs. Each configuration runs
on 655 manifest inputs. Counts are newly admitted programs that the
full checker rejects; all other outcomes are unchanged.}
\label{tab:ablation}
\centering
\small
\begin{tabular}{@{}llr@{}}
\toprule
Check removed & Newly admitted defect & Programs \\
\midrule
Obligation discharge & Incomplete support & 18 \\
Conflict coverage & Missing required attack & 5 \\
Attack typing and coverage & Invalid target or unlicensed attack & 32 \\
                          & Missing required attack & 5 \\
\bottomrule
\end{tabular}
\end{table}

%% file: src/related.tex
\section{Related work}
\label{sec:related}

\paragraph{Abstract and structured argumentation}
\Lara builds on the abstract frameworks of
\citet{Dung_1995} and the argument labellings of
\citet{Caminada_2006}. The \emph{Handbook of Formal
Argumentation} surveys the wider family of
formalisms~\citep{Baroni2018-BARHOF}. \Lara's contribution is a checked
language that generates such frameworks, with a proof that compilation
preserves a direct source semantics (\autoref{thm:preservation}). Its additional \stGap status
records claims with no complete support argument, before argument
labelling begins.

\aspic gives arguments rule structure and distinguishes rebut,
undercut, and undermine~\citep{Modgil_2014}. Work on its rationality
postulates studies closure and
consistency~\citep{Caminada_2007,cramer2026satisfyingrationalitypostulatesstructured}.
\Lara builds on this structure and adds evidence admission, replayable
strict certificates, and mandatory critical-question discharge
(\autoref{sec:policy}), following the argument-scheme account of
critical questions~\citep{Yu_2020}. Its consistency result
requires both attack completeness and the explicit strict-chain
restriction (\autoref{thm:rationality}). Thus the contribution is a
checked instantiation and compilation discipline, rather than a new
abstract defeat semantics.

\paragraph{Semantic publishing}
Micropublications represent claims, evidence, and support relations in
biomedical communication~\citep{Clark_2014}. The Argument
Interchange Format provides a vocabulary for inference, conflict, and
preference~\citep{CHES_EVAR_2006}. These representation models provide
objects on which reasoning tools can operate. \Lara specifies a
certificate language, a decidable acceptance judgment, and a compilation
whose claim statuses agree with the source judgment. Structured research
artifacts~\citep{liu2026humanwrittenpaperagentnativeresearch} are one source
of its declarations; their prose-to-formal bindings still require assessment.

\paragraph{Explicit-evidence logics}
The Logic of Proofs makes proofs explicit within formulas
\citep{Artemov_2001}. Its factive reading permits elimination
from a proof assertion to the asserted proposition. Dynamic logics of
evidence-based belief study changes in evidence and
belief~\citep{van_Benthem_2011,van_Benthem_2014}.
\citet{Pand_i__2020,Pand_i__2021} are particularly
close in structure: they represent defeasible arguments as
justification-logic terms with rebut, undercut, and undermine.
\Lara adopts this term structure and adds a certificate-checking
discipline with versioned policies, critical-question obligations,
provenance-based admission, and compilation to an argumentation
framework. The Logic of Proofs can serve as an optional strict backend;
its factive reasoning stays behind the interface
(\autoref{thm:isolation}). Its realization theorem does not establish
faithful formalization of a natural-language claim.

\paragraph{Possible-world semantics and bisimulation}
\Lara instantiates possible-world
semantics~\citep{kripke1963semantical,Blackburn_2001} with accepted
programs as worlds, declared applicable bridges as accessibility, and
claim statuses as the valuation (\autoref{def:outer-model}). Its modal
queries therefore describe support across declared checking contexts;
they do not establish the truth of a research claim.
\citet{grossi2010logic} connects the two fields from the other direction,
reading a single framework as a Kripke frame whose worlds are arguments
and whose accessibility relation is attack. \Lara places the modality one
level up, where a world is an accepted program and a bridge relates two
checking contexts; the two readings are compatible and answer different
questions.

\Lara uses total attack bisimulation to establish status preservation
across a bridge between compiled frameworks
(\autoref{thm:status-transport}).
\citet{Oikarinen_2011} characterize strong equivalence through
kernels, asking when two frameworks may replace each other under any
addition. \Lara asks a narrower question, whether a declared translation
between two given programs preserves a claim's grounded status, and its
checker decides the four clauses soundly and completely. Neither result
decides whether the compared settings are meaningfully comparable in the
domain being studied.

\paragraph{Proof-carrying architectures}
Proof-carrying code separates an untrusted producer from a trusted
checker~\citep{Necula_1997}. Foundational proof certificates support
multiple proof formats through a small checking
interface~\citep{miller2015foundational}. \Lara applies this discipline
to defeasible support. An accepted support term can later lose its
claim's justification, and evidence quarantine cannot create a public
justification (\autoref{thm:nonpromotion}). Different strict instances
may use different registered backends without a shared backend logic
(\autoref{thm:hetero}). Acceptance-preserving replacements leave claim
statuses unchanged; the appendix extends this result to admissible
linking contexts and partial-bijective certificate relations
(\appautoref{sec:context}). These results do not establish full
abstraction.

\paragraph{Verified and agentic claim checking}
EG-VAR verifies tool-attested empirical claims through Lean~4 kernel
proofs~\citep{ren2026evidencegroundedverifiedagenticreasoning}. \Lara
addresses a different part of the support relation: explicit attacks can defeat an otherwise checked
argument, and mandatory critical questions can prevent an argument
from entering the framework. Pramana attests claims at the protocol
layer for agent networks~\citep{https://doi.org/10.5281/zenodo.20283646}, whereas \Lara
computes claim statuses from support and attack declarations.
LLM judges produce assessments and rationales~\citep{Alvarez_2024};
such output can propose a \Lara certificate, but acceptance depends
on replaying the checker. Work on faithful
natural-language-to-Lean formalization~\citep{zhang2026compilationevaluatingfaithfulnaturallanguagetolean} addresses
whether formal statements capture their prose.
\Lara leaves that question outside its semantic guarantee.

%% file: src/conclusion.tex
\section{Conclusion}
\label{sec:conclusion}

We have presented \Lara, a kernel language built on the thesis that
support for research claims should be checked locally and transported across
contexts only under explicit bridge conditions. Locally, a decidable
checker compiles declared evidence, scheme instances, critical-question
answers, and typed attacks into a grounded status for every claim, and
compilation preserves the statuses of a direct source semantics. Across
contexts, structural bridges transport checked support with its
obligations, while preserving a claim's status requires stronger
correspondence conditions that an executable checker decides soundly and
completely; a translated argument can therefore remain well formed while
additional evidence defeats its claim.
The case studies show that this support need not come from measurements.
The same calculus represents objections and defenses in a philosophical
debate, and an axiom-withdrawal witness makes the assumptions required for
proof reuse explicit. \Lara thus checks how support depends on premises
and survives challenges across empirical and non-empirical arguments.
The calculus does not automate domain judgment: choosing a bridge
and defending its meaning remain the researcher's task, and
the checker makes the consequences of that choice inspectable.
Three directions follow: lowering research artifacts to \Lara programs
beyond the deterministic path exercised here, growing versioned policies
into community standards of review, and deciding the structural conditions
of candidate bridges mined across a literature. On this
substrate, we envision
autonomous research becoming more robust: an agent that builds on a
\stGap or \stDefeated claim inherits a status the checker reports, not
an error that propagates silently. As machines write and judge a
growing share of science, whether evidence supports a claim in a given
context can then become what well-typedness became for software: a
question a checker answers.

%% file: src/mechanization.tex
\section{Mechanization status}
\label{app:mechanization}

We map each formal result of \autoref{sec:metatheory} to its Lean~4
status. Five tables record the thirty-five mechanized specification
results, one per pipeline stage: checking and compilation
(\autoref{tab:mech-checking}), semantics-parametric observation and
aggregation (\autoref{tab:mech-observation}), strict-certificate
backends and linking (\autoref{tab:mech-backends}), source updates and
the surface layer (\autoref{tab:mech-updates}), and cross-context
guarantees, presentation, and cost (\autoref{tab:mech-context}). The
named guarantees G1--G6 all fall in the first three stages, and each
row there names the guarantee it serves. A sixth table
(\autoref{tab:admission}) lists the source-boundary admission family of
\autoref{pty:correspondence}--\autoref{pty:source-nonpromotion}. The
audited public theorems are \textsf{sorry}-free and stay within Lean's
standard axiom trio. The audit checks their transitive axiom dependencies
and requires coverage of public theorems and lemmas. Some private
executable examples use \texttt{native\_decide}, whose generated axioms
add trust in native evaluation; the three-axiom claim excludes those
examples. The local Lean gate and optional Lean CI workflow run the
audit. Required CI checks Haskell; cross-language checks run locally.
The pipeline
cost bound (\autoref{prop:complexity}), the NP-completeness bookkeeping
behind \autoref{thm:np-hardness}, and the closure and
indirect-consistency clauses of \autoref{thm:rationality} are argued on
paper (\appautoref{app:proofs}); the first two have no row.

\input{tables/mech-checking}
\input{tables/mech-observation}
\input{tables/mech-backends}
\input{tables/mech-updates}
\input{tables/mech-context}

\subsection{Narrower scopes and component-level results}
\label{app:narrow-scope}

Two guarantees have narrower scopes. The backend half of dependency
accountability (G2, \autoref{thm:accountability}) is an interface
obligation: we mechanize it for all four registered backends and
require future backends to discharge it. The direct-consistency clause of G6
(\autoref{thm:rationality}) is proved for the executable reference
checker over complete claims from an accepted unit, not over arbitrary
caller-supplied claims; its closure and indirect-consistency clauses
are paper arguments over the mechanized edge relation. The byte-exact
tests against that reference provide conformance evidence rather than
a replacement for the theorem.

Four further results are unnumbered because they concern components
rather than the calculus: each registered backend's soundness and
dependency accounting (\autoref{sec:strict}); the presentation-codec
round trip over the complete program and policy
syntax of \autoref{sec:language}, which a cross-language shape guard
pairs with a constructor-by-constructor comparison of the Lean and
Haskell abstract syntax, so neither can change alone; well-sortedness of
every atom an accepted unit reaches, which a substitution lemma carries
from the sorted policy patterns (\autoref{sec:leaves}); and the
direction-of-goodness contract for \texttt{comparison}, where the
generated \textsf{ord@1} goal holds exactly when the favored system is
better under the measurand's declared polarity, so a polarity error
requests the opposite comparison rather than a weaker one.

Two representation details fall outside the shape comparison: both
languages normalize sort names to bare strings, and each keeps its own
s-expression type for certificate payloads. The round trip concerns abstract syntax rather than parsing.
The surface calculus separately proves elaboration preservation,
reflection, and observation coherence above the parsed-AST boundary
(\appautoref{sec:surface}). The Haskell parser and elaborator remain
validated implementations of those definitions. Lean also verifies
the lowering of symbolic certificate premise references to numeric
slots, with conformance vectors for the Haskell implementation.

\subsection{The source-boundary admission family}

The admission judgment of \autoref{fig:admissionrules} carries eight
properties, numbered on their own counter and kept separate
from the specification results because they fix the checking
context rather than the core semantics.
The metatheory overviews the three that carry its results to the
boundary (\autoref{sec:metatheory}): \autoref{pty:pruning},
\autoref{pty:monotonicity}, and \autoref{pty:source-nonpromotion}. We state all eight here, each
citing its headline Lean declaration.

\begin{property}[Judgment/evaluator correspondence]
\label{pty:correspondence}
The declarative admission relation and the executable
evaluator agree in both directions
(\textsf{evaluateAdmission\_\allowbreak iff\_\allowbreak judgment}), so the relation
is functional (\textsf{admission\_\allowbreak deterministic}).
\end{property}

\begin{property}[Exact aligned checker context]
\label{pty:context}
Accepted outcomes require the metadata and declared leaf-id
sequences to agree position-by-position
(\textsf{accepted\_\allowbreak metadata\_\allowbreak aligned}). A leaf is
policy-admitted iff it is declared with decision \kw{admit}
(\textsf{policy\_\allowbreak admitted\_\allowbreak iff}), and it is in the final
checked context iff it is policy-admitted and not group-quarantined
(\textsf{checked\_\allowbreak admitted\_\allowbreak iff}). That context is exactly
the leaf table carried to the checker
(\textsf{accepted\_\allowbreak checked\_\allowbreak context\_\allowbreak exact});
policy-quarantined leaves are absent from it
(\textsf{policy\_\allowbreak quarantined\_\allowbreak absent}).
\end{property}

\begin{property}[Endpoint-safe attack pruning]
\label{pty:pruning}
An accepted outcome's semantic attack list is the successful resolution
of the declared raw attacks
(\textsf{accepted\_\allowbreak resolved\_\allowbreak aligned}). Resolution
first establishes that every raw attack names declared argument
identifiers. The retained semantic attacks are then exactly the
raw-endpoint-filtered aligned projection of that validated list
(\textsf{retained\_\allowbreak attacks\_\allowbreak selectAligned}), and
every retained raw attack names retained endpoints
(\textsf{retained\_\allowbreak attack\_\allowbreak endpoints}). Thus, two
arguments with equal support terms stay distinct.
\end{property}

\begin{property}[Canonical audit exactness]
\label{pty:audit}
The audit equals the combined prune's exact projection with
canonical multiplicity and order
(\textsf{admission\_\allowbreak audit\_\allowbreak exact}).
\end{property}

\begin{property}[All-admit checker-outcome identity]
\label{pty:all-admit}
When every table row admits (including an empty table), the prune, keep
predicates, and blocked queries coincide with the group-only path
(\textsf{policy\_\allowbreak all\_\allowbreak admit\_\allowbreak
group\_\allowbreak identity}). Applying the actual \textsf{checkUnit}
function produces the identical accept/reject outcome
(\textsf{policy\_\allowbreak all\_\allowbreak admit\_\allowbreak
checkUnit\_\allowbreak identity}).
\end{property}

\begin{property}[Restrictiveness monotonicity]
\label{pty:monotonicity}
Tightening decisions from \kw{admit} to \kw{quarantine} can
only remove leaves, arguments, and attacks
(\textsf{more\_\allowbreak restrictive\_\allowbreak cannot\_\allowbreak add\_\allowbreak structure}); claim
\emph{statuses} are deliberately non-monotonic, which is exactly the
hazard \autoref{thm:nonpromotion} manages.
\end{property}

\begin{property}[Rejection carries no checked unit]
\label{pty:rejection}
An R8 outcome names the first rejected leaf and carries no
checked unit (\textsf{source\_\allowbreak reject\_\allowbreak no\_\allowbreak checked\_\allowbreak unit}).
\end{property}

\begin{property}[Source non-promotion]
\label{pty:source-nonpromotion}
A query published \stJustified by the checked source result (with the
prune's exact keep predicates and aligned attacks) remains
\stJustified in the declared framework with the policy- and
group-quarantined material reinstated
(\textsf{source\_\allowbreak justified\_\allowbreak nonpromotion}). This
instantiates the production blocking theorem of
\autoref{thm:nonpromotion} from the admission carrier.
\end{property}

\input{tables/admission}

%% file: tables/mech-checking.tex
\begin{table}
\caption{Mechanization status, stage 1: checking and compilation.
Results 1--8 of the thirty-five mechanized specification results.}
\label{tab:mech-checking}
\centering
\footnotesize
\begin{tabular}{@{}rc >{\raggedright\arraybackslash}p{0.30\textwidth} >{\raggedright\arraybackslash}p{0.20\textwidth} >{\raggedright\arraybackslash}p{0.24\textwidth}@{}}
\toprule
\# & Guarantee & Result & Stated as & Lean status \\
\midrule
1 & G1 & Decidable checking & \autoref{thm:decidability} & mechanized \\
2 & -- & Conclusion uniqueness, support adequacy & \autoref{prop:support} & mechanized \\
3 & G2 & Dependency accountability & \autoref{thm:accountability} & leaf half mechanized; backend half mechanized for all four backends (\autoref{sec:strict}) \\
4 & G3 & Compilation soundness & \autoref{thm:compilation} & mechanized \\
5 & -- & Grounded termination, determinism & \autoref{thm:determinism} & mechanized \\
6 & G3 & Status preservation & \autoref{def:direct-status}, \autoref{thm:preservation} & mechanized \\
7 & -- & Image necessity, invariant transport & \autoref{thm:image-necessity} & mechanized \\
8 & -- & Fixed-policy non-sufficiency & \autoref{prop:non-sufficiency} & mechanized, with an executable accepted anchor \\
\bottomrule
\end{tabular}
\end{table}

%% file: tables/mech-observation.tex
\begin{table}
\caption{Mechanization status, stage 2: semantics-parametric
observation and aggregation. Results 9--13.}
\label{tab:mech-observation}
\centering
\footnotesize
\begin{tabular}{@{}rc >{\raggedright\arraybackslash}p{0.30\textwidth} >{\raggedright\arraybackslash}p{0.20\textwidth} >{\raggedright\arraybackslash}p{0.24\textwidth}@{}}
\toprule
\# & Guarantee & Result & Stated as & Lean status \\
\midrule
9 & -- & Parametric observation; the printed status is the grounded instance & \autoref{def:ext-semantics}--\autoref{thm:observe-grounded} & mechanized \\
10 & -- & \stGap{} independence; stable nonexistence; ten pairwise separations & \autoref{prop:gap-independent}, \autoref{prop:stable-noext} & mechanized, observation table generated from the definitions \\
11 & -- & Source-level observation transport & \autoref{thm:observation-transport} & mechanized, with a counterexample for the dropped support bound \\
12 & G6 & Consistency of the aggregation & \autoref{thm:rationality} & direct consistency mechanized, reference checker; closure clauses on paper \\
13 & -- & Quarantine non-promotion & \autoref{thm:nonpromotion} & mechanized \\
\bottomrule
\end{tabular}
\end{table}

%% file: tables/mech-backends.tex
\begin{table}
\caption{Mechanization status, stage 3: strict-certificate backends
and linking. Results 14--22.}
\label{tab:mech-backends}
\centering
\footnotesize
\begin{tabular}{@{}rc >{\raggedright\arraybackslash}p{0.30\textwidth} >{\raggedright\arraybackslash}p{0.20\textwidth} >{\raggedright\arraybackslash}p{0.24\textwidth}@{}}
\toprule
\# & Guarantee & Result & Stated as & Lean status \\
\midrule
14 & -- & Strict-backend isolation, non-factivity & \autoref{thm:isolation} & mechanized \\
15 & G4 & Strict-certificate soundness & \autoref{thm:strict-soundness} & mechanized \\
16 & G4 & Backend soundness, dependency accounting & \autoref{sec:strict} & mechanized \\
17 & -- & Heterogeneous compositionality, dependency union & \autoref{thm:hetero} & mechanized, with a mixed-backend witness \\
18 & G5 & Backend replacement & \autoref{thm:replacement} & mechanized, with transport \\
19 & -- & Contextual representation independence & \autoref{thm:ctx-independence} & mechanized, with a certificate-bearing witness \\
20 & -- & Contextual congruence at every semantics; grounded instance as an equivalence & \autoref{thm:ctx-sem} & mechanized, with an all-context separation \\
21 & -- & Typed term-level critical-question holes & \autoref{thm:holes} & mechanized, nested witness with a substituted attack \\
22 & -- & Relational parametricity over partial-bijective certificate relations & \autoref{thm:parametricity} & mechanized; partial bijectivity witnessed necessary at the observation \\
\bottomrule
\end{tabular}
\end{table}

%% file: tables/mech-updates.tex
\begin{table}
\caption{Mechanization status, stage 4: source updates and the
surface layer. Results 23--28; no named guarantee falls in this stage.}
\label{tab:mech-updates}
\centering
\footnotesize
\begin{tabular}{@{}r >{\raggedright\arraybackslash}p{0.34\textwidth} >{\raggedright\arraybackslash}p{0.20\textwidth} >{\raggedright\arraybackslash}p{0.26\textwidth}@{}}
\toprule
\# & Result & Stated as & Lean status \\
\midrule
23 & Update acceptance preservation & \autoref{thm:update-preservation} & mechanized \\
24 & Grounded transition matrices & \autoref{thm:update-matrices} & mechanized; matrices generated from the witnesses \\
25 & \stGap{} boundary; public layer; matrix non-promotion & \autoref{thm:gap-boundary}, \autoref{cor:matrix-nonpromotion} & mechanized, with the non-clean counterexample \\
26 & Surface elaboration preservation & \autoref{thm:elab-preservation} & mechanized \\
27 & Supported-fragment reflection & \autoref{thm:elab-reflection} & mechanized \\
28 & Surface observation coherence & \autoref{thm:surface-observation} & mechanized, five instantiated corollaries \\
\bottomrule
\end{tabular}
\end{table}

%% file: tables/mech-context.tex
\begin{table}
\caption{Mechanization status, stage 5: cross-context guarantees,
presentation, and cost. Results 29--35; no named guarantee falls in
this stage.}
\label{tab:mech-context}
\centering
\footnotesize
\begin{tabular}{@{}r >{\raggedright\arraybackslash}p{0.34\textwidth} >{\raggedright\arraybackslash}p{0.20\textwidth} >{\raggedright\arraybackslash}p{0.26\textwidth}@{}}
\toprule
\# & Result & Stated as & Lean status \\
\midrule
29 & Outer modal layer: normality, comparison adequacy, conservativity & \autoref{thm:pw-normal} & mechanized; T/D/B/4/5 refuted in generic models \\
30 & Exact cross-context support transport, path composition & \autoref{thm:support-transport} & mechanized, with the status-flip boundary witness \\
31 & Conditional cross-context status preservation, modal collapse & \autoref{thm:status-transport} & mechanized, with a sound and complete bridge decider \\
32 & Presentation-codec round trip & \autoref{sec:language} & mechanized, with a cross-language shape guard \\
33 & Well-sortedness of every atom an accepted unit reaches & \autoref{sec:language} & mechanized \\
34 & Direction-of-goodness contract for \texttt{comparison} & \autoref{sec:language} & mechanized \\
35 & Grounded query bounds, quartic tightness; 3SAT reduction correctness, realization, size & \autoref{thm:grounded-cost}, \autoref{thm:np-hardness} & mechanized; NP-completeness bookkeeping on paper \\
\bottomrule
\end{tabular}
\end{table}

%% file: tables/admission.tex
\begin{table}
\caption{Mechanization status of the source-boundary admission
family (\autoref{sec:semantics}, \autoref{fig:admissionrules}), stated
as \autoref{pty:correspondence}--\autoref{pty:source-nonpromotion}.
Every result has an audited public theorem that is \textsf{sorry}-free
and uses only Lean's standard axiom trio. The local Lean gate and
optional Lean CI workflow run the audit.}
\label{tab:admission}
\centering
\footnotesize
\begin{tabular}{@{}r >{\raggedright\arraybackslash}p{0.40\textwidth} l l@{}}
\toprule
\# & Result & Stated as & Lean status \\
\midrule
1 & Judgment/evaluator correspondence & \autoref{pty:correspondence} & mechanized \\
2 & Exact aligned checker context & \autoref{pty:context} & mechanized \\
3 & Endpoint-safe attack pruning & \autoref{pty:pruning} & mechanized \\
4 & Canonical audit exactness & \autoref{pty:audit} & mechanized \\
5 & All-admit checker-outcome identity & \autoref{pty:all-admit} & mechanized \\
6 & Restrictiveness monotonicity & \autoref{pty:monotonicity} & mechanized \\
7 & Rejection carries no checked unit & \autoref{pty:rejection} & mechanized \\
8 & Source non-promotion & \autoref{pty:source-nonpromotion} & mechanized \\
\bottomrule
\end{tabular}
\end{table}

%% file: src/supplement.tex
\section{Measurement inventories}
\label{app:measurements}

This appendix records the trusted base, implementation conformance
checks, ablations, and checker timings that support
\autoref{sec:evaluation}. Measurements are generated from the
implementation repository at separately recorded revisions. The current
implementation inventory and differential checks use \texttt{b62d33e0};
the ablations use \texttt{7d0f7384}, and the performance table uses
\texttt{5dd326b8}. Historical measurement-harness counts below retain
their \texttt{89c25ef} snapshot.

\subsection{The trusted base}
\label{app:tcb}

Acceptance depends on nine trusted components: the decode boundary
(presentation parser and wire decoder), the proposition normalizer and
its identity relation, the static checker for leaf admission, policy
instantiation, and support-term typing, the policy well-formedness
validator, the typed-attack checker, the strict-backend registry and
each shipped adapter, the compiler, the status engine, and the located
rejection diagnostics. The signature \Sig, the versioned policy \Pol,
the backend registry \Reg, and the policy-allowlisted theory digests
are trusted \emph{inputs} rather than trusted code: a program may
instantiate them but can never define them at run time
(\autoref{thm:isolation}).

Everything else is untrusted, including the elaborator, every backend
proof term, and any producer of the wire program. The strict kernel
performs normalization, mediates the backend interface, and runs the
four registered certificate checkers. Its guarantees depend on each
backend's encoding, consequence relation, and dependency laws.

\subsection{Conformance checks}
\label{app:conformance}

The production checker implements the complete pipeline in Haskell:
parser, policy validator, support and attack checker, the
\textsf{nd@1}, \textsf{ra@1}, \textsf{ord@1}, and \textsf{insp@1} strict backends,
compiler, grounded evaluator, and status engine. Its trusted base is a
strict subset of that pipeline, enumerated in
\appautoref{app:tcb}.
For the core calculus, compiled Lean semantics is the executable
reference interpreter, and a differential harness runs every committed
fixture, example program, and replay bundle through both drivers,
requiring byte-identical reports and exit codes.

A separate admission differential compares the Haskell pruning
primitives with Lean: all 15 semantic outcomes agree byte for byte,
and five malformed adapter fixtures fail at both codec boundaries.
Properties also exercise the integration of admission with checking.
The surface elaboration theorems start at a parsed AST
(\appautoref{sec:surface}); the Haskell parser and elaborator are
validated by tests.

The property suite and minimal negative programs exercise rejection
classes and acceptance invariants. At \texttt{b62d33e0}, the core
differential agrees on 673/673 verdict-bearing inputs and 66/66 malformed
inputs; three further depth-bound cases also agree. The verdict-bearing
inputs include both accepted programs and expected semantic rejections.
Seeded answer keys check diagnostic locations. A digest-pinned replay
bundle rejects tampered inputs, and a 78-row constructor inventory checks
agreement between the Lean and Haskell abstract syntax. A separate
25-case suite checks surface conformance. These are conformance checks on finite suites, not
an equivalence proof for the implementations.
The mutant and rejection-class inventory follows below.

\subsection{Mutants and rejection classes}
\label{app:mutants}

The historical measurement harness at \texttt{89c25ef} drove 541 mutant
programs and 60 corpus units, agreeing with expected rejection classes
and Lean on 601/601 inputs. Among its mutants, 113 are signature attacks, nine are
quarantine-attacker mutants, and 31 are localization mutants whose
defect manifests away from the edited site or at two sites at once. The
localization families carry a seeded answer key re-derived from the
mutant rather than copied from the checker, and a generation-time gate
compares the two, so a mislocating checker fails the build instead of
lowering a reported rate.

The checker has fourteen rejection classes, each with a runnable
minimal negative program: eleven use core fixtures, signature and codec
failures have separate fixture families, and admission rejects at the
source boundary. Both the production driver and the Lean reference
driver reject malformed inputs at the codec boundary.

\subsection{Ablating the checks}
\label{app:ablations}

Three ablations test whether obligations, typed attacks, and the
conflict scan affect acceptance. Each reruns the 655-input manifest
at \texttt{7d0f7384}
with a check removed: \textsf{no-CQ} skips the discharge of open
critical-question obligations, \textsf{no-conflict-scan} skips the
scan that demands an attack between contrary complete arguments, and
\textsf{no-typed} disables the typed-attack check together with that
scan, reducing the checker to nodes with arbitrary attack edges. Since
all three only remove rejections, every regression is a program that
the full checker rejects and the weakened checker accepts.

The ablations expose distinct failure classes
(\autoref{tab:ablation}). Removing obligation discharge admits
precisely the 18 incomplete-argument mutants. Removing the conflict
scan admits precisely the five mutants that drop the covering attack
between contrary complete arguments, the configuration
\autoref{thm:rationality} presupposes never survives checking. The
\textsf{no-typed} baseline admits those five plus the 32 bad-target or
unlicensed-attack mutants, 11 in class R10 and 21 in R11; no ablation changes any other
case. The admitted mutants span \stJustified, \stDefeated, and
\stContested outcomes, so the weakened checkers issue statuses for programs that should have
been rejected.

\subsection{Checker performance}
\label{app:performance}

\input{tables/performance}

We report wall-clock time for the 60 corpus units and the 601-record
harness at \texttt{5dd326b8}, measured on one machine with an Apple M5 Pro chip,
\qty{64}{\giga\byte} RAM, \texttt{darwin/aarch64}, GHC~9.14.1, and Lean~4.32.0. Each Haskell
time in \autoref{tab:performance} is an in-process median over
5~sections of 100~batched runs each. The median corpus unit takes about
\qty{630}{\micro\second} end to end, and the predecoded check-plus-render
kernel takes \qty{280}{\micro\second}. For each unit, the harness estimates
marginal decode cost as the nonnegative difference between these two
medians; the reported median of those estimates is about
\qty{352}{\micro\second}. This row is not a separately timed parser stage.
The check and evaluation phases include rendering their results to force
evaluation. The check phase includes signature and sort checking, but
the timings do not isolate their contribution. Validation, compilation,
and evaluation each remain below \qty{2}{\micro\second} in this protocol.

One pass over the 601-record harness takes about
\qty{250}{\milli\second}, so the differential suite runs interactively.
The Lean reference driver handles the same inputs in about
\qty{4}{\milli\second} of subprocess wall time, including startup and
decoding. This protocol is not comparable to the in-process production
measurements, but it is fast enough to adjudicate every test input.
Two independent benchmark runs agree within 3\% on every reported
median and within 1\% on most. A \texttt{make bench} target in the
supplemental artifact regenerates the table.

The compiled frameworks contain at most two nodes and one edge per
corpus unit. These timings establish low overhead on small inputs,
not scalability on large argumentation frameworks.

%% file: tables/performance.tex
\begin{table}[t]
\caption{Production-checker performance. End-to-end is
decode + check + verdict render. Marginal decode cost is estimated by
subtracting the predecoded kernel median from the end-to-end median per
unit. Other phases use pre-forced inputs; check and evaluation include
result rendering, so phases need not sum exactly.}
\label{tab:performance}
\centering
\small
\begin{tabular}{@{}lrr@{}}
\toprule
 & median & worst \\
\midrule
Corpus unit, end-to-end (\si{\micro\second}) & \num{629.6} & \num{675.3} \\
\quad estimated marginal decode (\si{\micro\second}) & \num{351.8} & \num{377.6} \\
\quad check + render, pre-decoded (\si{\micro\second}) & \num{281.3} & \num{316.4} \\
\quad validate (\si{\micro\second}) & \num{0.0} & \num{0.2} \\
\quad check (\si{\micro\second}) & \num{269.9} & \num{313.0} \\
\quad compile (\si{\micro\second}) & \num{0.0} & \num{0.9} \\
\quad evaluate (\si{\micro\second}) & \num{0.9} & \num{1.9} \\
Per-artifact total (27 artifacts, \si{\micro\second}) & \num{1278.9} & \num{2503.7} \\
Framework size (nodes / edges) & 0 / 0 & 2 / 1 \\
Lean reference driver, subprocess (\si{\milli\second}) & \num{4.0} & \num{5.0} \\
\midrule
601-record harness, one full pass (\si{\milli\second}) & \multicolumn{2}{r}{\num{246.8}} \\
\bottomrule
\end{tabular}
\end{table}

%% file: src/corpus.tex
\section{Exploratory corpus study}
\label{app:corpus}

\input{tables/cqs}
The corpus study is exploratory. Its statuses describe the encoded
artifacts under the selected policy and machine annotations, rather
than the validity of the underlying research claims.

We began with a 32-artifact corpus of paper-derived and agent-generated
research artifacts. From its 231 claims and 279 recorded dead ends we
drew a stratified 60-claim sample spanning 27 artifacts. One
machine annotation pass per artifact classified each claim's inference
scheme, leaf grain, critical questions, strict-certifier demand, and
attack candidates under a shared codebook; the supplemental artifact
ships that codebook and the stratification.
A second machine annotation pass, blind to the first, reclassified the claim type for
18 of the 60 claims and agreed on 14 ($\kappa = 0.73$). Three
measurements follow, each stated over its own denominator.

\paragraph{Missing support in the encoded artifacts}
Of the 60 sampled claims, 48 are \stGap, nine are \stJustified, and
three are \stDefeated. None is \stContested. Four claims in five lack complete support in
their encodings under the selected policy; this does not establish that
the original investigations lacked support. The 249
critical-question obligations those claims raise locate where the
support stops (\autoref{tab:cqs}): 71 of the 176 mandatory questions go
unmet, 68 of them as located gaps and three as defeaters. The verdicts
that do go through rest on 38 load-bearing leaves, all agent-produced,
of which 20 are \lkAttested and 18 \lkObserved; 15 point directly to an
equation, table, figure, or theorem in a paper.

\input{tables/schemes}
\paragraph{Scheme coverage}
Nine inference-scheme families cover all 60 sampled claims, and the top
three cover 46 of them (\autoref{tab:schemes}). The sample also demands
little heavyweight logic: 35 of the 60 claims call for a domain
checker, mostly to recompute arithmetic from reported tables, while
three call for the Logic of Proofs and one for natural deduction. That
distribution motivates a registry of small certified checkers rather
than one fixed calculus, and the corpus exercises exactly one of them
on an accept path. \textsf{adaptive-pruning}/C04 rechecks score cells
from Table~5 through the strict \texttt{rational\_drop\_recheck}
scheme, and a checked \textsf{ra@1} certificate representing the
claimed drop as an exact reduced fraction discharges it; the headline
claim nonetheless stays \stGap because its variance question is unmet.
Four of the five units documenting strict-flavored steps carry no
checked certificate in this study snapshot. A preliminary annotation
and design assessment judged roughly 90\% of the 60 sampled argument
shapes expressible under the proposed encoding conventions. Its six
residual cases concerned equivalence schemes, monotone functional
relationships, parametric growth rates, distribution-valued leaves,
negative existentials over code, and graded or undefined predicates.
This assessment measures judged expressibility, not compiler acceptance.
The nine scheme families classify all sampled claims without establishing
successful encoding of every argument shape. The later \textsf{insp@1}
backend checks a restricted form of negative existential over a declared
exhaustive inventory; it does not verify that the inventory faithfully
describes the source code.

\paragraph{Few recorded attacks}
Three typed attacks validate across the 60 claims, all of them
undercuts: two from exploration traces and one from a paper's own
table. Among 194 classified dead ends, four are undercut candidates and
only one executes as a claim-defeating attack, the rest being rejected
alternatives rather than defeaters. The sample does not establish why so few attacks are recorded: it
cannot separate omissions in the original record from annotation or
encoding limitations. Both trace-sourced undercuts defeat their target
claims. We use authored adversarial programs to exercise defeat
structures absent from the corpus, with further case studies in
\appautoref{app:casestudy}.

\subsection{Study limitations}
The study uses a single machine annotation pass per artifact, with human
review pending. We double-annotated only the claim-type field on 18 of
the 60 claims. The 32-artifact corpus was assembled rather than sampled
from a frame, so its distributions carry no external-validity claim.
The study does not evaluate bridge discovery or automated
literature-based review.

%% file: tables/cqs.tex
\begin{wraptable}{r}{0.46\textwidth}
\caption{Critical-question obligations generated by the 60-claim
sample. Unmet mandatory questions surface as located gaps; only three
act as defeaters: the \stGap-versus-\stDefeated split in the data.}
\label{tab:cqs}
\small
\centering
\begin{tabular}{@{}lrrrr@{}}
\toprule
 & Met & \shortstack{Unmet\\(gap)} & \shortstack{Unmet\\(defeater)} & Total \\
\midrule
Mandatory & 105 & 68 & 3 & 176 \\
Optional  & 34  & 39 & 0 & 73 \\
\midrule
Total & 139 & 107 & 3 & 249 \\
\bottomrule
\end{tabular}
\end{wraptable}

%% file: tables/schemes.tex
\begin{wraptable}[15]{r}{0.42\textwidth}
\caption{Inference-scheme families over the 60-claim sample; the top
three carry 77\%.}
\label{tab:schemes}
\footnotesize
\centering
\begin{tabular}{@{}lr@{}}
\toprule
Scheme family & Claims \\
\midrule
Controlled comparison & 25 \\
Intervention and ablation & 14 \\
Observation and measurement & 7 \\
Analytic argument & 4 \\
Benchmark or stress evaluation & 3 \\
Code inspection & 3 \\
Statistical correlation & 2 \\
Inductive generalization & 1 \\
Blind paired human evaluation & 1 \\
\midrule
Total & 60 \\
\bottomrule
\end{tabular}
\end{wraptable}

%% file: src/casestudy.tex
\section{Case studies}
\label{app:casestudy}

These case studies ask what researchers gain by expressing their claims
and evidence as \Lara programs. \Lara makes the connection between evidence,
inference, and objection explicit: the checker validates the declared
support and attacks, then computes which claims remain supported. We
first demonstrate three uses of this capability: replaying how a rebuttal changes
support, explaining a verdict as a review comment, and distinguishing
conflicting conclusions from conclusions about different settings.
Two further studies use no measurements: a philosophical debate shows how
assumptions and attributed premises support defeasible arguments, and an
axiom-withdrawal example shows when a checked deduction can be reused in
another context. Together, the studies demonstrate that \Lara checks the
structure and standing of support across different kinds of inquiry.

The sampled corpus retains little defeat (\appautoref{app:corpus}):
three undercuts, no rebuttals or underminings, no contested claims, and
no round in which new evidence changes a status. The worked-example
suite does exercise defeat, reinstatement, and contestation, but each of
those programs isolates one behavior in an authored setting.

The first three studies exercise these mechanisms over material from real
corpus units or families. The differential harness replays the underlying
checked programs byte-for-byte through the production checker and the Lean
reference driver. Below we
show only the declarations that distinguish the relevant rounds or
outcomes; the fuller listings are
\appautoref{lst:rebuttal-replay}, \appautoref{lst:mechanical-reviews}, and
\appautoref{lst:agreement-map}, and the anonymized supplemental artifact
carries the complete programs, policies, replay commands, and golden
reports.

\paragraph{Scope}
We hand-lowered the first three studies under the corpus-unit discipline and
anchored them to real units and families. We did not mine them from
review threads at scale. In particular, the first study derives its
reviews and rebuttal from a documented corpus gap rather than from an
OpenReview exchange. Scaling to real review threads and correlating
\Lara verdicts with human complaints require a separate empirical
study. The two non-empirical studies are authored demonstrations; we cite
their committed programs and distinguish their execution paths below.
Within that scope, the studies establish that \Lara's support checks and
attack semantics work together in these research situations. Researchers
still supply the formal claims, evidence declarations, and objections;
\Lara checks their consequences under the chosen policy. These studies
do not establish automatic extraction from papers, agreement with human
reviewers, or the truth of the encoded claims.

\input{tables/rebuttal}

\subsection{A rebuttal exchange as three checked programs}
\label{app:cs-rebuttal}

The first study asks whether a reader can trace exactly why a claim gains
or loses support during review. The submitted paper,
the review round, and the rebuttal form one framework whose argument
population grows: reviews add typed attacks to the paper's arguments,
and a rebuttal adds evidence leaves and counterattacks. \Lara
represents each round as a separate program, all three sharing one
artifact identity and one fixed policy because they concern the same
paper under the same standard. Grounded semantics re-adjudicates the
graph after each round, so claim statuses change non-monotonically
while the policy stays fixed.

\input{figures/rebuttal}

The exchange starts from \textsf{adaptive-pruning}/C04. Its
kurtosis-salience ablation has only one run, and no leaf discharges the
mandatory \textsf{variance\_reported} question. The corpus therefore
assigns the claim \stGap (\appautoref{app:corpus}). We lift three
claims from that paper into the exchange. We show the declarations that
change across rounds in \autoref{fig:rebuttal}; the fuller presentation
appears in \appautoref{lst:rebuttal-replay}.
At submission, \texttt{c\_bench} and \texttt{c\_measure} have complete
arguments. The missing variance leaf leaves \texttt{c\_kurt} without a
supporting argument and reproduces the corpus gap. We omit that argument
because an instance with an unanswered mandatory question would be
rejected; the claim itself can remain in the accepted program.

The reviews use different attack forms because their objections challenge
different parts of the support. The protocol objection disputes an
evidence premise, so an \atUndermine targets the benchmark's
fixed-protocol leaf. The replication supplies a conflicting conclusion,
so a \atRebut targets the benchmark argument's conclusion. The memory
objection challenges the inference from a measurement to the claimed
footprint, so an \atUndercut targets that argument's rule. These positions
tell the checker which support each objection can defeat.

The rebuttal supplies the variance leaf and adds an ablation argument that
uses it to discharge the obligation, so \texttt{c\_kurt} moves from
\stGap to \stJustified. The new leaf alone would not supply an argument. A
counter-\atUndermine and counter-\atUndercut defeat both benchmark
attackers, reinstating \texttt{c\_bench}. The author concedes the memory
point by adding no counterargument, so \texttt{c\_measure} remains
\stDefeated: concession is the absence of a defense rather than a
special construct. Every one of these moves comes from the growing
graph, since the standard never shifts.

The case shows why \Lara is useful for a rebuttal: it separates completing
missing support, defending existing support, and leaving a defeat
unanswered. A reader can replay each outcome from the declared changes
under the same policy. The status table therefore records the consequences
of the exchange, with the responsible arguments and attacks available for
inspection.

\subsection{Verdicts rendered as reviews}
\label{app:cs-reviewer}

The second study asks whether a checked verdict can explain what an
author needs to address. \Lara distinguishes a claim with no complete
supporting argument (\stGap) from one whose complete arguments have been
defeated (\stDefeated). This distinction matters for review: supplying a
missing argument and answering an objection are different tasks.
A deterministic renderer reads the checker's per-claim verdict and located diagnostic,
then fills a fixed phrasing table keyed on the diagnostic shape. It uses
no free-text generation and adds no trusted code. We condense the two
diagnostic shapes in \autoref{fig:reviewer} and
give the verbatim rendered output in \appautoref{lst:mechanical-reviews}.

\input{figures/reviewer}

Over the 60-unit corpus the renderer emits 51 review comments:
48 for \stGap claims and 3 for \stDefeated ones; the 9 \stJustified claims
raise none. The templates report the checked status and relevant
declarations, and a freshness test pins the rendered bytes. For the
ablation gap, the comment identifies the absence of a complete argument;
for the defeated claim, it names the attacker and the challenged question.

The case shows that \Lara can provide a traceable basis for review
comments: readers can follow a diagnosis back to the checked support.
The renderer does not discover scientific objections or assess novelty.
Its contribution is to communicate the consequences of the evidence and
objections already encoded in the program.

\subsection{Genuine versus apparent disagreement}
\label{app:cs-agreement}

The third study asks whether apparently opposing findings actually
concern the same comparison. We encode each conclusion with the systems,
metric, and experimental setting it concerns. The policy declares
\texttt{better} and \texttt{not\_better} contrary only when these arguments
match. Including the setting in the proposition lets \Lara check whether
a declared rebuttal concerns the same comparison.
Two proposition pairs from the
pruning family carry the same prose-level disagreement
(\autoref{tab:agreement}); we give the complete encoding in
\appautoref{lst:agreement-map}. P1 shares one setting atom, so its declared mutual \atRebut becomes
a 2-cycle and both claims are \stContested. P2 differs only at the setting
atom, so no contrary instance and no edge form; both claims remain
\stJustified. The mutual attacks leave P1 without an accepted argument
that resolves the conflict; \stContested preserves that unresolved state.

\input{tables/agreement}

Declining to attack is not asserting agreement. The
setting-mismatch non-attack leaves an open comparability critical
question naming the missing measurement, so non-comparability becomes a
concrete request: a bridging experiment must report both systems at a
shared setting. P1's headline conclusions rest on defeasible steps
only and satisfy the strict-chain restriction of
\autoref{sec:semantics}.

The case shows how \Lara can make the scope of disagreement inspectable.
Readers can see whether the encoded conflict concerns a shared setting
or whether comparison still requires additional evidence. This capability
depends on the supplied formalization: researchers must judge whether
the setting identifiers faithfully represent the experiments.

\subsection{A philosophical debate without measurements}
\label{app:cs-philmath}

Can \Lara check an argument whose premises are assumptions and attributed
positions rather than observations? This study reconstructs a small
philosophy-of-mathematics debate under an authored \textsf{philmath-v1}
policy~\citep{lara2026philmath}. Every evidence leaf is \lkAssumed or
\lkAttested, and the program uses no strict backend. The policy supplies
the inference schemes, mandatory questions, contraries, and exceptions;
the existing checker evaluates their instances.

The encoding makes the disputed commitments explicit. An
indispensability argument supports the existence of abstract mathematical
objects, but must discharge a mandatory question about why indispensability
warrants that commitment. A nominalist argument challenges its
indispensability premise, subject to a separate assumption that the
reconstruction covers the scientific fragment under discussion. We keep
this claim separate from the stronger claim that numbers are intrinsically
particular sets. The latter faces a mutual rebuttal from a structural-role
account of number identity. These two arguments remain \stContested:
the checker exposes the unresolved disagreement without choosing a side.

A second exchange shows how a defense restores support. A fictionalist
argument claims that mathematical usefulness need not require literal
mathematical truth. An objector undercuts that inference; a reply that
consistency suffices for the conservative applications at issue undercuts
the objector in turn. The reply is unattacked, so the objector becomes
\emph{out} and the fictionalist argument becomes \emph{in}.
The committed defense-removal test deletes just the reply's undercut and
checks that the fictionalist claim changes from \stJustified to
\stDefeated. We give the relevant declarations in
\appautoref{lst:philmath}; the complete program also defeats the
indispensability argument through both a premise challenge and an undercut.

This case demonstrates why \Lara is useful beyond empirical review:
it can require an argument to expose its assumptions, check that an
objection targets the stated premise or inference, and compute the effect
of a defense. The program and its golden verdict are registered in the
worked-example and Haskell--Lean differential suites. These checks concern
the encoded debate: the policy is authored, the prose-to-atom bindings
remain explicitly unreviewed, and the result does not settle the
philosophical positions it represents.

\subsection{Withdrawing an assumption blocks proof reuse}
\label{app:cs-axiom-withdrawal}

The final study asks what happens to a checked deduction when another
context withholds its premise. The source declares one assumed atom,
\texttt{parallel\_postulate()}, and a strict argument that reuses that
premise through the production \textsf{nd@1} backend~\citep{lara2026axiom}.
The certificate \texttt{(hyp 0)} selects premise zero; the backend theory
is empty, so the certificate cannot obtain the conclusion from an axiom
hidden in that theory. The atom names an assumption for the demonstration:
the program proves no geometric consequence. The source and admission
choices appear in \appautoref{lst:axiom-withdrawal}.

The three admission policies isolate different outcomes
(\autoref{tab:axiom-withdrawal}). Admission retains the premise and its
argument, making the claim \stJustified. Quarantine removes both but accepts
the remaining program, whose claim is \stGap because no support remains.
Rejection stops source admission and yields no checked claim status.
Withdrawing a premise therefore differs from defeating an argument:
this example contains no attack.

\input{tables/axiom-withdrawal}

A separate executable Lean witness checks the structural transport
boundary. When the assumption is retained at both ends, an identity bridge
preserves the leaf and the support-transport theorem applies to the
certificate-bearing argument. When it is withdrawn, the target has no
admitted leaves. The executable leaf-preservation check fails, and the
theorem \texttt{no\_withdrawn\_bridge} rules out any structural bridge to
that empty target environment: no mapping can supply the required image
of the source premise. This establishes a failure of the bridge contract,
in addition to the locally computed change in status. Support transport
alone does not assert status preservation.

This case demonstrates what \Lara adds around a deductive certificate:
it makes admission of the certificate's premises explicit and checks the
conditions under which that support can be reused. The Haskell tests
check source admission, local statuses, and the certificate's dependency
on premise zero; the Lean executable and proofs check the transport
boundary. The current \texttt{lara pw} command refuses the quarantined
world before loading bridges, so its refusal is not the structural-bridge
result. Both demonstrations use the existing calculus; the new policy
instances and Lean witness require no change to its kernel rules.

%% file: tables/rebuttal.tex
\begin{table}[t]
\caption{Per-claim status across the rebuttal exchange, in which only the
argument population grows. \texttt{c\_bench} is reinstated across an \atUndermine
and a \atRebut, \texttt{c\_kurt} discharges the variance gap the corpus
unit already records, and \texttt{c\_measure} is conceded.}
\label{tab:rebuttal}
\centering
\small
\begin{tabular}{@{}lccc@{}}
\toprule
Claim & Round 0 & Round 1 & Round 2 \\
      & submission & reviews & rebuttal \\
\midrule
\texttt{c\_bench} (APT vs.\ dense on OpenLLM) & \stJustified & \stDefeated & \stJustified \\
\texttt{c\_kurt} (kurtosis-salience term critical) & \stGap & \stGap & \stJustified \\
\texttt{c\_measure} (low training-memory footprint) & \stJustified & \stDefeated & \stDefeated \\
\bottomrule
\end{tabular}
\end{table}

%% file: figures/rebuttal.tex
\begin{figure}
\begin{lstlisting}[language=Lara,numbers=left,basicstyle=\ttfamily\casestudycodesize]
# ROUND 0 -- missing variance evidence means no a_kurt
claim c_kurt
  formal  = contributes(kurtosis_salience, apt_llama2_7b, openllm_avg)
# ROUND 1 -- one attack of each kind
undermine d_um a_bench.protocol_fixed.leaf
rebut d_rebut a_bench
rebut a_bench d_rebut
undercut d_meas a_meas.rule
# ROUND 2 -- new evidence completes a_kurt; counter-attacks reinstate a_bench
leaf rk_var : variance_reported(e03)
arg a_kurt : supports(c_kurt) by component_ablation(kurtosis_salience, apt_llama2_7b, openllm_avg, e03)
  discharge variance_reported with rk_var
undermine r_um d_um.leaf
undercut r_rebut d_rebut.rule
\end{lstlisting}
\caption{The language-level deltas across the three rounds (non-contiguous
excerpts). Round 2 adds no defense for the conceded measurement claim.
The supplemental artifact holds the repeated declarations, evidence
metadata, policies, and complete programs, which the differential harness
replays byte-for-byte through both drivers (\autoref{tab:rebuttal}).}
\label{fig:rebuttal}
\end{figure}

%% file: figures/reviewer.tex
\begin{figure}[H]
\begin{tcolorbox}[
  colback=black!3,
  colframe=black!45,
  boxrule=0.4pt,
  arc=1.5pt,
  left=6pt,right=6pt,top=5pt,bottom=5pt
]
\small
\textbf{\stGap: \textsf{adaptive-pruning}/C04.}
No checked argument assembles the unit's evidence; the complete-support set is
empty. The review asks for a support argument at the \textsf{corpus-v1}
evidential bar.

\medskip\hrule\medskip

\textbf{\stDefeated: \textsf{fre}/C04.}
The only support \texttt{a1} is \emph{out}: unattacked undercut \texttt{d1}
challenges its \textsf{variance\_reported} critical question. The review names
both the attacker and the failed question.
\end{tcolorbox}
\caption{Two diagnostic shapes rendered from checked verdicts. The box
compresses the renderer output to the located information a reviewer acts on;
we give it verbatim in \appautoref{lst:mechanical-reviews}.}
\label{fig:reviewer}
\end{figure}

%% file: tables/agreement.tex
\begin{table}[t]
\caption{Cross-paper agreement map. Splitting the shared arguments from the
setting atom exposes the semantic hinge: whether that one atom matches
decides whether a contrary instance, and so an edge, exists.}
\label{tab:agreement}
\centering
\small
\begin{tabular}{@{}ll>{\raggedright\arraybackslash}p{0.29\linewidth}
                    >{\raggedright\arraybackslash}p{0.22\linewidth}
                    >{\raggedright\arraybackslash}p{0.23\linewidth}@{}}
\toprule
Pair & Polarity & Shared arguments & Setting atom $D$ & Compiled outcome \\
\midrule
\multirow{2}{*}{P1}
  & \texttt{better} & \multirow{2}{=}{\texttt{apt, cofi, accuracy}}
  & \textcolor{OliveGreen}{\textbf{\texttt{roberta\_mnli\_s60}}}
  & \multirow{2}{=}{mutual \atRebut $\Rightarrow$ both \stContested} \\
  & \texttt{not\_better} &
  & \textcolor{OliveGreen}{\textbf{\texttt{roberta\_mnli\_s60}}} & \\
\midrule
\multirow{2}{*}{P2}
  & \texttt{better} & \multirow{2}{=}{\texttt{magnitude\_pruning, dense\_baseline, accuracy}}
  & \textcolor{Mahogany}{\textbf{\texttt{bert\_glue\_s50}}}
  & \multirow{2}{=}{no edge $\Rightarrow$ both \stJustified} \\
  & \texttt{not\_better} &
  & \textcolor{RoyalPurple}{\textbf{\texttt{llama\_openllm\_s90}}} & \\
\bottomrule
\end{tabular}
\end{table}

%% file: tables/axiom-withdrawal.tex
\begin{table}[t]
\caption{Assumption admission and proof reuse. The source files differ only
in policy selection; each policy makes a different admission decision for
the assumed leaf. Structural results come from the separate Lean witness.}
\label{tab:axiom-withdrawal}
\centering
\small
\begin{tabular}{@{}lp{0.28\linewidth}lp{0.34\linewidth}@{}}
\toprule
Decision & Admitted support & Local status & Structural result \\
\midrule
\texttt{admit} & Premise and argument retained & \stJustified
  & Identity bridge transports checked support \\
\texttt{quarantine} & Premise and argument removed & \stGap
  & No structural bridge to the empty target evidence environment \\
\texttt{reject} & Source admission stops & None
  & No target checking environment supplied \\
\bottomrule
\end{tabular}
\end{table}

%% file: src/linking.tex
\section{Contextual representation independence}
\label{sec:context}

Backend replacement compares two whole programs. Interchangeability,
however, is a claim about use: replacing a backend inside one
fragment of a program must go unnoticed by every program that links
against that fragment. This appendix adds the context
quantifier. We define fragments and linking contexts, observe
exported claim statuses, and prove that no admissible context can
observe an acceptance-preserving backend replacement
(\autoref{thm:ctx-independence}).

\begin{definition}[Fragments, contexts, and linking]
\label{def:fragment}
A \emph{fragment} declares a signature, a policy, evidence leaves, a
finite ground input, support terms, attacks, the leaf identifiers it
\emph{imports} from its environment, and the conclusions it
\emph{exports} for observation. A \emph{context} is a
fragment-shaped surrounding program. Linking a context $C$ with a
fragment $F$ first runs a hygiene guard that rejects duplicate or
clashing identifiers, unsatisfied imports on either side, and any
signature or policy mismatch, so a context may add evidence,
arguments, and attacks but may never redefine \Sig{} or \Pol, and
the registry \Reg{} stays a fixed parameter of observation. A
successful link merges the two sides, adds the cross-boundary
attacks that the attack-completeness scan (\autoref{sec:attacks})
requires, and submits the linked program to the unchanged checker.
\end{definition}

\begin{definition}[Observation and contextual equivalence]
\label{def:ctx-equiv}
The \emph{observation} $\mathrm{obs}_{\Reg}(C, F)$ is the link fault
when the hygiene guard rejects, the checker error when the linked
program is rejected, and otherwise the list of grounded claim
statuses (\autoref{def:status}) of $F$'s exported conclusions in the
linked program's compiled structured framework. Fragments $F_1$ and
$F_2$ are \emph{contextually equivalent} under \Reg{} when
$\mathrm{obs}_{\Reg}(C, F_1) = \mathrm{obs}_{\Reg}(C, F_2)$ for
every context $C$.
\end{definition}

A pair $(C, F)$ is \emph{admissible} under \Reg{} when the guard
passes, each side's terms and attacks check relative to the linked
leaf environment, the linked program passes the signature stage, and
the shared policy is well formed with distinct, in-scope rule
declarations. A \emph{backend relabeling} $f$ maps certificate
assurances to certificate assurances; $f \cdot F$ relabels $F$'s
assurances pointwise and touches nothing else, and $C$ is
\emph{fixed by} $f$ when relabeling changes none of $C$'s terms or
attacks.

\begin{theorem}[Contextual representation independence]
\label{thm:ctx-independence}
Let $f$ be an injective backend relabeling and $\Reg_1$, $\Reg_2$
registries such that whenever $\Reg_1$ accepts an assurance for a
rule instance, $\Reg_2$ accepts its relabeling under $f$. For every
pair $(C, F)$ admissible under $\Reg_1$ with $C$ fixed by $f$,
\[
\mathrm{obs}_{\Reg_1}(C, F) \;=\; \mathrm{obs}_{\Reg_2}(C, f \cdot F).
\]
In particular, a registry that accepts every assurance $\Reg_1$
accepts observes identically on every pair admissible under
$\Reg_1$, and the equation survives embedding $C$ into a composed
context that remains admissible and fixed by $f$.
\end{theorem}

Each hypothesis carries content. Acceptance preservation runs
forward only, so $\Reg_2$ may accept assurances $\Reg_1$ rejects;
because the observation separates a rejection from a status list, an
unconditional statement over every context would then be false.
Admissibility cannot be dropped either, since linking is partial and
a rejected pair observes its fault. The witnesses are not
degenerate: a fragment carrying a checked \textsf{nd@1} certificate,
relabeled by a non-identity injection into a registry that expects
the relabeled spelling, changes syntactically and keeps its
observation (\textsf{cert\_congruence\_witness}).

The theorem extends backend replacement (G5,
\autoref{thm:replacement}) along the context quantifier, and the two
results are incomparable: G5 quantifies over arbitrary whole checked
programs with no admissibility hypothesis, while the congruence
quantifies over admissible linking pairs, so the whole-program
instance cites G5 rather than re-deriving it.

\begin{theorem}[Contextual congruence at every semantics]
\label{thm:ctx-sem}
Replace the grounded status readout in
\autoref{def:ctx-equiv} by the claim observation of
\autoref{def:observation} at an arbitrary extension semantics. Then
\autoref{thm:ctx-independence} holds verbatim at every extension
semantics, with no additional hypothesis and in particular no
attack-extensionality, and so do its acceptance-profile,
composed-context, and whole-program forms. The grounded instance
recovers \autoref{def:ctx-equiv} exactly, as an equivalence of
relations.
\end{theorem}

The generalization is not vacuous. The congruence needs no new
hypothesis because it transports an equality of compiled carriers
rather than pointwise attack agreement, yet the parameter matters at
the context level: on a linked three-cycle, stable semantics
observes the no-extension report where grounded observes
\stContested, an arm the grounded reading cannot reach; and grounded
contextual equivalence does not imply equivalence at every adequate
semantics, since two fragments provably equivalent in \emph{every}
context under grounded are separated by an adequate
singleton-selector semantics at one accepted context. That
counterexample refutes the unrestricted implication only; it
separates none of the four standard non-grounded instances. \stGap
remains semantics-independent at the context level.

Contexts can also owe a fragment more than evidence: an answer to a
critical question is a support term inside a discharge map, which
leaf imports cannot supply. A \emph{hole template} extends support
terms with named holes at answer positions; two nested instances may
ask the same question yet demand different answers, and sharing a
hole identifier forces the demanded answers to agree through a
declared hole signature. A \emph{filling} maps hole identifiers to
core support terms.

\begin{theorem}[Typed hole substitution]
\label{thm:holes}
Erasing a template's unanswered holes to open questions yields a
typed source support term. Substituting a typed filling, complete
support terms whose conclusions are equivalent to the demanded
answers, derives support of the instantiated term with the same root
conclusion and residual obligations, and an obligation-free
instantiation crosses the unchanged compilation boundary.
Instantiated linking is checked, closed embeddings recover the
observations and contextual equivalence of \autoref{def:ctx-equiv}
exactly, and the congruences of \autoref{thm:ctx-independence} and
\autoref{thm:ctx-sem} pass through instantiation, provided the
relabeling also fixes the certificates in the context's fillings.
That proviso is substantive: a mechanized witness fixes a context's
frame while a certified filling moves.
\end{theorem}

The congruence quantifier itself also strengthens from a function to
a relation, which is the sense in which backend interchangeability
is parametric rather than merely representation-independent.

\begin{theorem}[Relational parametricity]
\label{thm:parametricity}
Let $R$ relate certificate assurances such that (i) $R$ is the graph
of a partial injection; (ii) acceptance transports from $\Reg_1$ to
$\Reg_2$ along the pairs $R$ relates, and only those; (iii) fragments $F_1$
and $F_2$ agree on signature, policy, declared leaves, ground input,
imports, and exports, with arguments and attacks pointwise
$R$-related; and (iv) $R$ relates the context's own material to
itself. Then for every pair $(C, F_1)$ admissible under $\Reg_1$,
\[
\mathrm{obs}_{\Reg_1}(C, F_1) \;=\; \mathrm{obs}_{\Reg_2}(C, F_2),
\]
at every extension semantics. The graph instance of this theorem
re-derives \autoref{thm:ctx-independence} exactly, and the
acceptance hypothesis can be localized to the fragment's own
certificate occurrences at the cost of two context-containment side
conditions, a trade rather than a strengthening.
\end{theorem}

Partial bijectivity is necessary, and the necessity is witnessed at
the observation: a relation that collapses two distinct certificate
payloads lets structural deduplication merge two wrappers of the
same conclusion, an undercut on one wrapper then reaches the merged
node, and the same exported claim changes from \stJustified to
\stDefeated with every other hypothesis intact. Finitely much
related material is always realized by some total injective
relabeling that also fixes the context, while extending an arbitrary
such relation to a total injection is refuted, so the residual gap
between the relational and functional forms is acceptance scope
alone. The honest name is therefore parametricity over
partial-bijective certificate relations, not over arbitrary
relations; and none of this is full abstraction, since we prove no
logical-relation characterization of contextual equivalence. Two
recorded obstructions gate that characterization. First, a
fragment's semantic interface is wider than its declared imports:
the attack-completeness scan ranges over all argument pairs, and
subargument closure feeds non-exported arguments back into exported
statuses, so a logical relation over declared imports alone is
refuted in both directions. Second, completeness is
policy-conditional: under a policy with no declared conflict, no
attack types and no context can separate fragments by forcing a
label, the same degenerate policy as \autoref{prop:non-sufficiency}.

%% file: src/realizability.tex
\section{The image of compilation}
\label{sec:image}

We next ask which frameworks a checked program can denote.
Every realizable structured framework satisfies the compiler invariant
below, but the invariant is not sufficient for realizability under an
arbitrary fixed policy. This result retains conclusion labels; it does
not characterize the class of label-erased Dung frameworks.

\begin{definition}[Structured framework and isomorphism]
\label{def:structured-af}
A \emph{structured framework} $F$ pairs a finite sequence of node
conclusions with a total attack relation
$\rightsquigarrow_F$ on positions. An \emph{isomorphism} from $F$ to
$G$ is a bijection $\sigma$ on positions such that position $i$
carries conclusion $c$ in $F$ exactly when $\sigma i$ carries $c$ in
$G$, and $i \rightsquigarrow_F j$ iff
$\sigma i \rightsquigarrow_G \sigma j$. Isomorphism is reflexive,
symmetric, and transitive. An accepted program compiles to the
structured framework whose conclusions are those of its complete
arguments in declaration order and whose attack relation is the edge
relation of \autoref{def:compile}, read positionally.
\end{definition}

\begin{definition}[Compiler invariant]
\label{def:compiler-invariant}
Fix \Pol's declared conflict data: its \kw{contrary} pairs and
\kw{exception} conditions. A structured framework $F$ satisfies the
\emph{compiler invariant} when (i) every edge stays in range:
$i \rightsquigarrow_F j$ implies positions $i$ and $j$ both carry
conclusions; and (ii) conflict is complete: whenever the conclusions
at positions $i$ and $j$ match under \Pol's declared \kw{contrary}
relation, $i \rightsquigarrow_F j$.
\end{definition}

\begin{definition}[Realizability]
\label{def:realizable}
Fix \Sig, \Pol, and \Reg. A \emph{realization} of a structured
framework $F$ comprises a leaf table, a finite ground input, and a
raw program that the executable checker accepts under them, such
that the accepted program carries exactly the fixed \Sig{} and \Pol,
the ground input covers every leaf a declared argument uses, and the
accepted program's compiled structured framework is isomorphic to
$F$. We call $F$ \emph{realizable} in that fixed context when a
realization exists.
\end{definition}

\begin{theorem}[Image necessity]
\label{thm:image-necessity}
For every fixed \Sig, \Pol, and \Reg: a realizable structured
framework satisfies the compiler invariant for \Pol's declared
conflict data. The invariant transports across isomorphism, so the
choice of realization witness does not matter.
\end{theorem}

Necessity also carries sorts: in any realization, every node
conclusion is well sorted under \Sig, by the coverage hypothesis and
the substitution lemma of \autoref{sec:leaves}
(\textsf{node\_conclusion\_wellSorted}).

\begin{proposition}[The invariant does not imply realizability]
\label{prop:non-sufficiency}
Consider the structured framework with one node and one self-edge.
For every \Sig, \Pol, and \Reg{} in which \Pol{} declares no
\kw{contrary} pair and no \kw{exception}, that framework satisfies
the compiler invariant, since conflict completeness is vacuous, yet
it is not realizable: no attack can type under such a policy, so
every compiled framework is edgeless, and an isomorphism preserves
edges. The hypothesis is executable rather than degenerate, because
the checker accepts a program under one such fixed context.
\end{proposition}

The counterexample refutes sufficiency over arbitrary fixed policies.
Sufficiency for policies with declared conflict remains open.

%% file: src/surface.tex
\section{The verified surface boundary}
\label{sec:surface}

The surface calculus formalizes the presentation syntax of
\autoref{sec:language} above the parser boundary. We define an independent judgment for the full
structured surface AST, prove that elaboration preserves and
reflects checking, and prove that a claim observation taken directly
on the surface output agrees with the core observation for every
attack-extensional extension semantics. Parsing bytes into the AST
stays validated, and every theorem begins at the parsed AST.

\begin{definition}[Supported fragment and surface judgment]
\label{def:surface-judgment}
A surface input pairs a complete presentation program with its
policy; the registry, identifier classifier, and proposition encoder
stay parameters. The \emph{supported fragment} is a decidable
structural predicate over the full AST: the program names the
supplied policy, and declaration identifiers, rule namespaces, value
bindings, comparisons, inferred arguments, named certificates,
attack paths, and premise labels satisfy the supported syntax restrictions.
The \emph{surface judgment} is a syntax-directed relation defined
independently of the elaborator: it carries the supported fragment,
independent expansion relations for values, prose interpolation, and
comparisons, role, status, and group well-formedness, and
declarative core obligations rather than a wrapper around executable
success. Its executable checker is sound, complete, and
deterministic for it.
\end{definition}

\begin{theorem}[Elaboration preservation]
\label{thm:elab-preservation}
A derivable surface input elaborates successfully to its output, and
the actual core checker accepts that output. Elaboration preserves
the authored obligation ledger exactly, retains exactly the resolved
attacks, preserves each reconstructed argument's conclusion, and
aligns the authored argument order with the core-visible question
sequence.
\end{theorem}

\begin{theorem}[Supported-fragment reflection]
\label{thm:elab-reflection}
If a supported input elaborates successfully and the core checker
accepts the result, then the surface judgment derives that
elaboration. Both hypotheses are load-bearing: successful
elaboration alone establishes neither the supported fragment nor
core acceptance. The theorem reflects the given surface input, not
arbitrary core units, so it claims neither surjectivity onto the
core nor unique source recovery.
\end{theorem}

Binding is verified at two grains: genuine binders (rule parameters
and named certificate lambdas) admit alpha-equivalence with
identical lowerings, and a typed global renaming of non-binder
identifiers is equivariant for the whole pipeline under explicit
structural and environment hypotheses, while replay-significant
identities (backend names and versions, digests, policy identifiers)
are not renamable at all.

\begin{theorem}[Surface observation coherence]
\label{thm:surface-observation}
For a derivable surface input whose output the core checker accepts,
a direct observation computed from the surface output alone, its
retained arguments, resolved attacks, and reconstructed claims,
without compiling, equals the core observation of the compiled
framework, for every extension semantics with an attack-extensional
specification, and hence for the five named instances. The stable
instance preserves the no-extension report, and a missing claim
identifier stays missing on both sides rather than becoming an
empty-support \stGap.
\end{theorem}

%% file: src/observations.tex
\section{Extension-semantics distinctions and source transport}
\label{app:observations}

These results supplement the observation definitions of
\autoref{sec:observation}. They vary the extension semantics within a
framework; the possible-world layer of \autoref{sec:worlds} instead
relates accepted programs and their checking contexts.

\begin{proposition}[\stGap alone is semantics-independent]
\label{prop:gap-independent}
For every extension semantics and framework, a claim's observation
is \stGap if and only if its support is empty, with no further
hypothesis. Each of the other three statuses varies with the
semantics: on a two-cycle whose two arguments support the same claim,
grounded observes \stContested where preferred observes \stJustified, and a claim supported by an
argument both cycle members attack moves from \stContested to
\stDefeated. All ten pairwise separations among the five instances
are witnessed in one mechanized statement.
\end{proposition}

\begin{proposition}[Stable nonexistence is reported, not resolved]
\label{prop:stable-noext}
The bare three-cycle has no stable extension, while the other four
instances enumerate at least one there, so nonexistence belongs to
stable semantics rather than to the framework; and it belongs to the
bare cycle, since one added argument attacking all three members
restores a stable extension. On the bare three-cycle, stable observation of any supported claim is
the no-extension report, never a four-state status. Preferred extensions, by contrast, exist for every finite
framework.
\end{proposition}

Two refutations bound what any observation function could be. No
four-state function realizes the credulous reading: under preferred
semantics on the two-cycle, one claim is accepted in some extension
and defeated in some extension at once, and the two cycle members
carry identical acceptance profiles, so no framework-internal
tie-break recovers a verdict
(\textsf{credulous\_not\_functional}); the credulous reading stays
exposed as per-argument bits, never as a status. The observation
also does not factor through per-argument acceptance data
(\textsf{observe\_not\_determined\_by\_profile}).

\begin{theorem}[Source-level observation transport]
\label{thm:observation-transport}
Call a specification \emph{attack-extensional} when two frameworks
with equal carriers that agree on attacks between carrier members
satisfy it for the same extensions; all five instances are. Fix such
a semantics, a checked program $P$, and a claim whose support
indices all lie among $P$'s arguments. Then a direct source-level
observation, defined over every edge oracle that agrees with the
declared checked attacks on declared arguments, holds of an
observation exactly when that observation equals the one computed on
the compiled framework.
\end{theorem}

The support-boundedness hypothesis is necessary: the observation
deliberately reads attacks at support positions outside the carrier,
exactly where oracle agreement says nothing, and a mechanized
counterexample refutes the hypothesis-free statement. Status
preservation (\autoref{thm:preservation}) is therefore not the
grounded instance of
this theorem and stands as proved, with no support hypothesis; a
separate equivalence connects the two under the support bound
(\textsf{srcStatus\_iff\_srcObservation}).

%% file: src/proofs.tex
\section{Proofs for the metatheory}
\label{app:proofs}

This appendix proves the results of \autoref{sec:metatheory} and of the
appendix sections that extend it, grouped by pipeline stage. Each proof
follows the structure of its Lean counterpart, whose status
\appautoref{app:mechanization} records; the artifact's name map lists
the declaration behind every displayed object. Three arguments are on
paper only: the pipeline cost bound
(\autoref{prop:complexity}), the NP-completeness bookkeeping behind
\autoref{thm:np-hardness}, and the subargument-closure, strict-closure, and
indirect-consistency clauses of \autoref{thm:rationality}. The admission
properties of \appautoref{app:mechanization} cite their declarations in
their statements and need no separate proof.

Throughout, we abbreviate $\Sig;\Pol;\Ctx;\Reg \vdash w : \supports(p)
\obl O$ to $\vdash w : p \obl O$ when the contexts are fixed, and we
write $\mathsf{G}$ for the grounded extension of the framework under
discussion.

\subsection{Checking}

\begin{proofof}{\autoref{thm:decidability} (decidability, G1)}
The support checker is defined by structural recursion on $w$ and
mirrors \autoref{fig:supportrules}. At a leaf it looks up $\Ctx(l)$. At
an instance it looks up $r$, instantiates the premise and conclusion
patterns under $\theta$ (which fails exactly when a parameter is
unbound; terms are ground, so no capture arises), recurses on the
premise and discharge subterms, and decides the remaining side
conditions: the domain of $\theta$, the premise count, the identities
$p_i \equiv P_i\theta$ and $p_q \equiv A_q\theta$, the partition
$\{q_j\} = D \uplus H$, and the assurance. Each test is finite; $\equiv$
is decidable (\autoref{sec:normalization}); and the assurance test on a
certified instance is one call to $\chk\backend$, total by
law~\ref{law:replay}. The checker therefore terminates, and it is sound
and complete for the judgment: soundness by induction on $w$, reading
each passed test as the matching premise of \rulename{Rule};
completeness by induction on the derivation, since each premise forces
the matching test to pass and the induction hypotheses force each
recursive call to return the derived conclusion and obligations.

The attack checker runs the support checker on the attacker, walks the
finite position $\pi$ in the target, and scans the finite \kw{contrary}
and \kw{exception} tables. The contrary test matches a declared pattern
pair against two ground atoms by first-order matching, which is
deterministic. Adequacy for \autoref{fig:attackrules} is a case analysis
on the attack kind, using \autoref{prop:support} to identify the
attacker's checked conclusion with the derivation's. Program checking
adds finitely many bounded scans over these two checkers: duplicates, $O
= \varnothing$, endpoint declaration, and attack completeness.
\end{proofof}

\begin{proofof}{\autoref{prop:support} (conclusion uniqueness and support adequacy)}
Uniqueness is by induction on the first derivation with inversion of the
second. At a leaf both read $\Ctx(l)$. At an instance both resolve the
same rule from the identifier, and the conclusion is $C\theta$ for the
instance's own explicit $\theta$; instantiation is a function, so the
conclusions agree without consulting the subterms. The obligation sets
agree because $O$ is a function of the subterms' obligation sets, equal
by induction, and of $r$ and $H$.

Hence $w \mathrel{\supports} c$ iff $\concl(w) \equiv c.\mathit{formal}$
for the unique $\concl(w)$: one run of the checker and one comparison of
normal forms. The relation respects $\equiv$ because $\equiv$ is
transitive.
\end{proofof}

\begin{proofof}{\autoref{thm:accountability} (dependency accountability, G2)}
\emph{Leaves.} $\leaves(w)$ is a leaf's own name, or the union over the
premise and discharge subterms of an instance. Every leaf of a checked
term is declared, by induction on the derivation: at a leaf the
\rulename{Leaf} premise is the witness; at an instance the leaf lies in
a subterm with its own derivation.

\emph{Certificates.} $\certDeps(w)$ collects, over the certified
occurrences $u$ of $w$, the slot set that $u$'s backend reports for its
certificate, resolving a slot $i$ below the premise count $n$ to premise
$i$ and a slot $i \ge n$ to theory entry $i - n$ of the digest-resolved
theory. A dependency is in $\certDeps(w)$ iff some certified occurrence
reports it, and every certified occurrence of a checked term is itself
checked, so each report comes from an accepted certificate. Each entry
is genuine: a premise entry names the premise subterm at its position,
up to $\equiv$, and a theory entry is in range of the resolved theory by
the validity clause of law~\ref{law:deps}. The coverage and accounting
clauses of law~\ref{law:deps}, discharged by each registered backend,
say that replay reads nothing outside the report and that the conclusion
follows from the reported entries alone. Because a digest resolves only
to theory data for one fixed checker, coverage also holds across
digests.
\end{proofof}

\subsection{Compilation and claim statuses}

\begin{proofof}{\autoref{thm:compilation} (compilation soundness, G3)}
Every node $w \in A(P)$ carries $\vdash w: \concl(w) \obl \varnothing$
by \rulename{Accept}. For the edges, the attacked occurrence is a
function of the attack: a rebut names its whole target, and an undercut
or undermine names the subterm at its recorded position. By
\autoref{def:compile}, $v \rightsquigarrow w$ iff some declared attack
has source $v$ and its occurrence is contained in $w$; \rulename{Accept}
types every declared attack, so no untyped attack contributes an edge,
and containment of the occurrence is the only way $w$ becomes a target.
The direct edge from each attack's source to its target is present,
since a target contains its own attacked occurrence.
\end{proofof}

\begin{proofof}{\autoref{thm:determinism} (determinism and termination)}
Let $n = \lvert\mathcal{A}\rvert$, $S_0 = \varnothing$, and $S_{k+1} =
\mathcal{F}_F(S_k)$. The characteristic function is monotone, since a
defender in $S$ is a defender in any $T \supseteq S$, so the chain
ascends. If $S_{k+1} \not\subseteq S_k$ then $\lvert\mathcal{A}
\setminus S_{k+1}\rvert < \lvert\mathcal{A} \setminus S_k\rvert$, so at
most $n$ rounds are strict, and once a round is stable all later rounds
are, by monotonicity. Hence $S_n = \mathcal{F}_F(S_n)$. It is the least
fixpoint, since every fixpoint contains each $S_k$ by induction on $k$.
Equivalently, $S_n$ is the least predicate $I$ closed under the defence
rule of \autoref{def:direct-status} read over $F$'s own edges: $S_k
\subseteq I$ by induction on $k$, and $I \subseteq S_n$ by induction on
the derivation, because $S_n$ is a fixpoint. The labelling and the
cascade of \autoref{def:status} are total functions of $S_n$, so
evaluation is deterministic.
\end{proofof}

\begin{proofof}{\autoref{thm:preservation} (status preservation)}
Index the declared arguments positionally; $A(P)$ is duplicate free, so
node $i$ of $\mathit{AF}(P)$ is the $i$-th declared term. The compiled
attack relation is decided by a structural scan for the attacked
occurrence inside the target. The scan agrees with positional
containment because checked terms have duplicate-free discharge keys
(\autoref{fig:supportrules}), so it decides the edge relation of
\autoref{def:compile} exactly on in-range indices and is false off
range.

The direct judgment (\autoref{def:direct-status}) and the same judgment
read over the compiled edges are least predicates closed under the same
two rules and differ only in the edge relation they read; the decider
translates each edge, so the two coincide by mutual induction on
derivations. On the compiled framework, directly in is membership in
$\mathsf{G}$ and directly out is being attacked by $\mathsf{G}$
(\autoref{thm:determinism}), so an argument is labelled \emph{in} iff it
is directly in, \emph{out} iff directly out and not directly in, and
\emph{undec} otherwise.

The four cases of \autoref{def:direct-status} are mutually exclusive,
each lower case carrying the negation of the higher ones, and the
cascade of \autoref{def:status} derives exactly one of them: a
successful scan of the support set yields the existential witness and a
failed scan the universal negation. A functional relation that holds at
the computed value is that value: the direct status equals the compiled
status.
\end{proofof}

\begin{proofof}{\autoref{thm:rationality} (rationality postulates, G6)}
Write $J$ for the set of conclusions of members of $\mathsf{G}$.

\emph{Closure under declared subarguments.} Let $s$ be a declared
argument and a subterm of $w \in \mathsf{G}$. Positions compose, so every occurrence
contained in $s$ is contained in $w$, and every attacker of $s$ attacks
$w$ (\autoref{def:compile}). $\mathsf{G}$ defends $w$, hence defends
$s$, and $s \in \mathsf{G}$ by the fixpoint property.

\emph{Closure under declared strict steps.} Let
$w = r\langle\theta; w_1,\dots,w_n; \varnothing; \varnothing;
\alpha\rangle$ be a declared strict instance with each $w_i$ declared
and in $\mathsf{G}$. No rule of \autoref{fig:attackrules} targets the
root of $w$: rebuttal and undercutting need a defeasible rule, and
undermining needs a leaf. Every attack on $w$ therefore targets an
occurrence inside some $w_i$ and so attacks $w_i$; $\mathsf{G}$
defends each $w_i$, hence $w$, and $w \in \mathsf{G}$. This is closure
for declared instances; the calculus never manufactures an undeclared
one.

\emph{Conflict-freedom.} No two members of $\mathsf{G}$ attack each
other, $a = b$ allowed. By induction on the round: if $a, b \in S_{k+1}$
and $a \rightsquigarrow b$, defence of $b$ yields $c \in S_k$ attacking
$a$, and defence of $a$ yields $d \in S_k$ attacking $c$, contradicting
the hypothesis for $S_k$.

\emph{Direct consistency.} Suppose contrary claims $p$ and $q$ are both
\stJustified. Then some $w_s, w_t \in \mathsf{G}$ have $\concl(w_s)
\equiv p$ and $\concl(w_t) \equiv q$, and since contrary matching
respects $\equiv$ their conclusions form an instantiated contrary pair.
The target $w_t$ is attackable at its root: a leaf can be undermined,
and if the root rule were strict its conclusion pattern would be
strict-reachable and would share a ground instance, up to $\equiv$, with
a contrary side, which the conservative overlap check reports and
\autoref{def:strict-wf} forbids; so the root is defeasible and can be
rebutted. Attack completeness (\rulename{Accept}) then supplies a
declared attack from $w_s$ onto an occurrence of $w_t$, so $w_s
\rightsquigarrow w_t$, contradicting conflict-freedom. Hence $J$ is
contrary free and no two contrary claims are both \stJustified.

\emph{Indirect consistency.} Let $\mathrm{Cl}(J)$ close $J$ under the
ground instances of \Pol's strict rules. Every atom of
$\mathrm{Cl}(J) \setminus J$ instantiates a strict-reachable pattern,
so by \autoref{def:strict-wf} it is a side of no instantiated contrary
pair. A contrary pair in $\mathrm{Cl}(J)$ therefore lies in $J$, which
is contrary free. Direct and indirect consistency thus coincide, in
the sense of \citet{Caminada_2007}.
\end{proofof}

\begin{proofof}{\autoref{thm:nonpromotion} (quarantine non-promotion)}
Let $G$ be the declared framework and $F$ the checked one. Both are
indexed by the declared argument list: $F$'s carrier is the retained
indices, and $F$'s attacks are those declared attacks whose endpoints
survive, aligned by declared identity rather than by term equality, so
$F$'s edges are a subset of $G$'s. Let the \emph{seed} be the removed
arguments together with the retained arguments that lost an incoming
edge, and let $B$ be its forward closure along $G$'s edges, computed by
the bounded iteration of \autoref{thm:determinism}. Outside $B$ the two
frameworks agree: an argument $y \notin B$ is retained and lost no
incoming edge, so its attackers in $F$ and in $G$ coincide.

\emph{Locality.} If $a \notin B$ is directly in for $F$ then it is
directly in for $G$, and if $b \notin B$ is directly out for $F$ then it
is directly out for $G$, by mutual induction on the derivation. For $a$:
a $G$-attacker $b$ of $a$ is retained, since a removed attacker would
seed $B$ and closure would put $a$ in $B$; so $b$ is an $F$-attacker, $b
\notin B$ by closure, and the induction hypothesis transfers its
out-derivation. For $b$: its $F$-defeater $c$ is a $G$-defeater by edge
agreement at $b$, and $c \notin B$ by closure. Grounded labelling reads
only transitive attackers, which is why forward closure suffices.

\emph{Conclusion.} The checker publishes \stJustified{} for $c$ only
when $F$ labels some $w \in \mathrm{supp}_F(c)$ \emph{in} and no member
of $\mathrm{supp}_F(c)$ lies in $B$ (\autoref{sec:semantics}). By
locality $w$ is \emph{in} for $G$, and $\mathrm{supp}_F(c) \subseteq
\mathrm{supp}_G(c)$ because reinstating removed material only adds
support; so $c$ is \stJustified{} in $G$. The converse fails because the
other three statuses read universal or emptiness conditions over a
support set that quarantine may have shrunk.
\end{proofof}

\subsection{The image of compilation}

\begin{proofof}{\autoref{thm:image-necessity} (image necessity)}
The compiled framework of every accepted program satisfies the
invariant. Edges join positions of declared complete arguments, which
carry conclusions. For conflict completeness, an accepted program's
policy is strict-chain well formed, so every contrary target is
attackable at its root (as in \autoref{thm:rationality}), and attack
completeness turns each contrary-matching pair into a declared attack
and hence a compiled edge. Both clauses are stated in terms of
conclusions and edges, which an isomorphism preserves, so the invariant
transports; composing along the realization's isomorphism gives the
claim. Sortedness of node conclusions follows from the coverage
hypothesis and the substitution lemma of \autoref{sec:leaves}.
\end{proofof}

\begin{proofof}{\autoref{prop:non-sufficiency} (non-sufficiency)}
Under a policy with no \kw{contrary} pair and no \kw{exception}, the
one-node self-edge framework satisfies the invariant, conflict
completeness being vacuous. But every attack rule demands conflict data,
so no attack types, every compiled framework is edgeless, and an
isomorphism would carry the self-edge onto a compiled edge. The context
is inhabited: the checker accepts the empty program under it, and a
nonempty accepted program realizes its own framework up to a
non-identity renaming.
\end{proofof}

\subsection{Semantics-parametric observation}

\begin{proofof}{\autoref{thm:observe-grounded} (the printed status is the grounded instance)}
On a duplicate-free carrier exactly one carrier subset is the least
complete extension, namely $\mathsf{G}$, so the grounded enumeration is
a one-element list. Unfolding \autoref{def:observation} at a singleton:
the \stGap{} arm is the emptiness test; the no-extension arm is
impossible; ``every extension contains a support member'' is ``some
support argument is \emph{in}''; ``every extension attacks every support
member'' is ``every support argument is \emph{out}''; and the residual
arm is \stContested{} on both sides. On a singleton enumeration the
skeptical and credulous quantifiers coincide.
\end{proofof}

\begin{proofof}{\autoref{prop:gap-independent} (\stGap alone is semantics-independent)}
The \stGap{} arm is the first guard and reads only the support set, and
every other arm lies past it. For the separations, take the two-cycle $0
\leftrightarrow 1$ with both nodes supporting $c$: its grounded
extension is $\varnothing$ and its preferred extensions are $\{0\}$ and
$\{1\}$, so grounded observes \stContested{} where preferred observes
\stJustified. Adding a node $2$, attacked by both and alone supporting
$c$, yields \stContested{} against \stDefeated. The ten pairwise
separations among the five instances are one mechanized statement.
\end{proofof}

\begin{proofof}{\autoref{prop:stable-noext} (stable nonexistence)}
No subset of the three-cycle is conflict free and attacks every outside
node, so its stable enumeration is empty, while the other four semantics
enumerate at least one extension there; a fourth node attacking all
three restores the stable extension $\{3\}$. For a supported claim the
\stGap{} guard passes and the empty-enumeration guard fires, so the
observation is the no-extension report. Preferred extensions exist for
every finite framework: $\varnothing$ is admissible, and an admissible
set of maximum size is maximal.
\end{proofof}

\begin{proofof}{\autoref{thm:observation-transport} (source-level observation transport)}
Each of the five specifications reads the attack relation only between
carrier members, so it is attack-extensional; conflict-freedom must be
relativized to the carrier for this to hold. Fix an edge oracle that
agrees with the declared attacks on declared arguments. Its framework
and the compiled one share the carrier and agree on attacks within it,
so they have the same extensions; the observation also reads attacks at
the claim's support positions, and the support bound keeps those in the
carrier. Existence and uniqueness of the source observation close the
equivalence. The oracle hypothesis is strictly weaker than faithfulness,
and dropping the support bound is refuted by a counterexample.
\end{proofof}

\subsection{Strict-certificate backends}

\begin{proofof}{\autoref{thm:isolation} (isolation and non-factivity)}
A registry entry exposes an encoding into an abstract formula type, a
consequence relation and an acceptance predicate over that type, a
Boolean replay, and a slot-index dependency report
(\autoref{def:registry-entry}). No operation returns a source
proposition, support term, or attack, and a certificate is an opaque
payload that replay receives unchanged; so no elimination of a backend
formula or proof term into a source judgment is expressible.

Non-factivity is refuted positively: no uniform projection sends every
certified step to a premise-free consequence of its encoded conclusion,
over all canonicalizers, backends, and theories. The identity step $p
\Rightarrow p$ is certified against \textsf{nd@1} by $\kw{hyp}\ 0$, yet
the all-false valuation refutes $p$ from no premises; the same step is a
consequence of its premise by \autoref{thm:strict-soundness}.

A program carries theory data addressed by digest and nothing else
backend-related. An unregistered name, version, or digest makes
acceptance false, and a digest resolves only to theory data for the
identity's one fixed checker, so a program selects among registered
backends but adds none.
\end{proofof}

\begin{proofof}{\autoref{thm:strict-soundness} (strict-certificate soundness, G4)}
The assurance test replays $\kappa$ against the goal
$\enc\backend(C\theta)$ under the encoded premises
$[\enc\backend(P_i\theta)]$ followed by the resolved theory $T$.
Law~\ref{law:replay} turns the verdict into acceptance, and
law~\ref{law:sound} then gives $T; [\enc\backend(P_i\theta)] \bmodels
\enc\backend(C\theta)$. A \kw{trusted} instance has no certificate, so
the hypothesis of law~\ref{law:sound} is unavailable. It remains to
discharge that law for each registered backend.

\emph{\textsf{nd@1}.} Consequence is propositional entailment. Replay
decodes $\kappa$ to a term $e$ and type-checks it by the structural
recursion of \autoref{fig:ndrules}, which agrees with the typing
relation. Soundness is induction on $\Delta \vdash e: \phi$ for a
valuation $v$ satisfying $\Delta$: \rulename{Hyp} reads a satisfied
context formula, \rulename{Abs} extends the satisfied context,
\rulename{App} is modus ponens, and \rulename{Abort} has an
unsatisfiable premise. The relevance form restricts the hypothesis to
the free indices of $e$, which is the accounting clause of
law~\ref{law:deps}.

\emph{\textsf{ra@1}.} A certificate names two premise slots and carries
a witness fraction in lowest terms. Replay checks, over exact integers,
that the cited cells carry the goal's numerals, that the full value is
nonzero, that the recomputed relative drop equals the witness, that the
witness equals the claimed drop, and that the claimed drop meets the
threshold. The consequence omits the witness; it cancels by multiplying
the claimed-drop identity through by the witness denominator, positive
by the grammar.

\emph{\textsf{ord@1}.} A certificate names two premise slots; both
numerals appear in the goal, so replay decides the order on them and
checks that the cited cells carry them. The decided order is the
consequence. Strict and non-strict order over transposed operands
exclude each other by arithmetic, which is why the strict and non-strict
order predicates declare no contrary pair.

\emph{\textsf{insp@1}.} A certificate selects one inventory premise,
or two for a planned-versus-shipped comparison. Replay checks that each
selected premise contains exactly one inventory for the source named
in the goal, then decides the goal's membership, absence, or uniqueness
condition. For a comparison it checks presence in the plan inventory
and absence in the implementation inventory. These conditions are the
backend consequence relation. Soundness is relative to the declared
inventories; it does not establish that an inventory exhaustively
describes the referenced source bytes.
\end{proofof}

\begin{proofof}{\autoref{thm:replacement} (backend replacement, G5)}
Let $f$ be an injective map on certificate payloads such that one
backend accepts a strict instance with payload $\alpha$ iff the other
accepts it with $f\alpha$; by law~\ref{law:replay} acceptance is a
function of the source, so this is what ``accept the same strict
instances'' means. Write $f \cdot w$ for $w$ with $f$ applied at every
assurance and everything else fixed.

\emph{Typing transports.} By induction on the derivation, $\vdash w: C
\obl O$ implies $\vdash f \cdot w: C \obl O$ with the same $C$ and $O$,
because only the assurance premise of \rulename{Rule} reads the payload.
Attack typing transports because positional lookup commutes with $f
\cdot$ and no other premise of \autoref{fig:attackrules} reads a
certificate. Since $f \cdot$ is injective on terms, the relabelled
program is duplicate free and is a checked program over the second
backend.

\emph{Frameworks coincide.} Nodes are list positions and $f \cdot$ maps
the argument list pointwise, so the carriers are equal. The edge test
between two positions compares attack sources for equality and scans the
target for the attacked occurrence; both are invariant under an
injective relabelling. Hence $\mathit{AF}(P) = \mathit{AF}(f \cdot P)$,
conclusions correspond index for index, and every claim has the same
status. Injectivity is load bearing: collapsing distinct payloads to one
marker can identify subterms and add closure edges.
\end{proofof}

\begin{proofof}{\autoref{thm:hetero} (heterogeneous backend compositionality)}
Fix a certified occurrence $o$ of $w$. Strict soundness lifts through
the derivation: at $o$'s node the derivation binds a registered backend
and a resolved theory and yields the rule, the instantiated premises and
conclusion, and the occurrence-local consequence. The node is itself
checked, and inverting its assurance premise is acceptance by its own
checker. The two inversions name the same rule, premises, and conclusion
because all three are values of the same lookups. The backend is bound
existentially inside each occurrence, so conjoining two occurrences
relates no formula types, theories, or consequence relations. The
dependency union is the collection identity of
\autoref{thm:accountability} applied per occurrence.

For the firewall, \rulename{Rule} reads a subterm only through its
conclusion, its obligations, and, for a discharge, its question key; so
replacing a subterm by one with the same conclusion and obligations
preserves the parent's typing. A witness performs such a swap under an
\textsf{nd@1} parent and leaves a term that uses two backends.
\end{proofof}

\subsection{Contextual representation independence}

\begin{proofof}{\autoref{thm:ctx-independence} (contextual representation independence)}
The observation is the hygiene guard, then the checker on the linked
program, then the status readout; each stage is blind to the
relabelling. The guard reads identifiers, signatures, policies, imports,
and exports, none of which $f$ changes. Since $C$ is fixed by $f$,
linking commutes with relabelling: the merged lists of $(C, f \cdot F)$
are the $f$-images of those of $(C, F)$, and the cross-boundary attacks
the link inserts depend only on conclusions. Acceptance transports along
$f$ as in \autoref{thm:replacement}, with well-sortedness and attack
completeness reading conclusions only. The compiled structured framework
reads conclusions and positional attacks and never an assurance, so the
two linked programs compile to the same framework and export the same
statuses. The acceptance-profile form is the case $f = \mathrm{id}$, and
the composed form instantiates the congruence at the composed context,
whose admissibility is assembled from its halves under explicit
cross-side coverage hypotheses.
\end{proofof}

\begin{proofof}{\autoref{thm:ctx-sem} (contextual congruence at every semantics)}
Both observations factor as one generic observation composed with a
projection of the compiled framework: the grounded status, or the claim
observation of \autoref{def:observation} at the chosen semantics. The
proof of \autoref{thm:ctx-independence} concludes an equality of
compiled frameworks and never inspects the projection, so it holds for
every projection at once, and the grounded instance is definitional. For
the separations: on a linked three-cycle, stable semantics reports no
extension where grounded reports \stContested; and two fragments that
differ only in their exported atom, each with unattacked leaf support
under a policy with no conflict, are grounded-equivalent in every
context yet separated by an adequate singleton-selector semantics at one
accepted context.
\end{proofof}

\begin{proofof}{\autoref{thm:holes} (typed hole substitution)}
Template typing factors the side conditions of \rulename{Rule} so that a
named hole receives its demanded answer from the hole signature. Erasing
the unanswered holes moves each question key into $H$, so the erased
term is a source term and its derivation follows by the same induction,
with mandatory questions becoming obligations. Substituting a typed
filling closes each discharge condition up to $\equiv$ and returns the
same root conclusion and residual obligations. Substitution may identify
templates, so the linker deduplicates the instantiated lists and
re-imposes attack coverage after substitution. Closed embeddings recover
the original observations definitionally, and the congruences pass
through because relabelling commutes with substitution. The context's
fillings become part of the substituted fragment, so a relabelling must
fix their certificates too; a witness fixes a context's frame while a
certified filling moves.
\end{proofof}

\begin{proofof}{\autoref{thm:parametricity} (relational parametricity)}
Lift $R$ structurally through terms, attacks, and fragments and re-run
\autoref{thm:ctx-independence} with the function replaced by the
relation. Support and attack typing transport along $R$ by acceptance
transport, and saturation commutes because conclusions are preserved.
Partial bijectivity is consumed exactly where the calculus compares
terms: deduplication in the merge, source comparison in the coverage
decider, and the checker's duplicate-freedom premise. The two linked
programs then compile to equal frameworks, after which the relation is
erased and every projection agrees. The graph of $f$ recovers
\autoref{thm:ctx-independence} definitionally.

Partial bijectivity is necessary. Dropping it refutes coverage
transport; and a relation that collapses two payloads merges two
wrappers of one conclusion under deduplication, so an undercut on one
wrapper reaches the merged node and an exported claim moves from
\stJustified{} to \stDefeated{}. Finitely much related material is
realized by a total injection that fixes the context, built by composing
one swap per constraint, while an arbitrary such relation need not
extend to a total injection.
\end{proofof}

\subsection{Source updates and status dynamics}

\begin{proofof}{\autoref{thm:update-preservation} (acceptance preservation)}
Each constructor edits one raw carrier and reruns admission and the
checker, so preservation is a matter of discharging both stages'
premises on the edited state. For \upAddLeaf, freshness keeps the
duplicate scans clean and the admitted key keeps the new row out of the
prune, so the retained arguments and attacks are unchanged and every
derivation transports by monotonicity in \Ctx{}. For \upTighten, the new
prune restricts the old one, retained-leaf agreement keeps the surviving
environments equal, and the coverage hypothesis re-closes attack
completeness. For \upAddAttack, the new attack resolves, types, and has
retained endpoints, so it extends an already typed list. For
\upAddInstance, freshness, well-sortedness, complete support, and
extended attack completeness are the checker's remaining premises. In
each case completeness of the checker (\autoref{thm:decidability})
accepts the edited state.
\end{proofof}

\input{tables/updates}

\begin{proofof}{\autoref{thm:update-matrices} (grounded transition matrices) and \autoref{thm:gap-boundary} (\stGap boundary)}
Each reachable cell is a fixture: an accepted source, an accepted
target, and a successful update between them, with both observations
evaluated by reduction. The matrices and their counts are computed from
the fixtures and pinned by a committed golden. Four lemmas exclude the
rest. A fresh admitted leaf changes no retained argument or attack, so
the framework and every status are fixed. Non-instance updates add no
argument, so empty support stays empty; additive updates remove none, so
nonempty support stays nonempty. These read only the support set and
hence hold for every extension semantics, which is
\autoref{thm:gap-boundary}. Finally, a fresh instance is a sink of the
old framework, since no old attack names it, so old labels persist: an
\emph{in} argument stays \emph{in} and an \emph{undec} argument never
becomes \emph{out}, which excludes the three cells from \stJustified{}
or \stContested{} into \stDefeated{}. This last lemma is what corrected
the predicted count of 13 reachable \upAddInstance cells to 10.
\end{proofof}

\begin{proofof}{\autoref{cor:matrix-nonpromotion} (matrix non-promotion) and the public layer}
On a clean source the seed of \autoref{thm:nonpromotion} is empty, so
the blocked set is empty and the public report is the core report. An
additive update removes nothing, so its target is clean and the additive
public transitions are the core ones, never reaching
\stEvidenceBlocked{}. Cleanliness is necessary: an accepted non-clean
source exists on which one added attack grows the blocked set and moves
a public \stJustified{} to \stEvidenceBlocked{} while the core status
reads \stDefeated{}. For the corollary, a public \stJustified{} target
under \upTighten forces a core \stJustified{} source, and on a clean
source the checked framework is the declared one. The two AGM probes are
fixtures: an \upAddInstance whose only new support is defeated on
arrival refutes success, and an \upAddAttack that removes a justified
claim refutes inclusion.
\end{proofof}

\subsection{The verified surface boundary}

\begin{proofof}{\autoref{thm:elab-preservation} (elaboration preservation)}
Elaboration runs a fixed sequence of passes: validation, value and prose
interpolation, comparison expansion, argument reconstruction with
substitution inference and certificate lowering, attack resolution, unit
assembly, and admission. A surface derivation carries, for each pass,
that pass's independent relation, and each pass has a completeness lemma
that drives it to success on the derivation's own output. Core
acceptance does not come from rerunning elaboration: the derivation's
declarative core obligations feed the core checker's completeness
theorem directly. The four projections are output-component equalities
the derivation carries.
\end{proofof}

\begin{proofof}{\autoref{thm:elab-reflection} (supported-fragment reflection)}
Invert a successful elaboration pass by pass: each adjacent pair of
passes has a boundary lemma recovering the earlier pass's output, and
each pass's soundness lemma yields its relation. The supported fragment
supplies the structural facts elaboration never checks, and the
successful core check supplies the core obligations. Both hypotheses are
necessary: inputs exist that elaborate while unsupported, and supported
inputs exist that elaborate while the core checker rejects the result.
Determinism of the judgment pins the derivation's output to the
elaborated one.
\end{proofof}

\begin{proofof}{\autoref{thm:surface-observation} (surface observation coherence)}
The direct framework has the retained argument positions as carrier and
the resolved attacks as edges, computed without the compiler. For an
accepted derivation it equals the compiled framework exactly, and the
direct and core claim ledgers agree, a missing identifier staying
missing on both sides. The direct claim's support lies in the carrier,
so \autoref{thm:observation-transport} transports the observation for
every attack-extensional semantics, and the five corollaries discharge
that hypothesis per instance. Carrier-locality is necessary.
\end{proofof}

\subsection{Cross-context comparison}

\begin{proofof}{\autoref{thm:pw-normal} (modal normality and comparison adequacy)}
Satisfaction of a box quantifies over accepted edges, and acceptance
does not depend on the formula, so K and necessitation follow by
unfolding and duality is classical negation over the same edge set. T,
D, B, and 5 fail on a two-world frame with one accepted edge and a
valuation separating the worlds, and 4 fails on a three-world chain
whose composite edge is not accepted. These countermodels inhabit the
generic frame type with arbitrary valuations, rather than the
checked-program instance with computed claim statuses. The executable comparison guards
translation, candidates, and acceptance in that order and returns either
a reason or the statuses of the accepted candidates; when its candidate
list and acceptance test present the bridge, a status is output iff the
diamond holds, and an incomparable query has no accepted witness. The
one-context, no-bridge instance is local evaluation by definition, and
\autoref{thm:preservation} at each world makes the direct and compiled
valuations agree, so they satisfy the same formulas.
\end{proofof}

\begin{proofof}{\autoref{thm:support-transport} (exact support transport)}
By induction on the derivation. The leaf case is the bridge's leaf
clause. In the instance case the rule clause supplies the translated
rule at the same identifier; instantiation commutes with translation,
and $\equiv$ survives it because normalization touches literals while
translation touches names; question keys, discharge keys, and holes
cross verbatim, so $O$ is unchanged; and the certificate clause handles
the assurance. Completeness and claim-level support follow, and the
certified occurrences of the transported term replay against the target
registry by \autoref{thm:hetero}. Definedness of the term's translation
is a genuine hypothesis: $\theta$ may bind an unused parameter outside
the bridge's vocabulary, so a translatable conclusion does not imply a
translatable term.

Symbol maps compose in Kleisli fashion with both unit laws, every
structural lift satisfies a composition law, and stepwise transport
along a path equals transport along the folded bridge. A direct bridge
commutes with a path iff their symbol and leaf maps agree, because
support terms expose both maps; two negative triangles, one disagreeing
only on a leaf renaming and one only on a predicate, show each component
necessary.
\end{proofof}

\begin{proofof}{\autoref{thm:status-transport} (conditional status preservation)}
Grounded status is invariant under a total attack bisimulation of finite
frameworks: directly in and directly out transfer in both directions by
induction on derivations, so labels correspond, and a support
correspondence then carries all four statuses. Nodes are positions, so
relate $i$ to $j$ when the $j$-th target argument is the transport of
the $i$-th source argument; then (i) is left-totality, (ii)
right-totality, and (iii) the forth and back conditions. The support
correspondence for a translatable claim follows: forward by
\autoref{thm:support-transport} and conclusion uniqueness, backward by
(ii) and reflection of $\equiv$ along the injective translation, which
is where injectivity is used; a non-injective bridge must supply the
correspondence itself. The statuses therefore agree, also at the source
observation through \autoref{thm:preservation} at both worlds.

Each hypothesis is load bearing: the witness of
\autoref{thm:support-transport} satisfies (i) and the forth half of
(iii), yet its target holds an argument that is no transport, (ii)
fails, and the status flips. The four clauses are bounded
quantifications over indices, so a Boolean checker decides them. A
declared-attack correspondence yields (iii) when the leaf renaming is
injective, since a collapsing renaming can merge subterms. The
hypotheses compose along paths, and statuses are values, so pathwise
preservation is transitivity of equality. When every accepted successor
is so related and one exists, the boxed and diamonded translations are
each equivalent to the local status atom.
\end{proofof}

\subsection{Cost}

\begin{proofof}{\autoref{thm:grounded-cost} (grounded query bounds)}
The instrumented evaluator counts one query per attack-relation
evaluation, in the reference evaluator's short-circuit order, and its
value projection equals the reference evaluator. Each of the $n$ rounds
tests defence for at most $n$ arguments; each test scans at most $n$
attackers, each costing one outer query and at most $n$ inner queries;
this gives $n^3(1+n)$. Each round issues at least one query per argument
per scan, which gives $n^2$. A status query adds one support scan and
one defeat scan over at most $n$ members, at most $2n^2$. The quartic
family places $k - 1$ neutral nodes and a defender before $k$ attacked
nodes and $k$ targets, with edges from the defender to each attacked
node and from each attacked node to each target; it is realizable in the
fixed context, and its repeated defence scans cost at least $k^4$
queries for $k \ge 2$. The two-node all-attacks framework costs exactly
four queries, so no instance-wise cubic floor holds.
\end{proofof}

\begin{proofof}{\autoref{thm:np-hardness} (hardness on the realizable class)}
\emph{Gadget.} For each variable, two literal roots that undermine
each other; for each clause, an instance undermined at each literal
position by the opposing literal root; and one query argument
undermined by every clause root. All leaves come from one closed
vocabulary with an injective spelling table, under one frozen
signature, policy, and registry.

\emph{Realizability and size.} The checker accepts every output: the
lemmas cover duplicate-freedom, sortedness, support for literal and
clause arguments, typed attacks, and attack completeness, the last using
injectivity of the spelling, without which two variables with colliding
payloads would demand an undeclared attack. With the compilation
isomorphism this is realizability in the fixed context; the node count
and byte bound hold by construction. The one-node self-edge framework is
not realizable there, since no policy row licenses its edge.

\emph{Correctness.} From a satisfying assignment, the set of the query
and the true literal roots is complete and contains the query. From a
complete extension containing the query, every clause root is attacked,
so each clause has a literal root in the extension, and the mutual
literal attacks keep the read-off assignment consistent.

\emph{On paper.} Constructing the reduction takes polynomial time:
the construction enumerates literal and clause nodes and their pairs,
decides each gadget edge by bounded scans, and writes the encoded
carrier. The node count and byte bound limit these operations
polynomially in the encoded formula's length. Membership of credulous
complete acceptance in NP is
guess-and-check: an extension has polynomial size, and completeness
and query membership are polynomial-time checkable. With the
polynomial-size reduction this gives NP-completeness on the realizable
class under standard encodings.
\end{proofof}

\begin{proofof}{\autoref{prop:complexity} (polynomial core checking and evaluation)}
The input contains explicit support trees and substitutions. At each
support node, the checker scans finite tables and lists, instantiates
policy patterns, checks normalized identities, and combines obligation
lists. A pattern instantiation copies only substitution terms already
present in the input, so its expanded size is polynomial in $N$.
Duplicate and membership checks may be quadratic in their list lengths;
the number and lengths of these lists are bounded by the explicit input
and its instantiated patterns. The structural recursion therefore
performs polynomial work, with backend encoding and replay charged as
opaque calls.

Attack checking adds bounded position walks and pattern matching.
Whole-program acceptance checks argument pairs for conflict coverage.
Compilation examines argument pairs, scanning declared attacks and
testing structural containment to decide each edge. These operations
are polynomial in $N$. Grounded evaluation makes polynomially many
edge queries by \autoref{thm:grounded-cost}; either recomputing an edge
or looking it up in a materialized relation preserves the bound.
Aggregation scans complete support sets and labels, so the stated core
pipeline is polynomial. This argument gives no bound in compact
surface-source size or on backend internals. The timings in
\appautoref{app:performance} measure overhead on small inputs and do
not test this bound.
\end{proofof}

%% file: tables/updates.tex
\begin{table}
\caption{The four grounded core transition matrices of
\autoref{thm:update-matrices}. Rows are source statuses and columns
target statuses: \stJustified{} (J), \stDefeated{} (D),
\stContested{} (C), \stGap{} (G). A \cmark{} cell carries a
mechanized witness, an accepted source and target program with a
successful update between them; a mechanized theorem excludes each
blank cell under the constructor's stated premises. Lean generates the
matrices from those witnesses and a committed golden pins them, so the
table cannot drift from the proofs.}
\label{tab:updates}
\centering
\footnotesize
\begin{tabular}{@{}l cccc @{\hspace{1.6em}} cccc @{\hspace{1.6em}} cccc @{\hspace{1.6em}} cccc@{}}
\toprule
& \multicolumn{4}{c@{\hspace{1.6em}}}{\upAddLeaf}
& \multicolumn{4}{c@{\hspace{1.6em}}}{\upTighten}
& \multicolumn{4}{c@{\hspace{1.6em}}}{\upAddAttack}
& \multicolumn{4}{c}{\upAddInstance} \\
\cmidrule(r{1.6em}){2-5}\cmidrule(r{1.6em}){6-9}\cmidrule(r{1.6em}){10-13}\cmidrule{14-17}
from & J & D & C & G & J & D & C & G & J & D & C & G & J & D & C & G \\
\midrule
J & \cmark &  &  &  & \cmark & \cmark & \cmark & \cmark & \cmark & \cmark & \cmark &  & \cmark &  &  &  \\
D &  & \cmark &  &  & \cmark & \cmark & \cmark & \cmark & \cmark & \cmark & \cmark &  & \cmark & \cmark & \cmark &  \\
C &  &  & \cmark &  & \cmark & \cmark & \cmark & \cmark & \cmark & \cmark & \cmark &  & \cmark &  & \cmark &  \\
G &  &  &  & \cmark &  &  &  & \cmark &  &  &  & \cmark & \cmark & \cmark & \cmark & \cmark \\
\bottomrule
\end{tabular}
\end{table}

%% file: src/appendix.tex
\section{Case-study and worked-example listings}
\label{app:case-study-listings}
\counterwithin{lstlisting}{section}

This appendix collects the longer listings behind the compact case-study
presentation in \appautoref{app:casestudy} and the certified-comparison worked
example of \autoref{sec:evaluation}. The accompanying supplemental artifact
contains the complete runnable programs, versioned policies, replay commands,
and golden reports; the listings below retain the paper-facing elisions noted
in their captions.

To read the programs, a \texttt{claim} names the proposition whose support
we want to assess, a \texttt{leaf} declares evidence, and an \texttt{arg}
applies a policy rule to construct support or a challenge.
The clause \texttt{discharge q with e} supplies evidence for a named
critical question; \texttt{open q} records an unanswered optional question.
Attack declarations name both the challenging argument and the part of
the target argument it disputes. A \texttt{status} query asks what support
survives these attacks under the declared policy.

\subsection{Certified comparison under attack}
\label{app:ord-under-attack}

This example asks what a certified numerical comparison establishes about
an empirical claim. We separate the arithmetic check from the inference
that the reported scores demonstrate a better system. The
\texttt{comparison} block generates both steps; its \texttt{binding}
premise declares that the scores are comparable. The settings audit
challenges that premise, so the attack targets the empirical bridge.
\Lara can therefore preserve the certified inequality while reporting
that the headline claim's support is defeated. The example demonstrates
how checked arithmetic and revisable empirical reasoning coexist in one
program without giving the empirical conclusion the certificate's guarantee.

\begin{lstlisting}[
  language=Lara,
  numbers=left,
  basicstyle=\ttfamily\casestudycodesize,
  caption={The certified comparison under attack, from the worked-example
  suite, authored in an earlier revision of the surface of
  \autoref{sec:language}. One \texttt{comparison}
  block generates the strict re-check \texttt{a1}, the defeasible bridge
  \texttt{a2}, and the arithmetic sub-claim \texttt{c2}; the author names
  only the system pair, measurand, dataset, relation, and evidence leaves.
  Claim prose and leaf metadata are elided except inside the block, whose
  sub-claim prose and attestation the block itself must declare. The
  audit's conclusion is the declared contrary of the binding leaf
  \texttt{e3}, so \texttt{x1} undermines the bridge \texttt{a2} at its
  binding premise; the label \texttt{binding} resolves to the same
  position as \texttt{a2.1.leaf} and never reaches the core. The grounded
  labelling puts \texttt{a1} and \texttt{x1} \emph{in} and \texttt{a2}
  \emph{out}: the headline claim \texttt{c1} is defeated while the
  certified inequality \texttt{c2} stays justified.},
  label={lst:ord-under-attack}
]
artifact ord_setting_demo at sha256:5454545454545454
policy ord-setting-v1
use backends [ord@1]
# the headline claim; the two measured cells below are not disputed
claim c1
  formal  = better(sys_new, sys_base, accuracy, imagenet_val)
leaf e1 : reports(exp1, score_cell(sys_base, accuracy, imagenet_val, 0.71))
leaf e2 : reports(exp1, score_cell(sys_new, accuracy, imagenet_val, 0.74))
# the binding: the only place the artifact asserts the two cells are comparable
leaf e3 : comparison_setup(sys_new, sys_base, accuracy, imagenet_val, 0.74, 0.71)
# the audit's evidence
leaf e4 : audits_settings(aud1, sys_new, sys_base, accuracy, imagenet_val)
# one block generates the strict re-check a1, the bridge a2, and the sub-claim c2
comparison : better(sys_new, sys_base) on accuracy @ imagenet_val
  relation = strictly-better
  recheck  = a1
  bridge   = a2
  result   = e2
  baseline = e1
  binding  = e3
  claims c2
    nl      = "The reported baseline accuracy {cell e1} is strictly below the reported system accuracy {cell e2}"
    binding = { author = alice, audit-status = reviewed }
  supports c1
# the critic: x1's conclusion is the declared contrary of the binding leaf
arg x1 : challenges(e3) by setting_audit(sys_new, sys_base, accuracy, imagenet_val, 0.74, 0.71, aud1)
# a2.binding.leaf names the disputed premise; a2.1.leaf is the same position spelled positionally
undermine x1 a2.binding.leaf
status c1   # defeated: every complete argument for c1 is out
status c2   # justified: a1 is in, untouched by the attack
\end{lstlisting}

\subsection{Rebuttal replay}
\label{app:rebuttal-replay}

Read the rounds as additions to the same dispute under a fixed standard.
The first round records what has support; the review identifies the
premises, conclusions, and inferences being challenged; the rebuttal adds
support and defenses. The three final queries distinguish a restored
benchmark claim, a conceded measurement claim, and a newly supported
ablation claim. \Lara computes these different outcomes from the declared
changes, so a reader can check why each claim's status changed or persisted.

\begin{lstlisting}[
  language=Lara,
  numbers=left,
  basicstyle=\ttfamily\casestudycodesize,
  caption={The rebuttal exchange as three checked \Lara programs over one
  artifact and one policy. Each comment header marks a program boundary;
  rounds 1 and 2 repeat the round-0 paper verbatim, elided here, as are claim
  prose/bindings and leaf metadata. The reviews add one attack of each kind;
  the rebuttal supplies the missing variance evidence and counter-attacks both
  benchmark attackers.},
  label={lst:rebuttal-replay}
]
# ROUND 0 -- submission: the paper alone
artifact adaptive-pruning at sha256:de3685165a53
policy rebuttal-v1
use backends [nd@1]
claim c_bench
  formal  = performs(apt, dense_baseline, openllm_avg)
arg a_bench : supports(c_bench) by benchmark_evaluation(apt, dense_baseline, openllm_avg, e02)
  discharge protocol_fixed    with pb2
  discharge benchmark_valid   with pb3
  open      environment_match as environment_match
# claim c_measure + arg a_meas: measurement, justified (elided)
claim c_kurt
  formal  = contributes(kurtosis_salience, apt_llama2_7b, openllm_avg)
leaf pk1 : reports(e03, ablation_effect(kurtosis_salience, apt_llama2_7b, openllm_avg, positive))
leaf pk2 : isolated(kurtosis_salience, e03)
leaf pk3 : matched_protocol(e03)
# no arg for c_kurt: variance_reported(e03) is mandatory, so the claim is gap
# ROUND 1 -- reviews: paper unchanged, three typed attacks added
leaf ru_np : not_fixed_protocol(apt, dense_baseline, e02)
arg d_um : challenges(pb2) by leaf(ru_np)
undermine d_um a_bench.protocol_fixed.leaf
arg d_rebut : supports(c_bench_neg) by null_benchmark(apt, dense_baseline, openllm_avg, e_rep)
  discharge protocol_fixed    with ru_nb2
  discharge environment_match with ru_nb3
rebut d_rebut a_bench
rebut a_bench d_rebut
leaf ru_ma : measurement_artifact(low_memory_footprint, apt)
arg d_meas : challenges(instrument_validity(a_meas)) by leaf(ru_ma)
undercut d_meas a_meas.rule
# ROUND 2 -- rebuttal: paper + reviews unchanged, author responds
leaf rk_var : variance_reported(e03)
arg a_kurt : supports(c_kurt) by component_ablation(kurtosis_salience, apt_llama2_7b, openllm_avg, e03)
  discharge single_variable   with pk2
  discharge protocol_parity   with pk3
  discharge variance_reported with rk_var
arg r_um : challenges(ru_np) by leaf(rk_pc)
undermine r_um d_um.leaf
arg r_rebut : challenges(environment_match(d_rebut)) by leaf(rk_em)
undercut r_rebut d_rebut.rule
# c_measure not defended (concession) => stays defeated
status c_bench
status c_measure
status c_kurt
\end{lstlisting}

\noindent\begin{minipage}{\linewidth}
\subsection{Mechanical-reviewer output}
\label{app:mechanical-reviewer}

These comments illustrate two different requests an author may need to
address: provide a complete argument, or answer an attack on an existing
argument. The first comment reports an empty support set; the second
names the defeated support and its attacker. \Lara supplies the checked
diagnosis, which a fixed template turns into review prose.

\begin{lstlisting}[
  numbers=left,
  basicstyle=\ttfamily\casestudycodesize,
  literate={—}{{\textemdash{}}}1,
  caption={Two of the mechanical reviewer's 51 rendered comments, verbatim
  from the renderer output except that each claim's natural-language statement
  is elided. The renderer uses the checker's verdict and located diagnostic;
  it performs no free-text generation.},
  label={lst:mechanical-reviews}
]
**51 review comments** — 48 gap, 3 defeated.

## adaptive-pruning.C04 — gap
Claim `c04`: `contributes(kurtosis_salience, apt_llama2_7b, openllm_avg)`
> No argument assembles the unit's evidence into a checked support for this claim; its complete-support set is empty, so at the `corpus-v1` evidential bar the claim is unsupported — status **gap**.

## fre.C04 — defeated
Claim `c04`: `contributes(reward_diversity, fre_all, antmaze_total_score)`
> The claim's only support `a1` is labelled `out`: the undercut `d1` (challenging the `variance_reported` critical question of `a1`) is unattacked and defeats `a1` — status **defeated**.
\end{lstlisting}

\end{minipage}

\subsection{Agreement map}
\label{app:agreement-map}

The final argument of each formal claim identifies its experimental
setting. Within P1, both conclusions concern the same setting and the
declared rebuttals express an unresolved conflict. Within P2, the settings
differ, so the policy does not license the corresponding conflict.
\Lara makes this distinction explicit in the verdicts: opposing wording
alone does not determine whether the encoded claims attack each other.

\begin{lstlisting}[
  language=Lara,
  numbers=left,
  basicstyle=\ttfamily\casestudycodesize,
  caption={The agreement map as one program. Claim leaves, discharges, and
  natural-language/binding fields are elided so that the formal conclusions
  and attack structure remain visible. The single setting-index difference
  flips the pair from contested to justified.},
  label={lst:agreement-map}
]
# P1 -- genuine disagreement: same atoms => contested
claim c_pos
  formal  = better(apt, cofi, accuracy, roberta_mnli_s60)
arg pa : supports(c_pos) by controlled_comparison(apt, cofi, accuracy, roberta_mnli_s60, paperA_exp1)
claim c_neg
  formal  = not_better(apt, cofi, accuracy, roberta_mnli_s60)
arg pb : supports(c_neg) by null_comparison(apt, cofi, accuracy, roberta_mnli_s60, paperB_exp1)
rebut pb pa
rebut pa pb
# P2 -- setting mismatch => no edge
claim c_low
  formal  = better(magnitude_pruning, dense_baseline, accuracy, bert_glue_s50)
arg pc : supports(c_low) by controlled_comparison(magnitude_pruning, dense_baseline, accuracy, bert_glue_s50, paperC_exp1)
claim c_high
  formal  = not_better(magnitude_pruning, dense_baseline, accuracy, llama_openllm_s90)
arg pd : supports(c_high) by null_comparison(magnitude_pruning, dense_baseline, accuracy, llama_openllm_s90, paperD_exp1)
# D differs, so no contrary instance and no edge
status c_pos
status c_neg
status c_low
status c_high
\end{lstlisting}

\subsection{Philosophical assumptions, objections, and defenses}
\label{app:philmath-listing}

The philosophy program makes two obligations visible: why indispensability
warrants belief in abstract objects, and whether a nominalist reconstruction
covers the fragment being discussed. The policy requires explicit answers
to both. The attacks then locate the disagreements: a challenge to the
indispensability premise, a dispute over number identity, and a chain of
objection and reply about usefulness and truth. The final undercut restores
the fictionalist argument by defeating its attacker.

\begin{lstlisting}[
  language=Lara,
  numbers=left,
  basicstyle=\ttfamily\casestudycodesize,
  caption={Selected declarations from the philosophy-of-mathematics
  program \citep{lara2026philmath}. Claim declarations and most leaves are
  omitted; all leaves in the full program are assumed or attested. The
  authored policy makes the commitment and scope questions mandatory.
  The number-identity pair remains contested; the reply reinstates the
  fictionalist argument. These are outcomes of the encoded debate.},
  label={lst:philmath}
]
artifact philmath_debate at sha256:p1-illustrative-snapshot
policy philmath-v1
use backends []

leaf commitment : ontological_commitment()
  kind = assumed
  provenance = ai-executed
  refs = []
leaf scope : scope_adequate()
  kind = assumed
  provenance = ai-executed
  refs = []

arg a_realism : supports(c_realism) by indispensability from [indispensability_premise]
  discharge commitment with commitment
arg a_field : supports(c_field) by nominalization from [field]
  discharge scope with scope
arg a_sets : supports(c_sets) by set_identification from [set_proposal]
arg a_structure : supports(c_structure) by structural_explanation from [benacerraf]
arg a_fiction : supports(c_fiction) by fictionalist_reading from [conservativity]
arg a_objection : supports(c_objection) by truth_objection from [truth_assumption]
arg a_reply : supports(c_reply) by leaf(reply)

undermine a_field a_realism.0.leaf
rebut a_structure a_sets
rebut a_sets a_structure
undercut a_fiction a_realism.rule
undercut a_objection a_fiction.rule
undercut a_reply a_objection.rule

status c_realism
status c_sets
status c_structure
status c_fiction
\end{lstlisting}

\subsection{A certificate that depends on an admitted assumption}
\label{app:axiom-withdrawal-listing}

The source below exposes the assumption used by its certificate. The
strict \texttt{citation} rule has that atom as both premise and conclusion,
and its backend theory is empty. The certificate \texttt{(hyp 0)} therefore
checks reuse of the declared premise. The three policy variants admit,
quarantine, or reject leaves with the pair
\texttt{(assumed, ai-executed)}. Quarantine removes the only premise and
its dependent argument, so the surviving claim has status \stGap.
The separate Lean witness proves why a structural bridge cannot preserve
that premise in the empty target evidence environment
(\appautoref{app:cs-axiom-withdrawal}).

\begin{lstlisting}[
  language=Lara,
  numbers=left,
  basicstyle=\ttfamily\casestudycodesize,
  caption={The admitted source of the axiom-withdrawal demo
  \citep{lara2026axiom}, with its introductory comment omitted.
  The withdrawn and rejected sources select their respective policies;
  the leaf, argument, and claim declarations are identical. The certificate
  establishes hypothesis reuse, not a geometric theorem or unconditional
  truth of the postulate.},
  label={lst:axiom-withdrawal}
]
artifact axiom_withdrawal at sha256:axiom-demo
policy admitted
use backends [nd@1]

claim c
  nl = "The parallel postulate is available under this admission policy"
  formal = parallel_postulate()
  binding = { author = demo_author, audit-status = unreviewed,
              rationale = "An assumed atom reused as itself, not a geometric consequence." }

leaf pp : parallel_postulate()
  kind = assumed
  provenance = ai-executed
  refs = []

arg a : supports(c) by citation from [pp]
  assurance = cert(nd@1, sha256:empty, (hyp 0))

status c
\end{lstlisting}